\documentclass[useAMS,usenatbib,usegraphicx]{mn2e}

\usepackage{graphicx}
\usepackage{floatflt}
\usepackage{multirow}
\usepackage[dvips]{color}

\protect\definecolor{green4}{rgb}{0.00,0.55,0.00}
\newcommand{\rev}[1]{#1}
\newcommand{\ionspeciesgap}{\ensuremath{\,}}
\newcommand{\fion}[2]{\ensuremath{[\rmn{#1} \ionspeciesgap \mbox{\textsc{#2}}]}}
\newcommand{\ha}{\ensuremath{\rmn{H} \alpha}}
\newcommand{\hab}{\ensuremath{\rmn{H} \alpha_B}}
\newcommand{\han}{\ensuremath{\rmn{H} \alpha_N}}
\newcommand{\hatt}{\ensuremath{\rmn{H} \alpha_T}}
\newcommand{\hb}{\ensuremath{\rmn{H} \beta}}
\newcommand{\nii}{\fion{N}{ii}}
\newcommand{\oi}{\fion{O}{i}}
\newcommand{\oiii}{\fion{O}{iii}}
\newcommand{\oiiic}{\ensuremath{\fion{O}{iii}_C}}
\newcommand{\oiiiw}{\ensuremath{\fion{O}{iii}_W}}
\newcommand{\oiiit}{\ensuremath{\fion{O}{iii}_T}}
\newcommand{\sii}{\fion{S}{ii}}
\newcommand{\fevii}{\fion{Fe}{vii}}
\newcommand{\feviic}{\ensuremath{\fion{Fe}{vii}_C}}
\newcommand{\feviiw}{\ensuremath{\fion{Fe}{vii}_W}}
\newcommand{\feviit}{\ensuremath{\fion{Fe}{vii}_T}}
\newcommand{\feviig}{\ensuremath{\fion{Fe}{vii}_{1G}}}
\newcommand{\fex}{\fion{Fe}{x}}
\newcommand{\fexi}{\fion{Fe}{xi}}
\newcommand{\scinot}[2]{\ensuremath{#1 \times 10^{#2}}}
\newcommand{\kms}{\ensuremath{\rmn{km \, s^{-1}}}}
\newcommand{\cgsflux}{\ensuremath{\mathrm{erg \, cm^{-2} \, s^{-1}}}}
\newcommand{\nodata}{---}
\newcommand{\critdens}{\ensuremath{N_{cr}}}
\newcommand{\gsimeq}
  {\hbox{\raise0.5ex\hbox{$>\lower1.06ex\hbox{$\kern-1.07em{\sim}$}$}}}
\title[FHIL-emitting AGN from SDSS]{AGN with strong forbidden
    high-ionisation lines selected from the Sloan Digital Sky Survey} 
\author[J. M. Gelbord, J. R. Mullaney and M. J. Ward]{
  Jonathan M. Gelbord$^{1}$\thanks{E-mail:
    j.m.gelbord\protect\linebreak[1]@\protect\linebreak[1]durham\protect\linebreak[0].ac.uk (JMG);
    j.r.mullaney\protect\linebreak[1]@\protect\linebreak[1]durham\protect\linebreak[0].ac.uk (JRM);
    m.ward\protect\linebreak[1]@\protect\linebreak[1]durham\protect\linebreak[0].ac.uk (MJW)},
  James R. Mullaney$^{1}$\footnotemark[1] and Martin J. Ward$^{1}$\footnotemark[1]\\
$^{1}$Department of Physics, Durham University, South Road, Durham, DH1 3LE, U.K.}
\begin{document}

\date{Accepted 2009 xxxxx xx.  Received 2009 April 20; in original form 2008 October 8}

\pagerange{\pageref{firstpage}--\pageref{lastpage}} \pubyear{2009}

\maketitle

\label{firstpage}

\begin{abstract}
We have defined a sample of 63 AGN with strong forbidden 
high-ionisation line (FHIL) emission. 
These lines, with ionisation potentials $\gsimeq 100$~eV, respond to 
a portion of the spectrum that is often difficult to observe directly, 
thereby providing constraints on the extreme UV--soft X-ray continuum. 
The sources are selected from the Sloan Digital Sky Survey (SDSS) on 
the basis of their \fex$\lambda$6374\AA\ emission, yielding one of the 
largest and the most homogeneous sample of FHIL-emitting galaxies. 
We fit a sequence of models to both FHILs (\fexi, \fex\ and \fevii) and 
lower-ionisation emission lines (\oiii, \oi, \ha, \nii, \sii) 
in the SDSS spectra. These data are combined with X-ray measurements
from \textit{Rosat}, which are available for half of the 
sample. The correlations between these parameters are discussed for 
both the overall sample and subsets defined by spectroscopic
classifications. 
The primary results are evidence that: 
(1) the \fex\ and \fexi\ lines are photoionised and their 
strength is proportional to the continuum flux around 250~eV; 
(2) the FHIL-emitting clouds form a stratified outflow in which 
the \fex\ and \fexi\ source regions extend sufficiently close to the 
BLR that they are partially obscured in Seyfert 2s whereas the \fevii\ 
source region is more extended and is unaffected by obscuration; 
(3) narrow-lined Seyfert 1s (NLS1s) tend to have the strongest \fex\ 
flux (relative to lower-ionisation lines); 
and (4) the most extreme \fex\ ratios (such as \fex/\oiii\ or 
\fex/\fevii) 
are found in the NLS1s with the 
narrowest broad lines and appear to be an optical-band indication of 
objects with strong X-ray soft excesses. 

\end{abstract}

\begin{keywords}
galaxies: active 
-- galaxies: Seyfert 
-- quasars: emission lines
-- line: profiles
-- X-rays: galaxies
-- galaxies: individual (KUG 1031+398, RBS 1249, SDSS J124134.25+442639.2)
\end{keywords}

\section{Introduction}
\label{sect:intro}

The forbidden high-ionisation emission lines (FHILs) have been known to
exist in the optical spectra of Seyfert galaxies for more than 40
years (see \citealp{Oke68} for the earliest published discussion of
the \fex\ species in NGC4151). 
Historically the \fex\ line and other highly ionised species
have been referred to in the literature as ``coronal lines'' because
these species were first identified in the spectra of the solar
corona.  However, in this paper we will refer to them in terms of
a specific physical property, i.e.\ as FHILs.  
We include in our definition of FHILs any forbidden line having
an ionisation potential (IP) $\gsimeq 100$~eV.

These lines are of interest for a variety of reasons.  They
define the nature of a galaxy in which they are detected as an AGN.
This is because stellar spectra do not have sufficient high energy
photons ($> 100 \ \rmn{eV}$) to produce such species at detectable
levels.  Although SNRs can exhibit FHILs, these are
distinguishable from those in AGN based on other properties, such as
their profiles and velocity shifts with respect to other emission
lines of lower ionisation.  In AGN the FHILs tend to be broader and sometimes
blueshifted compared to the profiles of the other emission lines
\citep*{Appenzeller88,Erkens97}.  Despite being well studied
some fundamental questions are still not settled, such as whether the
FHIL-emitting regions are powered by photoionisation or collisional
processes.  Also, the location of
the zone in which the bulk of their line flux is emitted is not well
established, although in some cases it has been possible to spatially
resolve the region (see \citealp{Rodriguez06} for a recent study).
Their velocity widths imply a possible origin in a region contiguous
with the innermost zones of the classic narrow line region (NLR), as
defined by \oiii$\lambda$5007.  This region can be spatially resolved
be to about 10 parsecs for some nearby AGN
\citep*[e.g.,][]{Kraemer08}, and extends out to 100's of parsecs and
even kiloparsecs in some cases \citep[e.g.,][]{Bennert06b,Bennert06a}.
By comparison the broad line region (BLR) is spatially unresolved, but
is inferred from emission line reverberation mapping to be less than
about a light month for a Seyfert of typical luminosity
\citep{Peterson04}.

On size scales between the BLR and the innermost resolved NLR lies the
putative dusty molecular torus.  This is believed to be a parsec-scale
structure, with an inner edge that depends on the dust sublimation
radius and may lie just beyond the BLR \citep{Barvainis87,Suganuma06}.
Based on these generic components it has been suggested that the
innermost FHIL-emitting regions lie somewhere between just beyond the
BLR and the dusty torus.
In addition to these qualitative considerations, the physical
conditions of the FHILs, such as the density and temperature of the
emitting gas derived from modeling their line flux ratios, has led to
claims of an association with the so-called X-ray warm absorber
\citep{Porquet99}.  This is an X-ray component very commonly
identified in the low energy X-ray spectra of Seyfert 1s
\citep{Blustin05}.  A consequence of these and other properties
is that FHILs
may be employed as diagnostics of outflows from the inner
regions of the AGN, and may be a useful ingredient of wind models
\citep{Rodriguez06,Mullaney09}.

Unfortunately a current limitation in the study of FHILs and their
potential use as diagnostics of the inner line emitting regions of AGN
is that the samples available are quite small and 
heterogeneous, being limited to bright, relatively nearby
AGN (\citealp{DeRobertis84b,Veilleux88,Erkens97,Murayama98},
hereafter MT98; \citealp*{Nagao00}, hereafter NTM00).
In this paper we introduce a new sample of significant size,
constructed using the FHILs themselves as a selection criterion, the
first time this technique has been employed.
This larger sample will enable meaningful statistical
tests to be performed on the FHILs and other properties of the AGN,
and also to consider the relevance of AGN classifications such as
Seyfert 1, Seyfert 2, and narrow line Seyfert 1 (Sy1, Sy2, and NLS1,
respectively).

The sample selection is presented in \S\ref{sect:sample}.
In \S\ref{sect:analysis} we describe the models and the fitting
procedure applied to the optical data, and how we interpret both these
models and X-ray count rates from \textit{Rosat}.
The results of these analyses are characterized in
\S\ref{sect:results}.
In \S\ref{sect:discussion} we consider the implications of these
results.
Our findings are summarized in \S\ref{sect:conclusions}.

We adopt cosmological parameters from WMAP
($H_\circ = 71 \  \kms \, \rmn{Mpc}^{-1}$, 
$\Omega_\rmn{M} = 0.27$ and $\Omega_\Lambda = 0.73$; 
\citealp{Spergel03}).
The wavelengths reported throughout this paper are those observed in air, 
not the vacuum wavelengths reported in the Sloan Digital Sky Survey
(SDSS; \citealp{SDSS}) catalog.
This is for consistency with the vast majority of previous literature
on this subject.
Reported errors for fitted parameters are $\pm 1 \sigma$ confidence
intervals; when average properties of samples are discussed, we
generally provide both the $\pm 1 \sigma$ confidence interval of the
mean value and the RMS scatter about the mean.

\section{Sample selection}
\label{sect:sample}
Our objective is to construct a sample of galaxies selected based on
their strong FHIL emission. 
We do not make the a priori assumption that such
emission will only be found in AGN.
Because FHIL emission is relatively rare, we require a large
spectroscopic survey from which to draw our sample.  
For this the SDSS is
ideally suited, in that the current release (DR6, \citealp{SDSS-DR6})
provides uniform quality spectra for nearly 900,000 galaxies and AGN
spanning more than 2 steradians (6860 square degrees).

The SDSS data processing pipeline \citep{SDSS-EDR} fits simple
Gaussian models to a set of features in each spectrum,
yielding a database of line parametrizations that
is readily searched.  
Unfortunately this database does not include any
optical FHILs.  However, it does include both lines of
the $\oi \lambda \lambda 6300,6364$ doublet.
$\oi \lambda 6364$ is close enough to $\fex
\lambda 6374$ that the SDSS line model will represent a blend of
these two features if this FHIL is present.
In such cases the pipeline models of the \oi\ lines will
neither share the same profile nor have a 3:1 flux ratio, 
as expected from atomic physics.
Thus \fex-emitting galaxies
may be found by identifying SDSS spectra whose
$\oi \lambda 6364$ models show evidence of
contamination.

We construct a sample of candidate \fex-emitting galaxies by applying
the following selection criteria to the SDSS database:
\begin{enumerate}
\item{identified as either a galaxy or a quasar by the SDSS
  spectroscopic pipeline (SPECCLASS = 2 or 3, respectively);}
\item{redshift $< 0.40$ to ensure that \ha\ lies within the
  observed 3800--9200 \AA\ wavelength range;}
\item{\ha\ Gaussian model width $\sigma < 50$~\AA\ to minimize
  the contamination of \oi\ by any broad H$\alpha$ wings\footnote{We note
    that the single-Gaussian model used by the SDSS pipeline may not
    be fitted to the broadest \ha\ component, so contamination at \oi\
    is still a possibility.};}
\item{\oi\ doublet models with significance greater than 10-$\sigma$
  in each line to ensure well-defined profiles;}
\item{$\oi \lambda 6364$ line centroids shifted by $\geq 1.5$~\AA\ 
  relative to $\oi \lambda 6300$ (in the emitted frame; equivalent to
  $> 1.0$ SDSS spectral bin), 
  which is indicative of \fex\ blends\footnote{The $\oi \lambda 6364$
    model centroid shift was used
    instead of criteria based upon either the line flux ratios or FWHM
    differences because it was found to be more robust.  The
    alternative tests were more likely to produce a false positive
    in response to relatively low S/N data, local inaccuracies in the
    SDSS continuum models, or \ha\ contamination.}.}
\end{enumerate}
%
%
The resulting $\sim$200 candidates were then visually inspected 
to identify the highest-quality sources
and
to eliminate spectra with obvious contamination around
6370~\AA\ (sky subtraction artefacts or 
any unmodeled \ha\ components
that introduce steep gradients).
Sources were selected for the final sample that have one or more
pieces of unambiguous corroborating evidence of FHIL emission: 
(1) a distinct emission feature near 6374~\AA, 
(2) $\oi + \fex$ blend significantly broader than $\oi \lambda 6300$,
(3) \oi\ doublet ratio inconsistent with 3, and/or 
(4) prominent $\fevii \lambda 6087$ emission.

\begin{table*}
\caption{SDSS galaxies selected by \fex\ emission (see \S\ref{sect:sample} for
    column descriptions).} 
\label{tab:sample}
{ \centering
\begin{tabular}{rc@{\ \ \ }c@{\ \ \ }ccc@{\ \ \ }rr@{$\pm$}lr@{$\pm$}lr@{$\pm$}ll}
\hline
ID	& RA $\pm$ Dec (J2000.0)	& SC	& Class	& r 	&$z_\rmn{\sii}$	& $v_\rmn{corr}$
                                                                        		& \multicolumn{2}{c}{$\Delta \lambda_{\oi 6364}$}
													& \multicolumn{2}{c}{$\Delta \rmn{FWHM}$}
															& \multicolumn{2}{c}{\oi\ ratio}
		                                                                                                                	& Alternative IDs	\\
\hline
1	& 00:18:52.47+01:07:58.5	& 2 	& S1.9	& 18.10	& 0.0640	& -47	& 7.	& 2.	& 12.	& 4.	& 0.70	& 0.09	& 	\\
2	& 01:10:09.01--10:08:43.4	& 3 	& NLS1	& 17.60	& 0.0583	& -66	& 3.	& 3.	& 15.	& 4.	& 0.57	& 0.07	& 1WGA J0110.1-1008	\\
3	& 02:33:01.24+00:25:15.0	& 2 	& S2	& 15.39	& 0.0224	& 19	& 4.3	& 0.4	& 2.9	& 1.0	& 0.80	& 0.08	& 1WGA J0232.9+0025	\\
4	& 07:31:26.69+45:22:17.5	& 3 	& S1.5	& 17.84	& 0.0921	& -4	& 3.	& 4.	& 11.	& 6.	& 1.06	& 0.13	& 	\\
5	& 07:36:38.86+43:53:16.5	& 2 	& S2	& 17.96	& 0.1140	& 37	& 3.5	& 1.3	& 8.	& 3.	& 1.55	& 0.17	& 	\\
6	& 07:36:50.08+39:19:55.2	& 2 	& S2	& 19.12	& 0.1163	& -13	& 5.8	& 1.8	& 13.	& 4.	& 0.98	& 0.10	& 	\\
7	& 08:07:07.18+36:14:00.5	& 2 	& S2	& 16.23	& 0.0324	& -21	& 8.5	& 1.0	& 19.0	& 1.6	& 0.25	& 0.03	& IC 2227	\\
8	& 08:11:53.16+41:48:20.0	& 2 	& S2	& 18.35	& 0.0999	& -96	& 5.6	& 0.7	& 1.6	& 1.9	& 1.45	& 0.14	& 	\\
9	& 08:29:30.59+08:12:38.1	& 3 	& S1.0	& 17.24	& 0.1295	& -95	& 6.5	& 1.0	& 15.	& 3.	& 0.54	& 0.05	& 	\\
10	& 08:30:45.37+34:05:32.1	& 3 	& S1.5	& 16.66	& 0.0624	& -39	& 5.0	& 0.9	& 13.	& 2.	& 0.89	& 0.09	& 	\\
11	& 08:30:45.41+45:02:35.9	& 3 	& S1.0	& 17.84	& 0.1825	& -49	& 3.	& 6.	& 16.	& 5.	& 0.22	& 0.02	& 	\\
12	& 08:36:58.91+44:26:02.4	& 3 	& S1.0	& 15.71	& 0.2544	& 15	& 2.0	& 1.2	& 10.6	& 1.3	& 0.60	& 0.05	& Q 0833+446; RBS 711	\\
13	& 08:42:15.30+40:25:33.3	& 2 	& S2	& 16.85	& 0.0553	& 2	& 8.	& 2.	& 15.	& 3.	& 1.05	& 0.12	& 	\\
14	& 08:46:22.54+03:13:22.2	& 3 	& S1.0	& 17.48	& 0.1070	& -22	& 8.8	& 0.5	& 7.9	& 1.2	& 0.67	& 0.06	& 	\\
15	& 08:53:32.22+21:05:33.7	& 2 	& S2	& 17.73	& 0.0719	& -32	& 6.2	& 1.1	& 7.	& 2.	& 1.07	& 0.11	& 	\\
16	& 08:57:40.86+35:03:21.7	& 3 	& S1.5	& 19.25	& 0.2752	& 46	& 3.4	& 1.6	& 10.8	& 1.6	& 0.77	& 0.08	& 	\\
17	& 08:58:10.64+31:21:36.3	& 2 	& S2	& 18.23	& 0.1389	& -15	& 6.1	& 1.5	& 13.4	& 1.7	& 1.03	& 0.10	& 	\\
18	& 09:17:15.00+28:08:28.2	& 3 	& S1.5	& 18.08	& 0.1045	& -37	& 10.0	& 1.2	& 19.	& 9.	& 0.35	& 0.04	& 	\\
19	& 09:18:25.79+00:50:58.4	& 3 	& S1.5	& 18.16	& 0.0871	& -43	& 6.6	& 0.7	& 11.	& 2.	& 0.79	& 0.09	& 	\\
20	& 09:23:43.00+22:54:32.6	& 3 	& NLS1	& 15.65	& 0.0330	& 51	& 8.2	& 0.6	& 16.8	& 0.9	& 0.37	& 0.03	& MCG +04-22-42	\\
21	& 09:42:04.79+23:41:06.9	& 3 	& S1.5	& 15.92	& 0.0215	& -49	& 7.0	& 0.7	& 4.5	& 1.8	& 1.19	& 0.14	& PGC 027720	\\
22	& 10:01:49.52+28:47:09.0	& 2 	& S1.9	& 17.32	& 0.1849	& 17	& 6.3	& 0.9	& 7.	& 2.	& 1.22	& 0.08	& 3C 234	\\
23	& 10:17:18.26+29:14:34.1	& 3 	& S1.5	& 16.71	& 0.0492	& -66	& 3.8	& 0.7	& 13.	& 2.	& 0.38	& 0.05	& 	\\
24	& 10:22:35.15+02:29:30.5	& 2 	& NLS1	& 18.71	& 0.0701	& -84	& 8.	& 5.	& 8.	& 9.	& 0.76	& 0.09	& 	\\
25	& 10:34:38.60+39:38:28.3	& 2 	& NLS1	& 16.80	& 0.0435	& -107	& 3.3	& 0.4	& 6.6	& 1.2	& 0.37	& 0.02	& KUG 1031+398	\\
26	& 10:55:19.54+40:27:17.5	& 3 	& S1.5	& 17.49	& 0.1201	& -27	& 2.3	& 1.0	& 16.	& 2.	& 0.90	& 0.09	& Mrk 1269	\\
27	& 11:02:43.20+38:51:52.6	& 2 	& NLS1	& 18.89	& 0.1186	& -19	& 8.3	& 1.3	& 18.8	& 1.7	& 0.84	& 0.09	& 1WGA J1102.7+3851	\\
28	& 11:07:04.52+32:06:30.0	& 3 	& S1.5	& 17.64	& 0.2425	& 104	& 2.	& 3.	& 2.9	& 1.3	& 1.25	& 0.06	& 	\\
29	& 11:07:16.49+13:18:29.5	& 3 	& S1.5	& 18.35	& 0.1848	& -44	& 3.5	& 0.7	& 10.	& 2.	& 0.56	& 0.07	& 	\\
30	& 11:07:56.55+47:44:34.8	& 3 	& S1.0	& 16.98	& 0.0727	& 3	& 5.	& 5.	& 16.	& 2.	& 0.40	& 0.05	& 	\\
31	& 11:09:29.10+28:41:29.2	& 2 	& S2	& 17.20	& 0.0329	& -77	& 5.	& 3.	& 18.	& 11.	& 0.55	& 0.05	& 	\\
32	& 11:26:02.46+34:34:48.2	& 2 	& S1.9	& 17.92	& 0.1114	& -78	& 1.9	& 0.7	& 4.1	& 1.9	& 1.18	& 0.14	& 	\\
33	& 11:31:07.10+11:58:59.3	& 3 	& S1.0	& 17.58	& 0.0910	& 2	& 4.3	& 0.9	& 10.	& 3.	& 0.78	& 0.09	& 	\\
34	& 11:39:17.17+28:39:46.9	& 2 	& S2	& 17.05	& 0.0234	& -11	& 9.5	& 1.5	& 1.	& 4.	& 0.81	& 0.09	& 	\\
35	& 11:42:16.88+14:03:59.7	& 2 	& S2	& 16.44	& 0.0208	& -48	& 6.4	& 0.2	& 6.9	& 0.5	& 0.95	& 0.04	& 	\\
36	& 11:52:26.30+15:17:27.6	& 3 	& S1.5	& 17.91	& 0.1126	& 2	& 2.	& 2.	& 12.	& 3.	& 1.15	& 0.12	& 	\\
37	& 11:57:04.84+52:49:03.7	& 2 	& S2	& 16.78	& 0.0356	& -39	& 6.3	& 0.8	& 12.1	& 1.7	& 1.32	& 0.10	& 	\\
38	& 12:04:22.15--01:22:03.3	& 3 	& S1.0	& 17.58	& 0.0834	& 2	& 6.	& 4.	& 17.	& 9.	& 0.53	& 0.07	& 	\\
39	& 12:07:35.06--00:15:50.3	& 2 	& S2	& 18.84	& 0.1104	& -28	& 2.6	& 1.9	& 7.	& 4.	& 1.46	& 0.17	& 	\\
40	& 12:09:32.94+32:24:29.3	& 3 	& NLS1	& 17.94	& 0.1303	& 97	& 3.5	& 0.8	& 11.	& 2.	& 1.32	& 0.13	& 	\\
41	& 12:10:44.28+38:20:10.3	& 3 	& S1.0	& 15.63	& 0.0230	& -43	& 10.0	& 1.4	& 15.	& 2.	& 0.82	& 0.08	& KUG 1208+386	\\
42	& 12:29:03.50+29:46:46.1	& 2 	& S1.5	& 18.45	& 0.0821	& -166	& 3.7	& 1.3	& 6.	& 3.	& 1.02	& 0.12	& 	\\
43	& 12:29:30.41+38:46:20.7	& 2 	& S2	& 18.20	& 0.1024	& -39	& 2.4	& 0.6	& 0.5	& 1.6	& 1.52	& 0.16	& 	\\
44	& 12:31:49.08+39:05:30.2	& 3 	& S1.5	& 17.50	& 0.0683	& -39	& 6.	& 6.	& 5.	& 7.	& 0.86	& 0.10	& 	\\
45	& 12:41:34.25+44:26:39.2	& 2 	& gal	& 18.87	& 0.0419	& 18	& 10.	& 5.	& 1.	& 10.	& 0.33	& 0.03	& 	\\
46	& 13:11:35.66+14:24:47.2	& 3 	& NLS1	& 17.03	& 0.1140	& -61	& 1.7	& 0.8	& 17.8	& 0.7	& 0.32	& 0.03	& 	\\
47	& 13:13:05.69--02:10:39.3	& 3 	& S1.0	& 16.70	& 0.0838	& -20	& 3.	& 3.	& 17.5	& 1.6	& 0.38	& 0.04	& 	\\
48	& 13:13:48.96+36:53:58.0	& 3 	& S1.5	& 17.59	& 0.0670	& -66	& 6.4	& 0.7	& 10.	& 2.	& 0.78	& 0.08	& 	\\
49	& 13:16:39.75+44:52:35.1	& 2 	& S2	& 17.37	& 0.0911	& -55	& 2.9	& 0.6	& 10.1	& 1.9	& 1.03	& 0.06	& 1WGA J1316.6+4452	\\
50	& 13:19:57.07+52:35:33.8	& 2 	& NLS1	& 17.92	& 0.0922	& -69	& 6.0	& 0.8	& 15.4	& 1.9	& 0.48	& 0.03	& RBS 1249	\\
51	& 13:23:46.00+61:04:00.2	& 2 	& S2	& 17.58	& 0.0715	& -36	& 2.5	& 0.8	& 4.7	& 1.0	& 2.37	& 0.13	& 	\\
52	& 13:46:07.71+33:22:10.8	& 3 	& S1.0	& 17.45	& 0.0838	& -44	& 5.	& 5.	& 17.	& 4.	& 0.60	& 0.07	& 	\\
53	& 13:55:42.76+64:40:45.0	& 3 	& NLS1	& 16.73	& 0.0753	& -42	& 3.	& 2.	& 15.	& 4.	& 0.48	& 0.02	& VII Zw 533	\\
54	& 14:34:52.46+48:39:42.8	& 3 	& S1.0	& 15.93	& 0.0365	& -8	& 6.	& 3.	& 17.3	& 1.5	& 0.60	& 0.06	& NGC 5683; Mrk 474	\\
55	& 15:32:22.32+23:33:25.0	& 2 	& S2	& 17.46	& 0.0465	& -50	& 8.	& 2.	& 14.	& 2.	& 1.37	& 0.14	& 	\\
56	& 15:35:52.40+57:54:09.5	& 3 	& S1.0	& 15.21	& 0.0304	& 15	& 4.7	& 1.0	& 17.1	& 1.1	& 0.62	& 0.06	& Mrk 290	\\
57	& 16:09:48.21+04:34:52.9	& 2 	& S2	& 17.77	& 0.0643	& -91	& 6.7	& 0.9	& 5.	& 2.	& 1.15	& 0.07	& 	\\
58	& 16:13:01.63+37:17:14.9	& 3 	& S1.0	& 16.46	& 0.0695	& 12	& 6.1	& 1.6	& 16.0	& 1.3	& 0.53	& 0.05	& KUG 1611+374B	\\
59	& 16:18:44.85+25:39:07.7	& 3 	& NLS1	& 16.94	& 0.0479	& 9	& 8.	& 3.	& 18.	& 7.	& 0.36	& 0.04	& 	\\
60	& 16:35:01.46+30:54:12.1	& 3 	& S1.5	& 17.33	& 0.0543	& -1	& 9.1	& 1.6	& 16.	& 4.	& 0.43	& 0.04	& 	\\
61	& 20:58:22.14--06:50:04.4	& 3 	& NLS1	& 18.22	& 0.0740	& -58	& 4.8	& 0.7	& 8.1	& 1.7	& 0.86	& 0.10	& 	\\
62	& 22:02:33.85--07:32:25.0	& 3 	& NLS1	& 17.05	& 0.0594	& -6	& 4.0	& 0.6	& 11.3	& 1.6	& 0.76	& 0.06	& 	\\
63	& 22:15:42.30--00:36:09.8	& 3 	& S1.0	& 17.30	& 0.0994	& -30	& 3.7	& 1.0	& 11.	& 2.	& 0.98	& 0.10	& 	\\
64	& 23:56:54.30--10:16:05.5	& 3 	& S1.5	& 16.59	& 0.0740	& -77	& 2.2	& 0.3	& 14.6	& 0.9	& 0.42	& 0.02	& 	\\
\hline
\end{tabular}
}
\end{table*}

The final sample, consisting of 64 strong \fex-emitting galaxies, is
presented in 
Table~\ref{tab:sample}.
The columns are: 
(1) ID number;
(2) J2000.0 coordinates;
(3) SDSS SPECCLASS value (``SC''; 2 = galaxy, 3 = quasar);
(4) our spectral classification (see \S\ref{sect:specTypes});
(5) $r$ magnitude integrated over a 3-arcsec diameter region (SDSS
fiber magnitude); 
(6) redshift defined from the observed wavelengths of the
\sii\ doublet (see \S\ref{sect:redshiftRedef}); 
(7) correction to recessional velocity due to difference between SDSS-
and \sii-defined $z$ (\kms; $\Delta v_\rmn{corr} < 0$ if $z_\rmn{SDSS}
< z_\rmn{\sii}$); 
(8--10) SDSS \oi\ model parameters that provide evidence of blended
\fex: 
(8) offset of $\oi \lambda 6364$ model centroid from expected position
($\Delta \lambda_{\oi 6364} = \lambda_{\oi 6364} - \lambda_{\oi 6300}
- 63.472$~\AA, where wavelengths are measured in the emitted
frame in \AA);
(9) difference in widths of \oi\ doublet line models ($\Delta
\rmn{FWHM} = \rmn{FWHM}_{6364} - \rmn{FWHM}_{6300}$, in \AA), and 
(10) flux ratio of $\oi \lambda 6300 / \oi \lambda 6364$; and
(11) alternative common names (including WGA catalogue IDs for the four
sources detected in pointed \textit{Rosat} observations but not in the
\textit{Rosat} All-Sky Survey).  
%
These represent a diverse range of Seyfert objects:
26 type~1s (including 12 NLS1s), 
19 intermediate types (16 of type 1.5 and 3 of type 1.9), and 
18 type~2 Seyferts (our working definitions for these
spectral types are described in \S\ref{sect:specTypes}).  
In addition we found one galaxy with an unusual spectrum that is not
Seyfert-like.  
It is listed in Table~\ref{tab:sample} (object 45, spectral type
``gal'') but is not included in the rest of this paper; 
it is discussed in detail in a separate study 
(Ward et~al.\ in preparation). 
We note that 27 out of these 64 objects are spectroscopically
classified by the SDSS pipeline as galaxies and not quasars: the one
non-Seyfert galaxy, all of the Sy2 and Sy1.9 sources, plus one Sy1.5 
(source 42) and four NLS1s (sources 24, 25, 27, and 50).

This is one of the largest and by far the most homogeneous
sample of AGN with strong FHILs to date.
However, we note that this sample is by no means complete.  
There are certainly many other galaxies with strong \fex\ emission in
the SDSS catalogue that have been excluded by the \oi\ significance
or the \ha\ width criteria.
Our selection criteria introduce biases against Seyfert~1s with the
broadest permitted lines and any FHIL-emitting sources dominated by
lines with lower IPs (e.g., with low $\fex / \fevii$
ratios).
This is in addition to any biases inherited from the SDSS
spectroscopic survey (e.g., the photometric criteria used by SDSS to
select spectroscopic targets coupled with the fact that we directly
observe the AGN continuum flux in Sy1s and not in Sy2s
means that SDSS-selected type 1 objects will extend to higher
redshifts and may include less intrinsically powerful nuclei).
Nevertheless, the size and relative homogeneity of this sample makes
it useful for testing correlations between FHIL features and other
properties.

\section{Analysis}
\label{sect:analysis}

In this paper we focus our attention on the strongest FHILs and the
most prominent other lines available for low redshift objects within
the SDSS spectral band.  In the wavelength range $3500 < \lambda <
9000$~\AA, the strongest observed lines with IPs $\gsimeq 100$~eV are
all species of iron: $\fevii \lambda 6087$, $\fex \lambda 6374$ and
$\fexi \lambda 7892$.  Other FHILs in this range include
$\fion{Fe}{xiv} \lambda 5303$, $\fexi \lambda 3987$, and several
\fevii\ lines (notably ones at 5721, 5276, 5159, 4893, 3967 and 3586
\AA; see, e.g., \citealp{Osterbrock81,Appenzeller88,Erkens97}), but we
do not measure these because they are not as strong, some are blended
with other features,
and the shorter wavelength lines will be more strongly affected by any
dust that is present.  In addition to the three strongest iron FHILs,
we also measure \ha\ and the following lower-ionization forbidden
lines: $\oiii \lambda 5007$, $\oi \lambda \lambda 6300,6364$, $\nii
\lambda \lambda 6548,6583$, and $\sii \lambda \lambda 6718,6730$.

\subsection{Spectral fitting procedure}
\label{sect:routine}

We designed a set of \textsc{idl} scripts to fit a series of models to
the SDSS spectra.
In order to minimize the complexity of these models, we made separate
analyses of six bands around \oiii, \fevii, \oi+\fex, \ha+\nii,
\sii\ and \fexi.
In each interval we first fit a low-order polynomial to establish the
local continuum.
This may include both true continuum and other
broad components, notably the wings of any broad \ha\ components that
sometimes extend to wavelengths near \fex\ and \sii.
We therefore used a third-order polynomial in
these two bands to allow for some curvature in the continuum model and
a second-order polynomial in the other bands.
This local continuum is subtracted from the data before applying a
sequence of emission line models.

\begin{figure}
  \centering
    \resizebox{8.4cm}{!}{\includegraphics{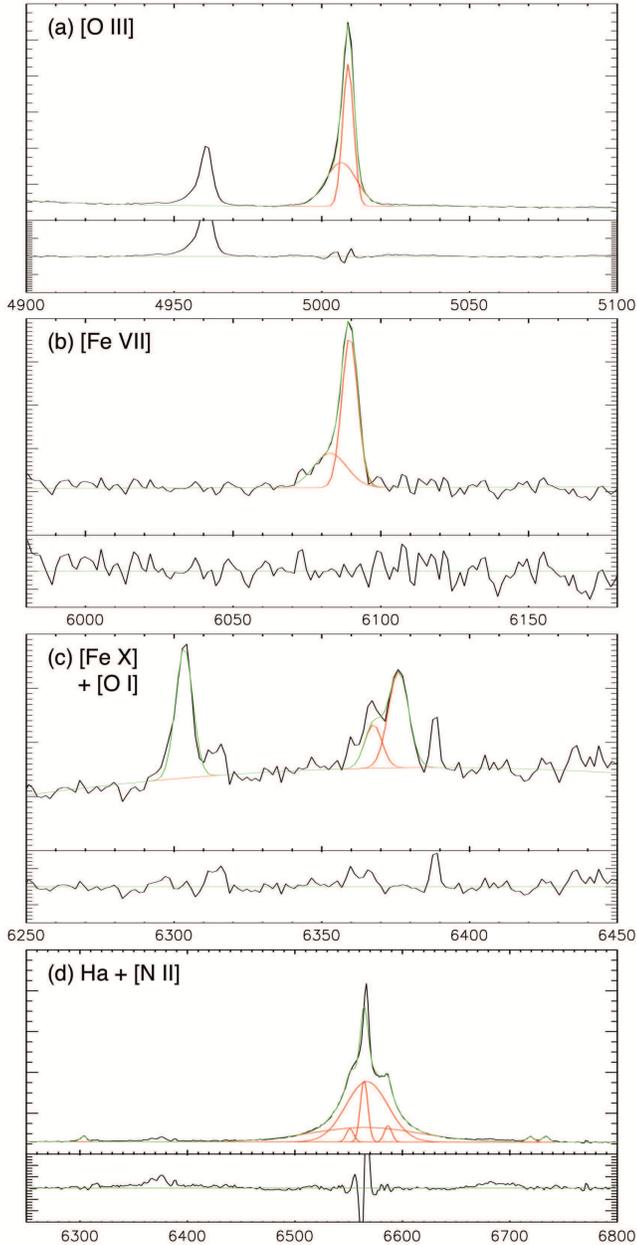}}
  \caption{A demonstration of the multi-component line models as
  applied to 
  the NLS1 VII Zw 533 (source 53).
  Pairs of panels
  show the data with the best-fitting models overlaid and the
  residuals after the models have been subtracted.  From the top down
  are the \oiii, \fevii, \fex, and \ha\ bands.  Red curves are the
  individual model components; green curves are either the total model
  (in the data plots) or the zero line (in the residual plots).  Note
  the distinct blue wing in both the \oiii\ and \fevii\ line
  profiles.  Wavelengths are in the emitted frame
  (as defined by SDSS redshifts)
  and the flux density scale is arbitrary.}
  \label{fig:lineModels}
\end{figure}

The sequence of models applied to each emission line starts with a
single Gaussian. 
The initial fitting is done with the line profile parameters (width
and centroid) fixed at assumed values in order to minimize the number
of free parameters.
Next, the profiles are allowed to vary (within limits: doublets are
assumed to have the same profiles; additional constraints are
described below)
and the best-fitting parameters are determined.
We then try adding a second Gaussian, using $\chi^2$
statistics to test whether the added component provides a
significantly better description of the data\footnote{We consider a
  model to be significantly better if the probability of
  obtaining the same reduction in $\chi^2$
  by chance is $< 1$ per cent.}.
In the case of \ha\ we also try adding a third Gaussian to the model,
once again testing whether the added component is significant.
The most detailed model that offers a significantly improved fit is
adopted as the best-fitting model, provided that it passes a final
visual inspection (described in the next section).
The best-fitting models are presented online in
Tables~\ref{tab:FexData}--\ref{tab:HaData}.
In Fig.~\ref{fig:lineModels} we demonstrate the best-fitting line
models for one of our sample members in the \oiii, \fevii,
\fex(+\oi), and \ha(+\nii) bands (the \sii\ and \fexi\ line models are
not shown because these are fitted with only a single Gaussian and are
not blended).

Some notes on details of the line-fitting process:

\begin{itemize}

\item \textit{Two-component forbidden line models:}
  The double-Gaussian model provides a significantly improved fit to
  the \oiii\ line for every member of our sample.
  The same cannot be said of the \fevii\ line, which is found to
  require a second component in only 14 cases.
  The other forbidden line data generally do not require a
  second component so only single Gaussian models are used.
  We note that the use of a mix of single- and double-Gaussian models to
  fit \fevii\ does not introduce a strong systematic effect, as the
  fluxes of most compound models differ from those of best-fitting
  single Gaussians by no more than 15 per cent, which is comparable to
  the uncertainty in the measurements.  The advantage of sometimes
  using the two-component model is a more accurate characterization of
  the profile of the \textit{core} of the \fevii\ lines.

\item \textit{Blended lines:}
  In order to separate blended lines, additional constraints were
  applied.
  In the case of the \oi+\fex, we first determined the
  best-fitting parameters for the \oi$\lambda$6300 line. 
  We then assumed an \oi$\lambda$6364 model with the same width and
  velocity shift and 1/3 the intensity of \oi$\lambda$6300 and fitted
  the \fex\ model to the remaining flux.
  A similar process was used to deblend \ha\ and the \nii\ doublet.
  \rev{If the \nii\ models are not constrained, they will sometimes
    include a significant amount of \ha\ flux.  This is especially
    problematic when the width of \ha\ is $\sim 1500 \  \kms$ because
    the \nii\ models can effectively slice off the broad line wings,
    resulting in \ha\ models that are systematically too narrow.
    To mitigate this tendency, a}
  Gaussian fitted to 
  \oi$\lambda$6300 was used as a template for the narrow line emission
  profile.  The \nii\ models were assigned the same width and
  velocity shift as \oi, with only the intensity allowed to vary.
  These intensities were re-fitted with each successive \ha\
  model.

\item \textit{\ha\ model components:}
  Up to three Gaussian components are used to model the \ha\ lines;
  the third is discarded as unnecessary in 21 instances.
  One of the Gaussians is forced to have the same width and velocity
  shift as the narrow line template defined by \oi\ (see previous
  comment on blended lines).
  Thus, the \ha\ model is guaranteed to have at least one component
  with the characteristics of the NLR.

\item \textit{\fexi\ measurements:}
  SDSS does not provide spectral coverage at the wavelength of \fexi\
  for eight of our sample members (six are at $z > 0.16$; two others
  have gaps in their data).
  Of the remaining 55 AGN spectra, 36 provide \fexi\ detections with $>
  99$ per cent confidence despite the increased noise due to OH sky
  lines at these wavelengths.

\end{itemize}

\subsection{Optical parameter analysis}
\label{sect:interp}

\subsubsection{Rejection of spurious models}
\label{sect:interpSpurious}

A final check of the best-fitting models was made to identify any
dubious line model components.  
The most common mode of failure was when a second (or third) component
of \fevii\ or \ha\ became extremely broad to account for residual
continuum flux.
In such cases we reverted to the best-fitting
parameters of a previous, less complex model.
Apart from these cases,
there were four other best-fitting models that were
revised after inspection.
In two instances (sources 40 and 48) \fexi\ models 
were discarded because they
were found to be fitted to sky line residuals.
For source 28, a Sy1.5 with a very broad \ha\ component,
the \fex\ Gaussian was used to model part of the
continuum
and had to be re-fitted manually.
In one other object (source 5, a Seyfert~2 with relatively simple line
profiles) one of the \ha\ components was fitted to $\nii \lambda
6583$; to correct this, its flux has been reassigned to the \nii\ line.

\subsubsection{Line velocity shifts and redshift redefinition}
\label{sect:redshiftRedef}

The velocity shifts ($v_\rmn{sh}$) of the emission lines are measured
relative to the recession velocities of their host galaxies.  Hence,
a negative velocity shift indicates a reduced recession velocity of
a line-emitting region.  We interpret $v_\rmn{sh} < 0 \ \kms$ as
outflow velocities along our line of sight, although we cannot rule
out the possibility of emission from infalling clouds approaching from
the far side of the AGN.

In order to measure $v_\rmn{sh}$ with the highest possible precision,
care must be taken in how the systemic redshifts are defined.
The SDSS pipeline uses multiple methods to measure redshifts;
the final value is defined by the method that yields the
highest confidence measurement.
However, there can be systematic differences between the redshift
values determined by these methods (e.g., when emission lines are
dominated by outflows and are used as the basis of the redshift
measurement).
Such systematics may be present at the few hundred \kms\ level even
when the SDSS pipeline reports that the measurements are
mutually consistent (as is the case for 46 out of 63 sample members).
Therefore, if we are to measure $v_\rmn{sh}$ with precision better
than a few hundred \kms\ and avoid systematic effects, we must
ensure that a consistent definition of redshift is used for our entire
sample.

Ideally we would define systemic redshifts from the stellar properties
of the host galaxies, but not all of our sample members have
measurable absorption features.
Instead, we base our definition upon the centroid wavelengths of the
\sii\ doublet.
This doublet is chosen because it has a low \rev{IP} and a
low critical density (\critdens).  
Therefore, amongst the emission lines available in all our spectra,
this is the feature most likely to be dominated by 
\rev{regions
  farthest from the central engine, where the
  influence of the AGN is expected to be weakest. 
  Thus, the \sii\ lines should be more closely related than any other
  available feature to the kinematics of the ISM of the host galaxy. 
  However, it is possible that these lines may sometimes arise in
  outflowing gas.
  In such cases the line centroid will not accurately indicate the
  systemic velocity of the host
  if the observed velocity structure is not symmetric
  (e.g., if there is a bi-conal outflow and we do not
  integrate comparable amounts of flux from both sides).
}

The change from SDSS-defined to \sii-defined redshifts
results in a range of $v_\rmn{sh}$ adjustments from 
-166 to +$103 \  \kms$.
The median adjustment is $-34 \  \kms$, so 
the redshift redefinition
tends to increase the outflow velocities.
All $v_\rmn{sh}$ values reported in this paper are
measured relative to the \sii-defined redshifts.
It is worth noting that the $v_\rmn{sh}$ values of \oi\ and the
\textit{core} of \oiii\ (defined in the next section) agree very well
with those of \sii, with mean values of $0 \pm 4$ and $-5 \pm 5 \
\kms$, respectively.
This agreement is somewhat surprising because the critical densities
of these lines are larger by factors of $10^3$ and $10^{2.5}$, so
their 
emission is expected to be dominated by different regions.

\subsubsection{\oiii\ and \fevii\ components}
\label{sect:OiiiInterp}

For every source in this sample the two-component line model provided
a significantly better fit to the \oiii\ profile than did the
single-Gaussian model.
In most cases the second component was used to represent
asymmetric line profiles.
We therefore adopt the nomenclature of ``core'' and ``wing''
components of the line 
(\oiiic\ and \oiiiw, respectively),
with the core defined to be the component
whose centre is closer to the peak of the overall profile. 
The core is invariably the Gaussian with the smaller FWHM.
With this definition, 53 \oiii\ lines have blue wings, 6 have red
wings, and 5 have components separated by less than 1/4 pixel ($< 17
\ \kms$; see Table~\ref{tab:OiiiData}).

Unlike \oiii, double-Gaussian models seldom provide a 
significantly better fit to the \fevii\ lines.  
When two Gaussians are used, we again divide them into core and wing
components (\feviic\ and \feviiw).
In all cases we find that \feviiw\ is both broader and bluer than
\feviic.
In the remaining 49 cases the single Gaussian model is used as
\feviic\ (hence $\feviig = \feviic$ for most sample members).
\rev{The best-fitting \fevii\ models are presented in
  Table~\ref{tab:FeviiData}.  
  This table includes the best-fitting single Gaussian models for all
  sample members, even those for which the double Gaussian model
  provides a significant improvement, in order to provide a model that
  is uniform across the entire sample. 
  We use the notation \feviig\ when we refer to the single Gaussian
  models for the entire sample.  
}

\rev{
  We note that there is a potential degeneracy when deblending the
  core and wing line components.  If too much or too
  little flux is attributed to the core, then the width and
  velocity shift of the wing can be affected.  At best this degeneracy
  introduces an additional uncertainty in the parameters of the
  individual components, contributing noise to the data; at worst it can
  introduce systematic effects that skew the average values across the
  sample.  Thus, the core and wing components may be thought of as a
  possibly non-physical parametrization of an asymmetric line
  profile.  As such, the properties of these individual components,
  especially those of the wing, may best be interpreted qualitatively.
  Although the exact details of these components may be
  suspect, they do provide a more accurate description of the shape of
  the line profile than a single Gaussian model could.  Moreover, the
  flux of the combined model provides a better measure of the actual
  flux from the line.  We therefore consider the sum of the component
  fluxes to be a robust measure of the line strength, and will
  emphasize this in the ensuing discussion.
  Hereafter, we use the subscript ``$_T$'' when referring to this
  combined flux.
}

\subsubsection{\ha\ components}
\label{sect:HaInterp}

\begin{figure}
  \centering
  \rotatebox{270}{\resizebox{!}{8.4cm}
		{\includegraphics{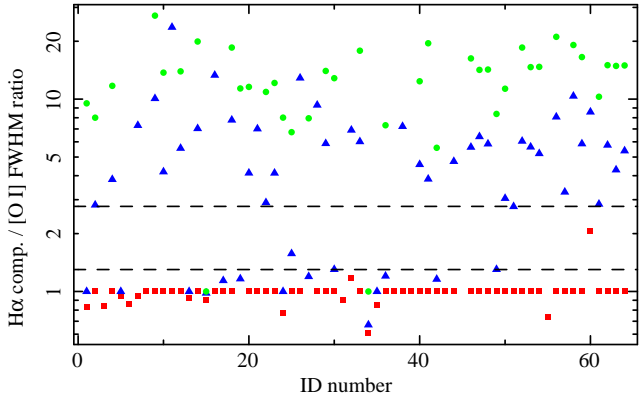}}}
  \caption{\rev{Ratios of the \ha\ component widths over \oi\ FWHM.  Red
    squares, blue triangles and green circles represent the FWHM ratios of
    progressively broader \ha\ Gaussian components over \oi $\lambda$6300.
    Dashed lines are drawn at ratios of 1.3 and 2.8.  Components that
    fall below the lower line are assigned to \han\ whilst those above
    the higher line are assigned to \hab; the few points between are
    considered individually.  By construction, all \ha\ models include
    one Gaussian with the FWHM of \oi\ unless this component is found
    to not be significant.}}
  \label{fig:HaOiFWHMrat}
\end{figure}

The \ha\ lines are fitted with up to three Gaussian components.
These models generally have one narrow, one intermediate, and one
broad component (a third component is rejected in only 21
instances).
However, sometimes more than one of these had a FWHM
comparable to that of \oi.
Clearly, the intermediate-width component did not consistently
represent a distinct region: sometimes it was unambiguously broad
($\rmn{FWHM} > 2000 \ \kms$), other times it was used to
fine-tune the modeling of the NLR contribution to \ha.
We therefore decided to combine the Gaussians into two \ha\
components:
\han, interpreted as the NLR contribution to \ha, combines all
components not much broader than \oi, and 
\hab, the BLR contribution, includes all Gaussians substantially
broader than \oi.
The ratios of \ha\ component FWHM over \oi\ FWHM almost always had
values $\leq 1.3$ or $\geq 2.8$ 
\rev{(Fig.~\ref{fig:HaOiFWHMrat})},
so these thresholds were used to
define \han\ and \hab\ components\footnote{Just two
  best-fitting \ha\ models had components with velocity widths between
  these limits and therefore required special consideration.
  The intermediate-width \ha\ component of the NLS1 galaxy
  KUG~1031+398 (source 25) has $\rmn{FWHM = 1.6 \times FWHM \, \oi}$;
  this is considered part of \hab.
  The narrowest \ha\ component fitted to the Sy1.5 IRAS~F16330+3100
  (source 60) has $\rmn{FWHM = 2.1 \times FWHM \, \oi}$, but this is a
  reasonable fit to what is clearly a narrow component of \ha\ with a
  double peak, so it is assigned to \han.}.
When \han\ or \hab\ consist of more than one component, the
flux used is the sum of the components and the profile is that of the
highest-flux Gaussian.
\rev{As with \oiiit\ and \feviit, we use \hatt\ when referring to the
  combined flux of all the \ha\ components.}

\subsubsection{Spectroscopic classifications}
\label{sect:specTypes}

\begin{figure}
  \centering
  \rotatebox{270}{\resizebox{!}{8.4cm}
		{\includegraphics{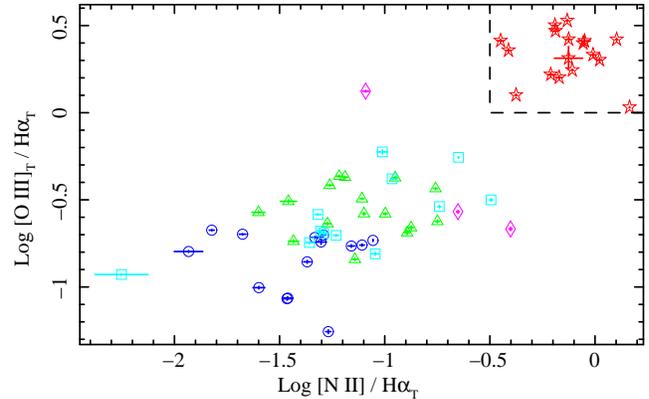}}}
  \caption{\rev{Line diagnostic ratios used to separate Sy2s from
      other Seyferts.  Sy2s (red stars) are bounded by
      $\log(\oiiit/\hatt) > 0$ and $\log(\nii/\hatt) > -0.5$.  The
      other sample members are classified as Sy1.9s (magenta diamonds),
      Sy1.5s (green triangles), NLS1s (cyan squares), and
      Sy1.0s (blue circles) according to their H line profiles.}}
  \label{fig:lineDiag}
\end{figure}

The first challenge in establishing the spectroscopic classifications
of our sample members is to separate the Seyfert 2s from the other
Seyfert types.
Simply using the FWHM of the broadest \ha\ component is
problematic because in several Sy2s (the ones with strong 
\oiii\ wings), 
the \ha\ models include a weak but
statistically significant broad component.
This component appears to be real but its width (FWHM sometimes $2000
\  \kms$ or more; 
\rev{cf.\ Table~\ref{tab:HaData}})
is non-physical, in that it is fitted to the
blended, otherwise-unmodeled blue wings of \ha +\nii\ emission from
the NLR.
Another complication is that
the \hab\ FWHM of the narrowest NLS1s is comparable to
the broadest \han\ lines amongst the Sy2s.
Instead of using \ha\ FWHM, we use two line flux ratios,
\nii/\hatt\ and \oiiit/\hatt.
The total \ha\ flux is a strong discriminator between Sy2 and other
Seyfert types because lines that include a broad component usually
have much more flux,
but neither ratio alone is sufficient as a single parameter to
separate the Sy2s because there is some overlap with
other Seyfert types.
However, following the approach of \citet*{Baldwin81}, we find that
the Sy2s are very well separated from all other AGN in the
plane defined by these two ratios \rev{(Fig.~\ref{fig:lineDiag}; see
  also \citealp{Zhang08})}. 
All of the Sy2s in the present study are bounded by 
$\log (\oiiit /\hatt) > 0$ and $\log (\nii /\hatt) > -0.5$.
\rev{These ratios have not been corrected for reddening, which would
  tend to increase the \oiiit/\hatt\ values.
  However, Sy2s generally require the largest corrections, so
  accounting for reddening would strengthen their separation in this
  plane.}

Once the Sy2s are separated, the remaining type 1 AGN are classified
based on the properties of their Balmer lines.
The 12 sources with $\hab\ \rmn{FWHM} < 2200 \  \kms$ are
defined to be NLS1; the other type 1 Seyferts collectively comprise
the broad-lined Seyferts (BLS1).
Note that the division between NLS1s and BLS1s is somewhat
arbitrary\footnote{The specific FWHM threshold adopted to define NLS1s
  is the same as that used by \citet{Zhou06}, but is a bit broader
  than the more commonly used value of $2000 \  \kms$ (e.g.,
  \citealp{Goodrich89b}).  Our choice is motivated by the data, in
  that there is a break in the distribution of \hab\ FWHM values
  around $2200 \  \kms$.}
but we apply it here to compare with previous work and to see if other
properties distinguish this sub-class.
The BLS1s are subdivided into three groups: 
the ones with very little \han\ emission are Sy1.0s (based upon the
fraction of the overall \ha\ flux in the \han\ component, with a
subjectively-chosen threshold of no more than 8 per cent);
most of the rest, with larger \han /\hab\ ratios, are Sy1.5s;
the few with no clear broad component at \hb\ are classified as
Sy1.9s.
\rev{The final tally of spectral types is 14 Sy1.0s, 12 NLS1s, 16
  Sy1.5s, 3 Sy1.9s, and 18 Sy2s.}

\subsubsection{Estimates of dust reddening}
\label{sect:reddening}

The measured line fluxes are not corrected for reddening.
The impact of dust is generally small, as
most sample members appear to have very little reddening (based upon
both the Balmer decrements and the shapes of the continua).
Furthermore, it is difficult to measure the Balmer decrement reliably
for many type~1 Seyferts because of the difficulty in separating
\han\ from \hab, rendering any correction uncertain.
\rev{Amongst the few spectra that do appear to be affected by dust, the
  most reddened objects are sources 7, 13, and 32; two Sy2s and a Sy1.9
  for which we estimate $A_V = 3.3$, 3.5 and 2.9, respectively.
  In no other Seyfert have we found $A_V > 2.2$.}
To mitigate the effect of any uncorrected reddening, we will place
some emphasis on comparisons of lines that are close in wavelength.
For instance, with $A_V < 2.2$ for the rest of the sample, the effect 
upon \fevii/\fex\ ratios will be $\leq 10$ per cent and is
always smaller than the measurement uncertainty.

\subsubsection{Summary of the final line profile models}
\label{sect:finalModels}

The final set of models adopted for the discussion that follows is
summarized in Table~\ref{tab:finalModels}.
The \oiii\ and \fevii\ lines fitted with two Gaussians are organized
into core and wing components, as described in
\S\ref{sect:OiiiInterp}.
The \ha\ lines are fitted with up to three Gaussians which are then
combined into two components, ``broad'' and ``narrow'' (\hab\ and
\han; \S\ref{sect:HaInterp}).
\rev{For these lines, which at least sometimes include more than one
component, we use the subscript ``$_T$'' (i.e., \oiiit, \feviit\ and
\hatt) to refer to the combined fluxes of all components.}
The remaining lines (the \oi, \nii\ and \sii\ doublets, \fex\ and
\fexi) are fitted with single-Gaussian models.
For many of the NLR lines, the model parameters are linked to each
other.  
\rev{
  In particular, the $\oi \lambda 6364$ model is completely defined by
  the fit to $\oi \lambda 6300$.  Because of this, unless
  $6364$~\AA\ is specified, any mention of \oi\ will implicitly refer
  to the stronger line at $6300$~\AA.}

\begin{table}
\caption{Final spectral line models used for analysis.}
\label{tab:finalModels}
{\centering

\begin{tabular}{@{}l@{\ \ }c@{\ \ }c@{\ \ }c@{\ \ }l@{}}
\hline
Emission & IP   & Log $N_{cr}$      & Comp  & \multicolumn{1}{c}{Details of} \\
line     & (eV) & ($\rmn{cm}^{-3}$) & count & \multicolumn{1}{c}{final fit}  \\
\hline
$\oiii  \lambda 5007$ &  35.1 &  5.8 & 2    & Core \& wing components \\
\multirow{2}{*}{$\fevii \lambda 6087$}
                      & \multirow{2}{*}{99.1}
                              & \multirow{2}{*}{7.6}
                                     & \multirow{2}{*}{1--2}
                                            & Only 14 final models \\
                      &       &      &      & \ \ \ include wing component \\
$\oi    \lambda 6300$ &  ---  &  6.3 & 1    & All parameters free \\
$\oi    \lambda 6364$ &  ---  &  6.3 & 1    & Pars.\ defined by $\oi \lambda 6300$ \\
$\fex   \lambda 6374$ & 233.6 &  9.7 & 1    & All parameters free \\
$\nii   \lambda 6548$ &  14.5 &  4.9 & 
\multirow{2}{*}{$\left. \rule{0mm}{2.65ex} \right\}1 \ \ \ $}
                                            & \nii\ $v_\rmn{sh}$ \& FWHM \\
$\nii   \lambda 6584$ &  14.5 &  4.9 &      & \ \ \ defined by $\oi \lambda 6300$ \\
\ha                   &  ---  &  --- & 2    & Broad \& narrow comps. \\
$\sii   \lambda 6716$ &  10.4 &  3.2 & 
\multirow{2}{*}{$\left. \rule{0mm}{2.65ex} \right\}1 \ \ \ $}
			                    & \sii\ doublet fitted with \\
$\sii   \lambda 6731$ &  10.4 &  3.6 &      & \ \ \ single $v_\rmn{sh}$ \& FWHM \\
$\fexi  \lambda 7892$ & 262.1 & 10.4 & 1    & 36 \fexi\ det., 19 non-det. \\
\hline
\end{tabular}

}

\medskip
$\oi \lambda 6300$ is used as a template for the parameters of the
$\oi \lambda 6364$ model in order to deblend \oi+\fex, and for the
profile of the NLR when fitting \ha+\nii .  
Columns are (1) line ID;
(2) ionisation potential (eV);
(3) log of the critical density at $T = 10^4$~K for the forbidden
transition lines (cm$^{-3}$);
(4) number of model components used; 
(5) details regarding parameter constraints and (in the cases of
\fevii\ and \fexi ) the frequency with which models are considered
significant. 
\end{table}

\subsection{\textit{Rosat} X-ray data}
\label{sect:RosatData}

Half of our sample members (32 out of 63) appear in catalogues of
soft X-ray sources detected with the \textit{Rosat}
PSPC\footnote{Twenty-eight were detected in the \textit{Rosat} All-Sky
  Survey (RASS; \citealp{RASS-BSC,RASS-FSC}). 
  Four others appear in the \citet*{WGACAT2000} catalogue of sources
  found in pointed \textit{Rosat} observations}.
We convert the 
PSPC count rates to fluxes
using the PIMMS tool from HEASARC\footnote{Online at
  {http://\protect\linebreak[1]heasarc.gsfc\protect\linebreak[0].nasa.gov\protect\linebreak[1]/Tools\protect\linebreak[1]/w3pimms.html}.},
assuming a power law model with a photon index $\Gamma = 1.5$
and an intervening absorption column of
$n_H = \scinot{3}{20} \  \rmn{cm}^{-2}$ at $z=0$.
It is well established that NLS1s often have much steeper soft X-ray
spectral indices than other Seyfert galaxies \citep{Boller00}, so we
also consider the effect of using $\Gamma = 3.0$ for the NLS1s.
Given these assumed models, 
the conversion from RASS count rates 
to fluxes in the full 0.1--2.4~keV
\textit{Rosat} band is equivalent to
multiplicative factors of 
\scinot{1.30}{-11} and 
$\scinot{8.12}{-12} \ \rmn{erg \, cm^{-2} \, count^{-1}}$
for $\Gamma = 1.5$ and 3.0, respectively.
In \S\ref{sect:FHILxrays} we make use of the estimated flux of these
two models in 
a narrow band starting at the IP of \fex\ (233--300~eV).
The fractions of the overall \textit{Rosat} flux that lies in this soft
X-ray band are 0.0216 if $\Gamma = 1.5$ and 0.108 if $\Gamma = 3.0$.
Thus, the count rate conversion factors for this soft band flux are
\scinot{2.81}{-13} and
$\scinot{8.75}{-13} \  \rmn{erg \, cm^{-2} \, count^{-1}}$ (the
\textit{Rosat} PSPC count rates, RASS hardness ratios and estimated
fluxes are provided online in Table~\ref{tab:XrayFluxes}).
%

\section{Results}
\label{sect:results}

\subsection{Sample statistics}
\label{sect:sampleStats}

The final sample consists of 63 Seyfert galaxies, all 
with measured \fex, \fevii, and \oiii\ fluxes.  
Included amongst these are 36 sources with measured \fexi,
twice as many as any previous study.
\rev{The only earlier sample that has a comparable number of AGN with
  FHIL measurements is that of NTM00.
  We will therefore use their sample for comparison in this and
  the following section.
  We note that NTM00 has 65 sample members with measured
  ratios of both \fex/\oiii\ and \fevii/\oiii;
  of these, only 47 have reported \oiii\ fluxes and 17 have
  \fexi\ detections.}

\rev{We have cross-correlated the SDSS \fex\ sample with X-ray and radio
  catalogues.  As noted in \S\ref{sect:RosatData}, 32 sample members
  have been detected by \textit{Rosat}.  These include 11 out of 14
  Sy1.0s, 10/12 NLS1s, 8/16 Sy1.5s, 0/3 Sy1.9s, and 3/18 Sy2s.}
\rev{To determine which of the \fex\ sources are radio loud, we
  obtained fluxes or flux limits from the FIRST \citep*{FIRST} and
  NVSS \citep{NVSS} surveys.
  These are combined with the rest-frame flux density at $4400
 $~\AA\ (including an estimated reddening correction) to define the
  radio--to--optical flux density ratio $R$.
  Only one sample member is unambiguously radio loud (source 22,
  3C~234, with $R \sim 1.7 \times 10^4$).
  Several others are borderline radio loud objects, with
  $R$ values around 10, the strongest of
  which are sources 25 ($R = 17$), 6 ($R = 10$), and 51 ($R = 8$).
}
%

\begin{figure}
  \centering
  \resizebox{8.4cm}{!}{
	\includegraphics*[0.235in,0.68in][6.25in,2.95in]{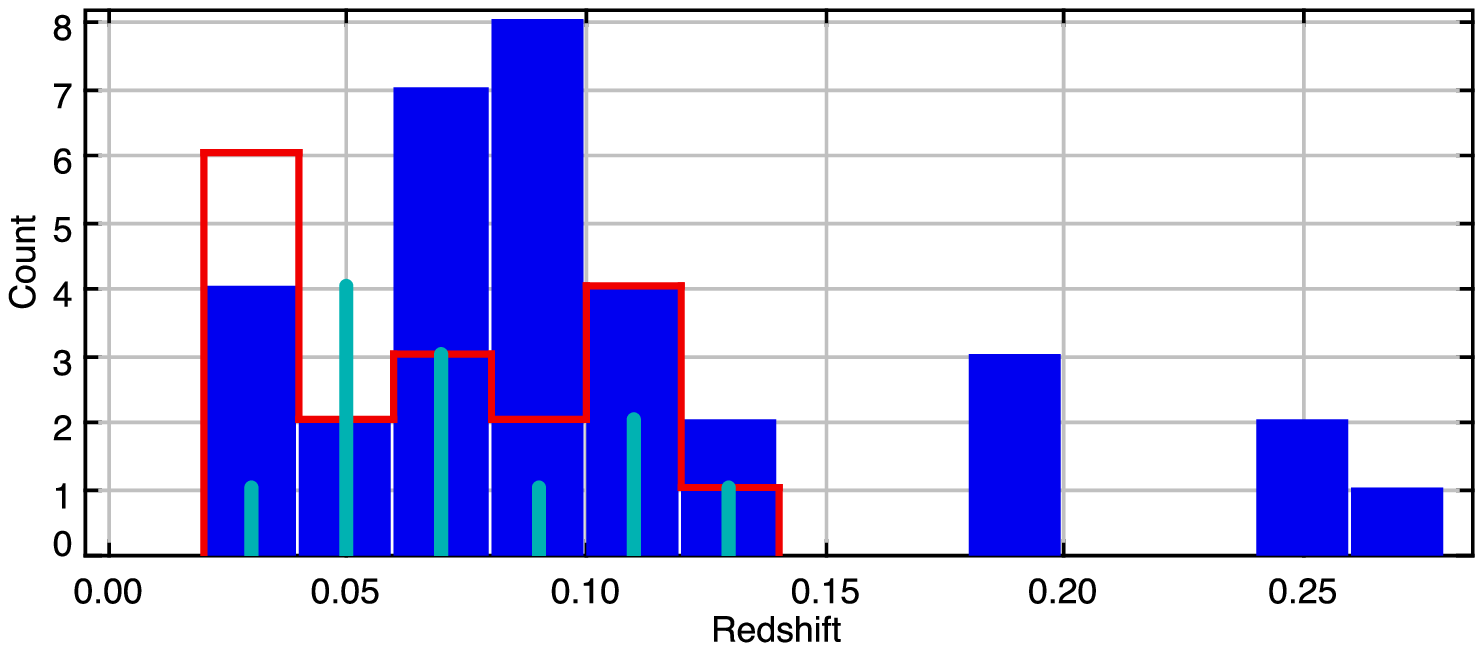}}
  \resizebox{8.4cm}{!}{
	\includegraphics*[0.45in,0.75in][6.25in,2.92in]{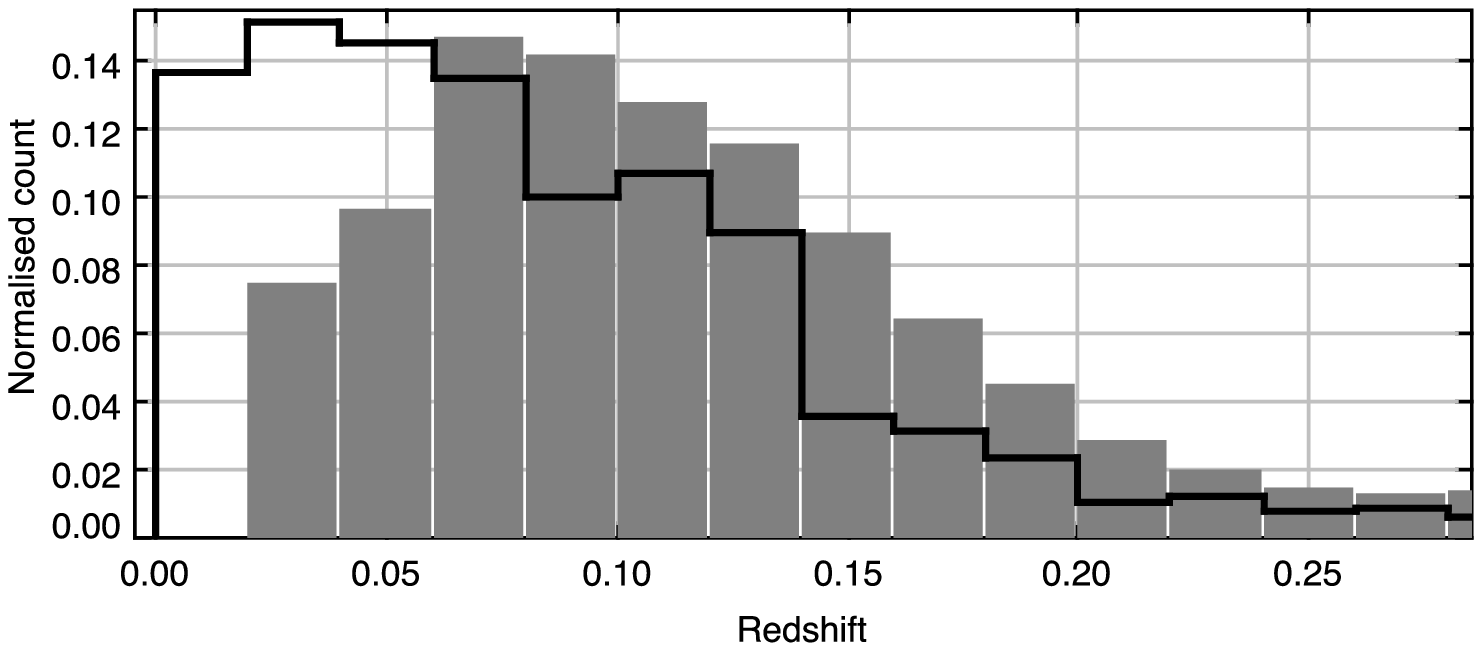}}
  \resizebox{8.4cm}{!}{
	\includegraphics*[0.32in,0.35in][6.25in,2.92in]{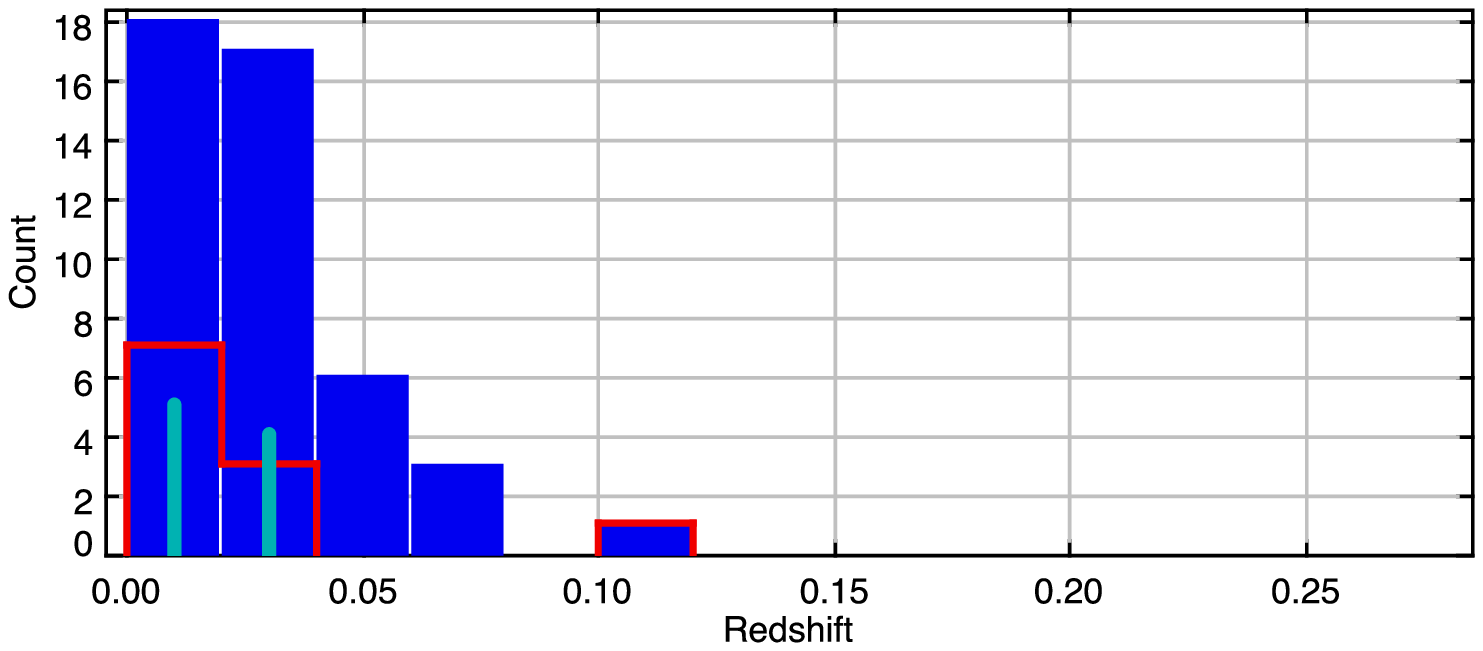}}
  \caption{Redshift distributions.
    \rev{The \fex-selected sample (top)
      is subdivided by spectral type:}
    broad blue bars represent broad-line Seyferts including
    all intermediate types (Sy1.0--1.9), 
    narrow light blue spikes depict the NLS1s, and 
    red outlines indicate the distribution of Seyfert 2.0s.  
    \rev{For comparison, we present broader samples drawn from the
      SDSS-DR6 (middle):
      all $\sim$600,000 galaxies and quasars with $z < 0.28$, and
      the 1098 \oi-emitting sources that satisfy 
      criteria (i) through (iv) of \S\ref{sect:sample}
      (grey filled-in histogram and black outline, respectively, with
      a normalized scale).
    }
    \rev{The 65 NTM00 sources with reported \fex/\oiii\ and
      \fevii/\oiii\ ratios (bottom) are subdivided between BLS1s,
      NLS1s and Sy2s with the same symbols as the \fex\ sample.}
    }
  \label{fig:z}
\end{figure}

We present the redshift distribution 
\rev{of the SDSS \fex\ sample}
in Fig.~\ref{fig:z} (top).
There are no significant differences between the distributions of the
Sy1.0, 1.5 and 1.9 populations, so these are combined as
the BLS1 distribution.
\rev{The principle difference between the Seyfert types is that only
  BLS1s are found beyond $z > 0.14$.
  Amongst the $z < 0.14$ sources, the median redshifts of the BLS1s,
  NLS1s and Sy2s are 0.082, 0.072 and 0.068, respectively.
  The Sy2s are the most locally-concentrated group,
  revealing a selection bias against these sources.
  The greater incompleteness of type~2
  Seyferts at even modest redshifts 
  is suggestive of a flux limit affecting our source selection.
  This is likely a bias inherited from the SDSS.
  Sy2s both lack a strong AGN
  contribution to their
  optical continuum and are more likely to be reddened, so 
  their photometric magnitudes are
  more likely to fall below the $r = 17.77$~mag 
  completeness limit of the SDSS spectroscopic survey.
  Indeed, most (five out of six) Sy2s in our sample with $0.09 < z <
  0.14$ have magnitudes within a few tenths of this limit,
  whereas only one third
  (four out of twelve) Sy1.0 and 1.5s in the redshift range $0.09 < z
  \leq 0.18$ appear this faint.
  However, it is also possible that Sy2s are weaker sources of
  \fex\ emission, which would contribute to the dearth of these
  objects at higher redshifts.
}

\rev{The \fex-selected sample redshift distributions may be compared
  to more broadly-defined samples of SDSS sources.
  The distribution of BLS1s, and of the overall sample,
  are consistent with the distribution of all galaxies and quasars in
  the SDSS-DR6 with $z < 0.28$ (gray histogram, Fig.~\ref{fig:z}, middle).
  We note, however, that the set of all 1098 quasar and galaxy spectra
  that satisfy our redshift, \ha\ and \oi\ emission line criteria
  (with or without evidence of \fex\ emission; i.e., 
  criteria i--iv in \S\ref{sect:sample}) is more strongly skewed
  toward low redshifts (black outline histogram, Fig.~\ref{fig:z}, middle).
  Thus the \oi-emitting galaxies must consist of two
  populations: one in which \fex\ is not detected that is very local
  and the \fex-selected sample that more closely follows the
  redshift distribution of the overall SDSS parent population.
  Lastly, we note that the \fex- and \fevii-detected portion of the
  NTM00 sample (Fig.~\ref{fig:z}, bottom) is much more locally concentrated
  than any of the other samples.
}

\subsection{Line strengths}
\label{sect:lineStrengths}

\begin{figure}
  \centering
  \resizebox{8.4cm}{!}{
	\includegraphics*[0.235in,0.75in][6.25in,2.95in]{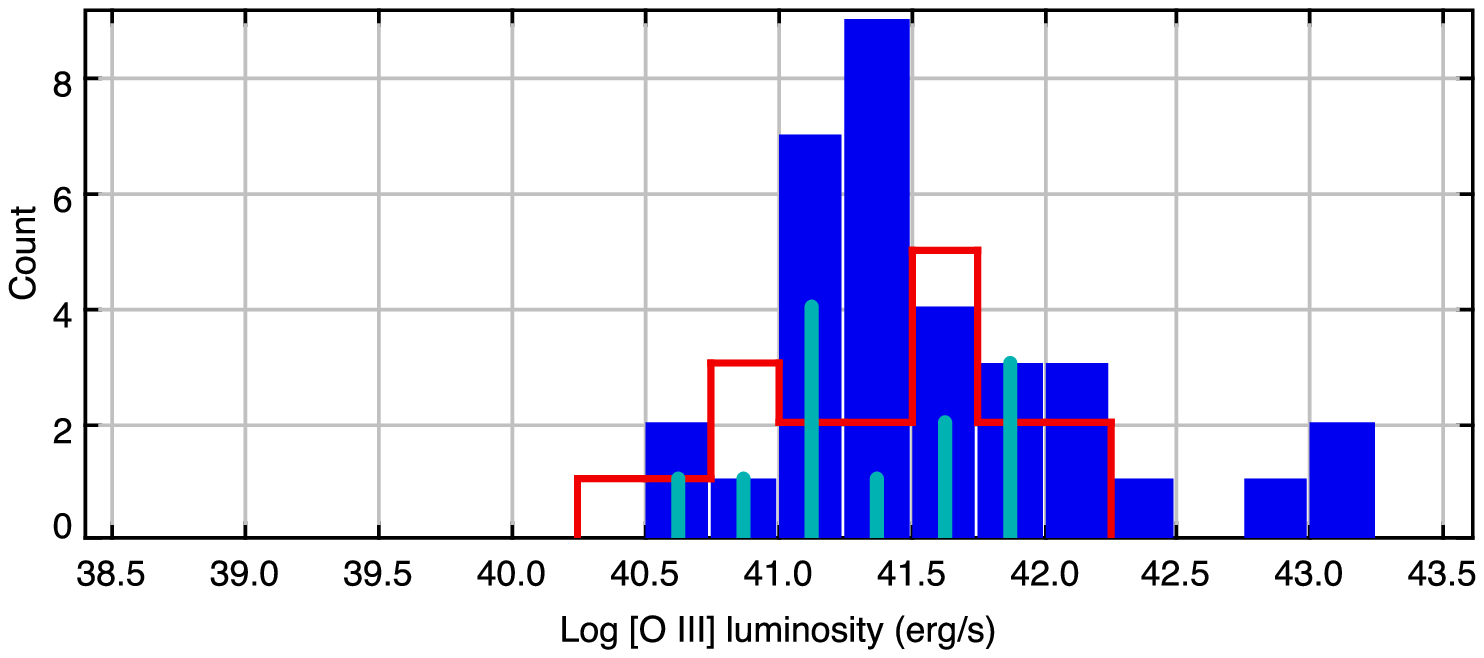}}
  \resizebox{8.4cm}{!}{
	\includegraphics*[0.32in,0.35in][6.25in,2.92in]{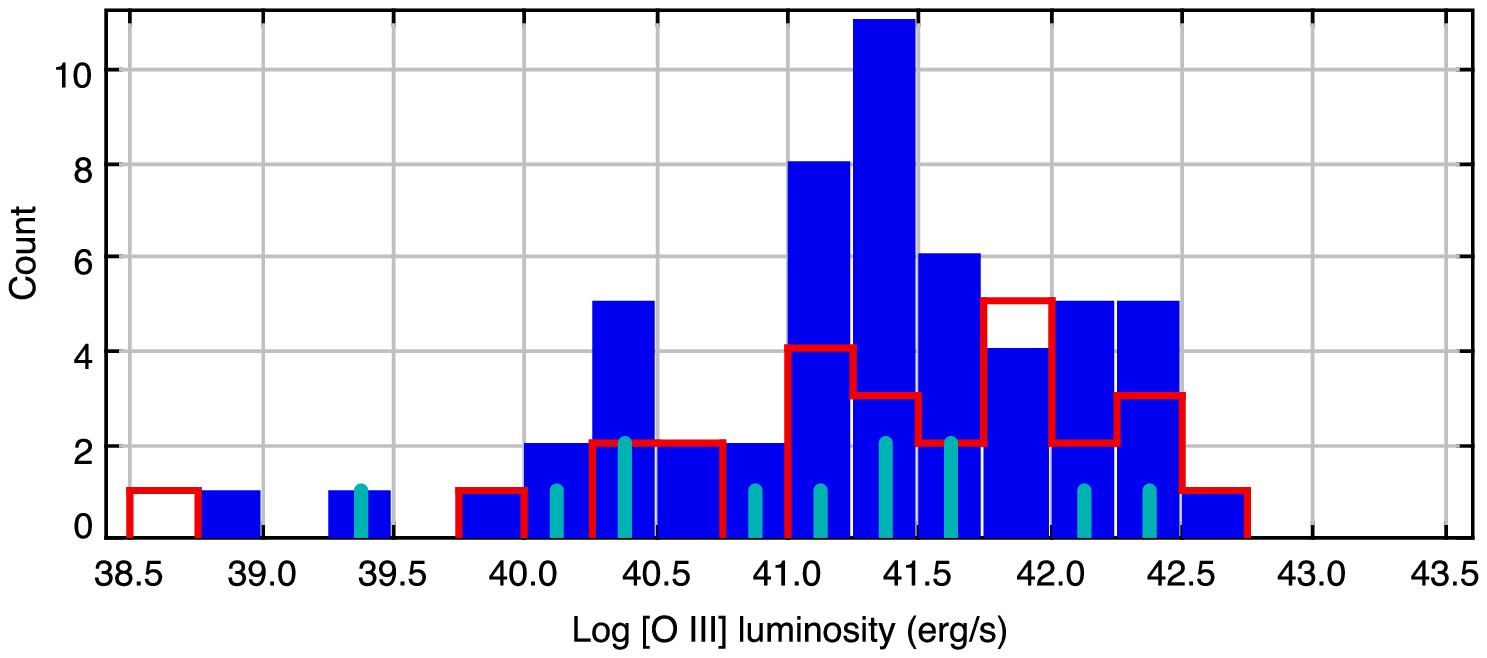}}
  \caption{Distributions of \oiii\ luminosities amongst BLS1s
    NLS1s, and Sy2s (symbols as in Fig.~\ref{fig:z}, top).
    \rev{The upper panel includes \oiiit\ luminosities for the 63 SDSS
      sample members; the lower panel presents the distributions for
      the 92 NTM00 objects with reported
      \oiii\ luminosities.  These luminosities have
      been re-evaluated using WMAP cosmological parameters.}
  }
  \label{fig:OIII_lum}
\end{figure}

The distribution of \oiiit\ luminosities ($L_{\oiiit}$) is similar
amongst the BLS1s, NLS1s, and Sy2s in the \fex-selected SDSS sample
(Fig.~\ref{fig:OIII_lum}, top).
The only clear difference is that the BLS1s extend to
higher $L_{\oiiit}$; these correspond to the highest-redshift
members of the sample.
\rev{The distributions of NTM00 are comparable except that a few of
  their lowest-redshift objects are found at considerably lower
  $L_{\oiii}$ values (Fig.~\ref{fig:OIII_lum}, bottom).
  Thus, apart from a few sources at the extremes of the redshift
  distributions, the \oiii\ power is reasonably well matched between
  the SDSS sample members of different spectral types and between the
  SDSS and NTM00 samples.
}

In Table~\ref{tab:fluxCorrel} we present the correlation statistics
between the fluxes of selected lines, line components, and the X-ray
band
\rev{(scatter plots demonstrating many of these correlations are
  provided online in
  Figs.\ \ref{fig:many-OIA_Lum}--\ref{fig:many-Rosat_Lum})}.
We find that all fluxes are positively correlated.
This is to be expected because in general the more powerful AGN should
be stronger emitters of both X-ray continuum and various lines.
The more interesting questions to ask are which parameters correlate
most strongly, 
\rev{with the least scatter,} 
and whether any trends are present when ratios are used
to normalize fluxes.
\rev{These are summarized in Table~\ref{tab:fluxRatios}, in which the
  average flux ratios and the RMS about each average are presented for
  the same pairs of fluxes appearing in Table~\ref{tab:fluxCorrel}.}

\begin{table*}
\begin{minipage}{175mm}
\caption{Correlation statistics amongst selected line and X-ray fluxes.}
\label{tab:fluxCorrel}
{\centering
\begin{tabular}{lcccccccc}
\hline
		& \oi		& \han		& \hatt$^a$ 	& \oiiit	& \feviit	& \fex		& \rev{\fexi$^b$}	& ${F_x}^c$	\\
\hline
\oi		& \nodata	& 0.760	(-12.2)	& 0.760	(-8.9)	& 0.856	(-18.4)	& 0.739	(-11.3)	& 0.650	(-8.1)	& 0.590	(-3.8)	& 0.603	(-3.3)	\\
\han		& 0.760	(-12.2)	& \nodata	& 0.559	(-4.2)	& 0.752	(-11.9)	& 0.628	(-7.4)	& 0.659	(-8.4)	& 0.666	(-5.0)	& 0.539	(-2.6)	\\
\hatt$^a$	& 0.760	(-8.9)	& 0.559	(-4.2)	& \nodata	& 0.824	(-11.5)	& 0.748	(-8.4)	& 0.590	(-4.7)	& 0.640	(-3.0)	& 0.514	(-2.4)	\\
\oiiit		& 0.856	(-18.4)	& 0.752	(-11.9)	& 0.824	(-11.5)	& \nodata	& 0.820	(-15.7)	& 0.647	(-8.0)	& 0.587	(-3.8)	& 0.465	(-2.0)	\\
\feviit		& 0.739	(-11.3)	& 0.628	(-7.4)	& 0.748	(-8.4)	& 0.820	(-15.7)	& \nodata	& 0.704	(-9.9)	& 0.617	(-4.2)	& 0.383	(-1.4)	\\
\fex		& 0.650	(-8.1)	& 0.659	(-8.4)	& 0.590	(-4.7)	& 0.647	(-8.0)	& 0.704	(-9.9)	& \nodata	& 0.884	(-12.0)	& 0.646	(-3.8)	\\
\rev{\fexi$^b$}	& 0.590	(-3.8)	& 0.666	(-5.0)	& 0.640	(-3.0)	& 0.587	(-3.8)	& 0.617	(-4.2)	& 0.884	(-12.0)	& \nodata	& 0.600	(-1.6)	\\
${F_x}^c$	& 0.603	(-3.3)	& 0.539	(-2.6)	& 0.514	(-2.4)	& 0.465	(-2.0)	& 0.383	(-1.4)	& 0.646	(-3.8)	& 0.600	(-1.6)	& \nodata	\\
\hline
\end{tabular}
}

~$^a$ \hatt\ correlations include only the 45 type~1 Seyferts, as the
Sy2s have only \han\ components.

\rev{~$^b$ Only the 36 detections are used to evaluate most
  \fexi\ correlations; 23 type~1 Seyferts are used for \hatt--\fexi, 14
  for $F_x$--\fexi.}

~$^c$ $F_x$ (0.1--2.4~keV) correlation statistics are based upon
the 29 type~1 Seyferts with \textit{Rosat} detections 
\rev{(14 sources in the case of \fexi)}.

\medskip
Spearman rank correlation coefficient, $\rho$,
and the log of the null
hypothesis probability (i.e., the probability 
\rev{of finding by chance a correlation this strong between two
  parameters that are intrinsically not correlated};
in parentheses) for selected emission
line and X-ray fluxes.  Probabilities are based upon the full sample
of 63 objects unless noted otherwise.  Values are repeated in the
upper-right and lower-left halves of the table for convenience.
\end{minipage}
\end{table*}

\begin{table*}
\begin{minipage}{168mm}
\caption{\rev{Log of the average flux ratios, with RMS.}}
\label{tab:fluxRatios}
%
%
{ \centering
\begin{tabular}{lcccccccc}
\hline
		& \oi		& \han		& \hatt$^a$	& \oiiit	& \feviit	& \fex		& \fexi$^a$	& ${F_x}^a$	\\
\hline
\oi		& \nodata	& 1.00 (0.24)	& 2.01 (0.26)	& 1.38 (0.18)	& -0.02 (0.30)	& -0.25 (0.29)	& -0.18 (0.34)	& 3.28 (0.41)	\\
\han		& -1.00 (0.24)	& \nodata	& 1.00 (0.36)	& 0.37 (0.24)	& -1.02 (0.36)	& -1.26 (0.31)	& -1.19 (0.34)	& 2.33 (0.45)	\\
\hatt$^a$	& -2.01 (0.26)	& -1.00 (0.36)	& \nodata	& -0.64 (0.24)	& -2.07 (0.29)	& -2.22 (0.37)	& -2.08 (0.42)	& 1.22 (0.48)	\\
\oiiit		& -1.38 (0.18)	& -0.37 (0.24)	& 0.64 (0.24)	& \nodata	& -1.40 (0.27)	& -1.63 (0.31)	& -1.54 (0.36)	& 1.91 (0.45)	\\
\feviit		& 0.02 (0.30)	& 1.02 (0.36)	& 2.07 (0.29)	& 1.40 (0.27)	& \nodata	& -0.23 (0.33)	& -0.23 (0.37)	& 3.37 (0.51)	\\
\fex		& 0.25 (0.29)	& 1.26 (0.31)	& 2.22 (0.37)	& 1.63 (0.31)	& 0.23 (0.33)	& \nodata	& -0.02 (0.18)	& 3.52 (0.38)	\\
\fexi$^a$	& 0.18 (0.34)	& 1.19 (0.34)	& 2.08 (0.42)	& 1.54 (0.36)	& 0.23 (0.37)	& 0.02 (0.18)	& \nodata	& 3.43 (0.55)	\\
${F_x}^a$	& -3.28 (0.41)	& -2.33 (0.45)	& -1.22 (0.48)	& -1.91 (0.45)	& -3.37 (0.51)	& -3.52 (0.38)	& -3.43 (0.55)	& \nodata	\\
\hline
\end{tabular}
}
~$^a$~As~noted~in~Table~\ref{tab:fluxCorrel}, these ratios are based
upon subsets of the \fex\ sample: 45 type~1 Seyferts for ratios
involving \hatt, the 29 type~1 Seyferts with X-ray detections for
ratios involving $F_x$, and the 36 \fexi-detected sample members (of
any spectral type) for the \fexi\ ratios.

\medskip
\rev{Log of the average flux ratios with RMS scatter about each mean
(within parenthesis, reported in dex).
Fluxes in the numerator are given across the top and those in the
denominator are listed at the left.  Thus, the entries in the first
row are log(\han/\oi), log(\hatt/\oi), log(\oiiit/\oi), etc.
Formal statistical uncertainties for the average ratios are $\pm
0.03$--0.05 for ratios defined by the full \fex\ sample and can be as
large as $\pm 0.10$ for ratios involving \fexi, \hatt, or $F_x$ due to
the smaller subsamples upon which these are based (in the case of the
\fexi/$F_x$ ratios, which involve only 14 data points, the uncertainty
is $\pm 0.15$).
Refer to the null hypothesis probabilities in
Table~\ref{tab:fluxCorrel} to assess whether these correlations are
real.}
\end{minipage}
\end{table*}

\begin{figure}
  \centering
  \rotatebox{270}{\resizebox{!}{8.4cm}{\includegraphics{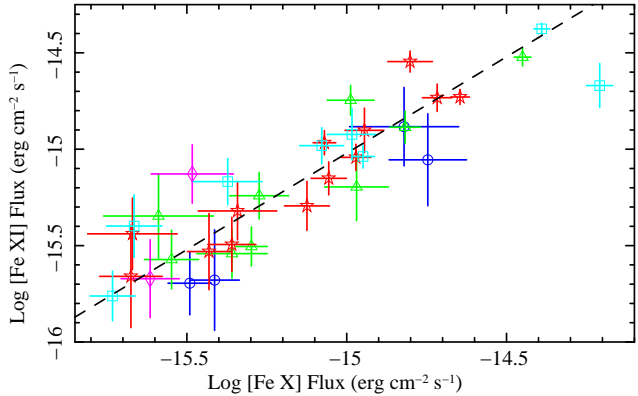}}}
  \caption{\rev{Fluxes of \fexi\ and \fex.
    The dashed line represents the average \fexi/\fex\ ratio, 0.95.
    Only two sample members have ratios that differ from this by
    $>2.3$-$\sigma$.
    Point styles are the same as in Fig.~\ref{fig:lineDiag}.}
 }
  \label{fig:FeXI-FeX_Flux}
\end{figure}

\rev{Not unexpectedly, one of the best correlations, both in terms of strength (as
  measured by the Spearman rank correlation coefficient, $\rho$) and
  scatter (RMS about the mean ratio) is between \fex\ and
  \fexi\ (Fig.~\ref{fig:FeXI-FeX_Flux}).}
The mean
\fexi/\fex\ ratio is $0.95 \pm 0.08$\footnote{\rev{Here we have not
    accounted for the effect of dust, which is generally small
    (\S\ref{sect:reddening}).
    When we apply estimated reddening corrections based upon the
    observed Balmer decrements, we obtain a mean \fexi/\fex\ ratio of
    $0.81 \pm 0.07$.}}
for the 36 sources in which we detect \fexi.
Henceforth, we focus upon \fex\ and not \fexi\ when discussing the
flux of the highest ionisation lines, as this line provides higher
S/N, offers more detections, and is closer in wavelength to the
other measured lines.

Apart from \fexi, the flux that correlates most strongly with \fex\ is
\feviit.  The correlations with \oiiit, \oiiic, \han, and X-ray fluxes
are comparable to each other and not much weaker than that with \feviit.

The next-highest ionisation species in this study is \fevii.  Its
ionisation potential is mid-way between those of \fex\ and \oiii\
(2.4$\times$ lower than the former and 2.8$\times$ higher than the
latter), so it is unclear a priori whether \fevii\ will have
properties more like the higher-IP FHILs or the traditional
NLR.
As noted above, the flux of \feviit\ correlates well with \fex.
However, the strongest correlation 
\rev{and the smallest RMS scatter amongst the \feviit\ flux ratios is
  provided by \oiiit.}
\rev{We note that of the two \oiii\ components, the flux of \oiiiw\ 
  is more closely correlated with \feviit, with $\rho = 0.762$ and RMS
  = 0.29.}

\begin{figure}
  \centering
  \resizebox{8.4cm}{!}{
    	\includegraphics*[0.32in,0.68in][6.25in,2.95in]{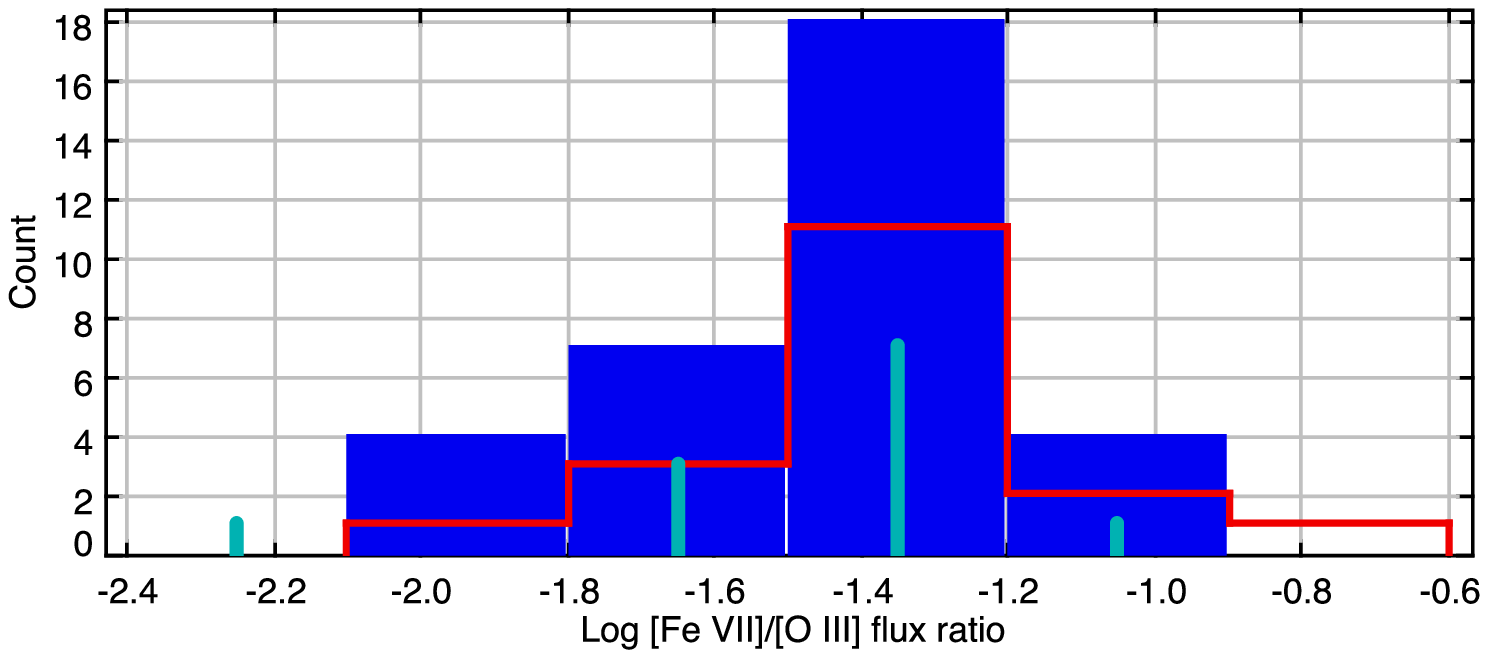}}
  \resizebox{8.4cm}{!}{
	\includegraphics*[0.32in,0.35in][6.25in,2.92in]{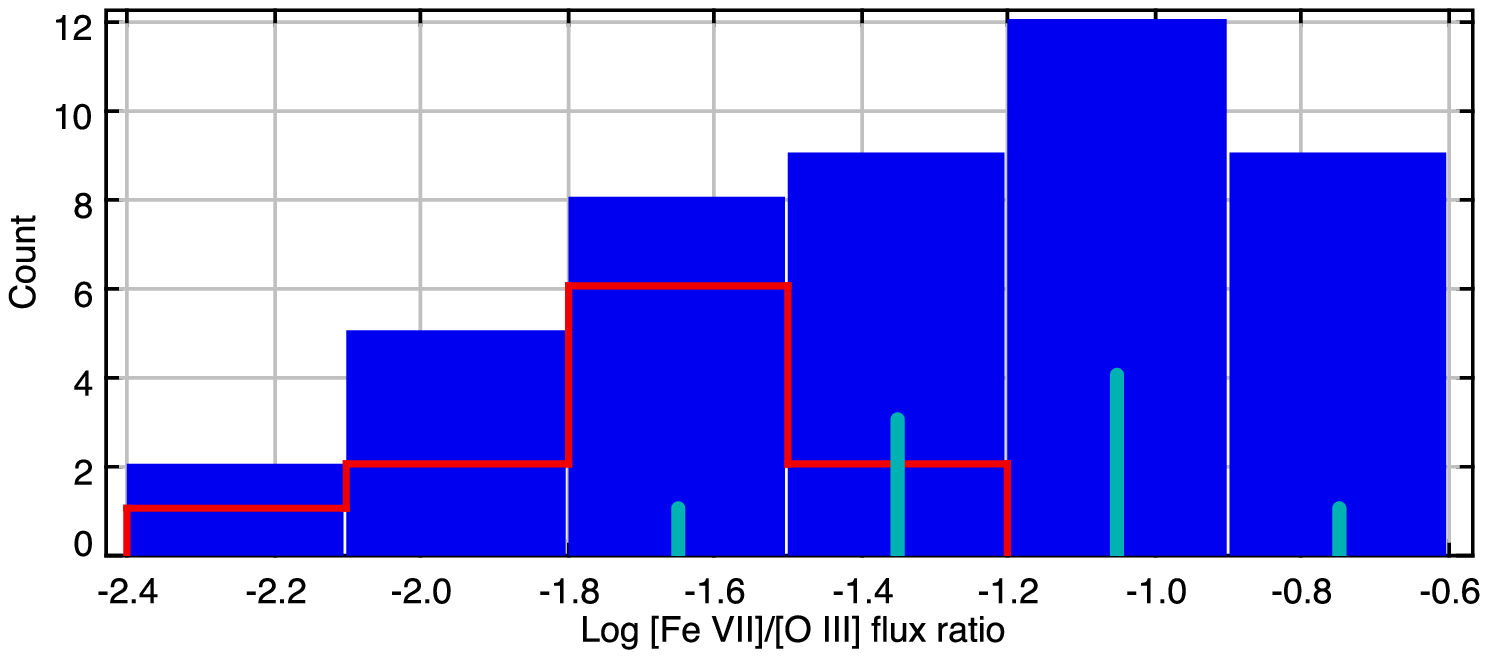}}
  \caption{Distributions of \feviit /\oiiit\ flux ratios amongst NLS1,
    BLS1 and Sy2 populations
    in the \fex-selected and NTM00 samples (upper and lower panels,
    respectively).
    In contract with NTM00, we find no significant difference between
    the BLS1 and Sy2 distributions.
  }
  \label{fig:FeVII-OIII_rat}
\end{figure}

It is worth noting that
when we use \oiiit\ strength to normalize \feviit, 
we do not find any significant differences between
the NLS1, BLS1, and Sy2
populations (Fig.~\ref{fig:FeVII-OIII_rat}, top).
\rev{This contradicts MT98 and NTM00, who find that Seyfert 2s have
  substantially lower \fevii/\oiii\ ratios
  (Fig.~\ref{fig:FeVII-OIII_rat}, bottom).}
In addition, NTM00
report a similar,
albeit weaker, disparity in the distributions of \fex/\oiii\ ratios
between type 1 and type 2 Seyferts.
Our \fex\ data show something qualitatively similar, in that the
\fex/\oiii\ ratios of Sy2s tend to be lower than those of Sy1s,
but quantitatively the difference we find is neither as large nor as
significant as that reported by
NTM00\footnote{The
  peaks in our type~1 and type~2 Seyfert \fex/\oiii\ distributions are
  separated by only 0.2--0.4 dex whereas NTM00
  shows a separation of 0.6--0.8 dex.  For our sample the
  separation is not significant: a KS test
  indicates a $>$10 per cent chance that these
  \fex/\oiii\ ratios are drawn from the same parent population.}.
These conflicting results are likely a consequence of how the samples are
selected.
The earlier studies used
heterogeneous collections of nearby, well-known Seyferts with data
drawn from the literature.
\rev{In particular, 
  31 out of 65 sources for which NTM00 report both \fex/\oi\ and
  \fevii/\oi\ ratios have lower redshifts than any member of the SDSS
  \fex\ sample (Fig.~\ref{fig:z}, bottom).
  This contrast is most pronounced for the obscured Seyferts: 18 out
  of 28 Sy1.9 and Sy2s are closer than any of the SDSS sources.
  As a result of this proximity and consequently the larger angular
  scale of the galactic structures, the
  observers may have been better able to isolate the unresolved
  nuclear emission from the extended flux.
  Thus the contamination due to stellar light may be reduced,
  allowing for the discovery of weaker nuclear features in a subset of
  their sample.
  Additionally, the area integrated might not have covered the full
  extent of the NLR, thereby affecting the observed flux ratios in
  some systems.}
The \fex-selected sample is more homogeneous, but 
\rev{our selection by \fex\ 
  creates a bias favoring sources with generally strong FHIL emission,
  reducing our sensitivity to low \feviit/\oiii\ ratios as well as
  \fex-weak spectra.}
Moreover, we note that 38 per cent of the \fevii-detected sources in
the NTM00 sample do not have \fex\ detections.
It is possible that there is a significant population of \fex-weak
Seyferts that are missing from the present sample.

\begin{figure}
  \centering
  \rotatebox{270}{\resizebox{!}{8.4cm}{\includegraphics{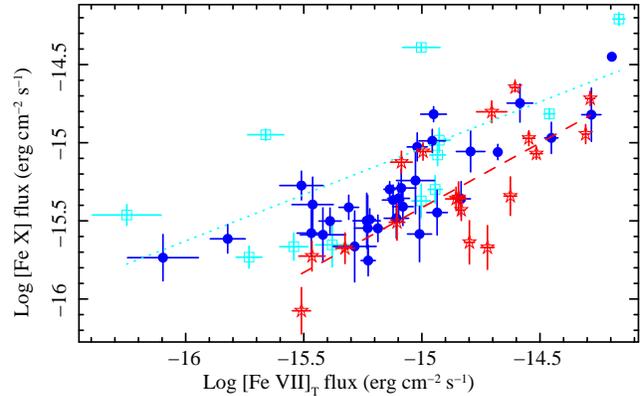}}}
  \caption{\rev{\fex\ vs.\ \feviit\ flux.
    Blue filled circles are BLS1s (Sy1.0, 1.5 and 1.9),
    cyan open squares are NLS1s, and red stars are Sy2s;
    the cyan dotted line is a linear regression fitted to the NLS1s
    and the red dashed line is a line fitted to the Sy2s.
    The distribution of NLS1s and Sy2s overlap, but the
    \fex/\feviit\ ratios of the NLS1s are systematically higher than
    those of the Sy2s: not one NLS1 lies below the line fitted to the
    Sy2s.}
    }
  \label{fig:FeX_vs_FeVII_flux}
\end{figure}

For the overall sample, 
the next-best \feviit\ correlation after \oiii\ is with \fex\ 
\rev{(Fig.~\ref{fig:FeX_vs_FeVII_flux}).
  However, 
  we find that there is a systematic offset in this plane between the
  NLS1s and the Sy2s.  Comparing linear regressions fitted to these two
  subsamples, we find that the \fex/\feviit\ ratios of NLS1s tend to be
  2--3 times higher.  Whilst the line fitted to the NLS1s is influenced
  by a few notable outliers and there is some overlap between these
  subsamples, not one NLS1 lies below the line fitted to the Sy2s (red
  dashed line).}
Table~\ref{tab:fluxCorrel} shows that \feviit\ also correlates very
well with \hatt, but this is only true when Sy2s are excluded from
the analysis (\hatt\ is dominated by \hab\ in the type 1 Seyferts,
whereas by definition the \ha\ lines in Sy2s do not have any BLR
contribution; \rev{Fig.~\ref{fig:many-HaT_Lum}c}).
The weakest \fevii\ correlation and the worst scatter is provided by
X-ray flux \rev{(Fig.~\ref{fig:many-Rosat_Lum}d)}, in contrast
to the case of \fex\ and X-rays.

The most prominent of the lower-ionisation forbidden lines is \oiii. 
The \oiiit\ flux correlates well with all of the other lines with
$\rmn{IP} < 100$~eV ($\rho \geq 0.75$).
However, it is \oi\ that correlates best when 
\rev{all spectral types are}
considered, followed by \feviit.
When we consider only the type~1 Seyferts, we find that \oiiit\
correlates 
\rev{comparably well with \hatt\ as with \oi\ and \feviit\ ($\rho =
  0.82$, 0.82, and 0.80, respectively).}

\subsection{Line profiles}
\label{sect:lineProfiles}

The widths and velocity shifts ($v_\rmn{sh}$) of the best-fitting
line components are summarized in Table~\ref{tab:avgLineProfiles} for
both the overall sample and subsets defined by Seyfert spectral
classification.
Note that \fexi\ and \feviiw\ 
are included in only some of final models.
Consequently, the values of these parameters are more strongly
affected by small number statistics, especially for some subsamples,
and should be interpreted with caution.
Likewise, there are only three Sy1.9s 
so the typical line properties of this class of Seyfert are not very
well constrained; 
their average best-fitting parameters have large uncertainties and are
consistent with those of both the Sy1.5s and the 2.0s.
We therefore exclude Sy1.9s from Table~\ref{tab:avgLineProfiles}.
\rev{Caution is also advised when comparing lines fitted with
  different models, as this can introduce systematic effects.  To
  facilitate the comparison of \fevii\ with higher-ionisation FHILs,
  we include statistics for the best-fitting single-Gaussian models,
  \feviig, as well as the double-Gaussian components.
}

\begin{table*}
\begin{minipage}{162mm}
\caption{Average parameters of fitted line profiles.}
\label{tab:avgLineProfiles}
{
\begin{tabular}{lccccc}
\hline
Line	& 	Overall			& 	NLS1			& 	Sy1.0			& 	Sy1.5			& 	Sy2			\\
model	& Mean			(RMS)	& Mean			(RMS)	& Mean			(RMS)	& Mean			(RMS)	& Mean			(RMS)	\\
\hline
\multicolumn{6}{c}{Line velocities measured relative to \sii\ ($v_\rmn{sh}$, \kms)}	\\
\hline
\oiiic	& -5	$\pm$	5	(40)	& -8	$\pm$	6	(21)	& -8	$\pm$	6	(22)	& -15	$\pm$	13	(50)	& 12	$\pm$	11	(47)	\\
\oiiiw	& -134	$\pm$	19	(147)	& -112	$\pm$	30	(100)	& -65	$\pm$	24	(87)	& -134	$\pm$	57	(221)	& -196	$\pm$	29	(121)	\\
\oi	& 0	$\pm$	4	(30)	& 7	$\pm$	7	(23)	& -1	$\pm$	6	(20)	& -2	$\pm$	12	(46)	& 2	$\pm$	5	(21)	\\
\feviic	& -74	$\pm$	16	(128)	& -99	$\pm$	41	(136)	& -54	$\pm$	28	(101)	& -93	$\pm$	42	(163)	& -44	$\pm$	25	(103)	\\
\feviiw$^a$
	& -296	$\pm$	28	(102)	& -331	$\pm$ \nodata\ (\nodata)& -213	$\pm$	63	(63)	& -305	$\pm$	63	(63)	& -308	$\pm$	42	(118)	\\
\rev{\feviig}
	& -86	$\pm$	16	(126)	& -100	$\pm$	41	(135)	& -57	$\pm$	27	(99)	& -95	$\pm$	42	(162)	& -83	$\pm$	25	(102)	\\
\fex	& -84	$\pm$	18	(143)	& -128	$\pm$	44	(147)	& -39	$\pm$	36	(130)	& -107	$\pm$	44	(171)	& -61	$\pm$	30	(124)	\\
\fexi$^a$
	& -212	$\pm$	32	(192)	& -258	$\pm$	39	(103)	& -41	$\pm$	19	(33)	& -217	$\pm$	76	(216)	& -186	$\pm$	52	(181)	\\
\han	& -34	$\pm$	11	(85)	& -69	$\pm$	21	(68)	& -27	$\pm$	33	(120)	& -32	$\pm$	24	(93)	& -22	$\pm$	13	(52)	\\
\hab	& -3	$\pm$	26	(206)	& 52	$\pm$	21	(71)	& -6	$\pm$	70	(253)	& 24	$\pm$	57	(222)	& 	\nodata			\\
\hline
\multicolumn{6}{c}{Line widths (FWHM; \kms)}																				\\
\hline
\sii	& 322	$\pm$	12	(90)	& 284	$\pm$	22	(73)	& 314	$\pm$	24	(88)	& 330	$\pm$	25	(98)	& 346	$\pm$	23	(90)	\\
\oiiic	& 319	$\pm$	12	(96)	& 267	$\pm$	14	(48)	& 315	$\pm$	17	(63)	& 363	$\pm$	36	(140)	& 316	$\pm$	19	(80)	\\
\oiiiw	& 779	$\pm$	33	(257)	& 801	$\pm$	59	(196)	& 758	$\pm$	48	(175)	& 777	$\pm$	77	(298)	& 798	$\pm$	72	(296)	\\
\oi	& 380	$\pm$	20	(154)	& 299	$\pm$	21	(70) 	& 354	$\pm$	19	(70)	& 437	$\pm$	55	(213)	& 404	$\pm$	41	(169)	\\
\feviic	& 500	$\pm$	29	(230)	& 502	$\pm$	74	(245)	& 489	$\pm$	53	(190)	& 561	$\pm$	60	(234)	& 433	$\pm$	57	(236)	\\
\feviiw$^a$
	& 860	$\pm$	91	(327)	& 652	$\pm$ \nodata\ (\nodata)& 696	$\pm$	357	(357)	& 459	$\pm$	279	(279)	& 1009	$\pm$	94	(267)	\\
\rev{\feviig}
	& 571	$\pm$	29	(230)	& 507	$\pm$	73	(241)	& 526	$\pm$	43	(156)	& 567	$\pm$	59	(227)	& 645	$\pm$	64	(263)	\\
\fex	& 556	$\pm$	32	(251)	& 591	$\pm$	73	(241)	& 508	$\pm$	35	(125)	& 690	$\pm$	84	(326)	& 447	$\pm$	56	(231)	\\
\fexi$^a$
	& 643	$\pm$	65	(383)	& 864	$\pm$	176	(467)	& 463	$\pm$	104	(180)	& 749	$\pm$	161	(455)	& 488	$\pm$	78	(272)	\\
\han	& 415	$\pm$	20	(157)	& 335	$\pm$	29	(95)	& 405	$\pm$	25	(91)	& 498	$\pm$	57	(219)	& 406	$\pm$	35	(143)	\\
\hab	& 3475	$\pm$	273	(1949)	& 1515	$\pm$	150	(498)	& 4718	$\pm$	523	(1887)	& 4491	$\pm$	377	(1458)	& 	\nodata			\\
\hline
\multicolumn{6}{c}{Model counts}																					\\
\hline
Most features	& 	63		& 	12			& 	14			& 	16			& 	18			\\
\feviiw 	& 14			& 1				& 2				& 2				& 9				\\
\fexi   	& 36			& 8				& 4				& 9				& 13				\\
\hline
\multicolumn{6}{l}{$^a$ Statistics are limited (especially for some
	spectral types) because not all final models include \feviiw\
	and \fexi.} \\
\end{tabular}
}

\medskip
Forbidden lines are listed in order of increasing \critdens.
Columns are:
(1)	Line model component;
(2--6)	average value $\pm$ the 1-$\sigma$ statistical uncertainty of
  mean and the RMS distribution (in parentheses) of model parameters
  for overall sample and subsets of each spectral type:
(2)	all Seyfert types, including Sy1.9s;
(3)	NLS1;
(4)	Sy1.0;
(5)	Sy1.5;
(6)	Sy2.
Sy1.9s are not presented because there are only three in our sample
(two with detected \fexi, none with \feviiw), so the average model
parameters are not very well constrained.  The Sy1.9 averages are
consistent with those of both Sy1.5s and Sy2s.
\end{minipage}
\end{table*}

\rev{Several things may be demonstrated with the average profile
  properties in Table~\ref{tab:avgLineProfiles}.
  We find that the cores of lines with the lowest IPs
  have velocity shifts that are generally consistent with \sii, whilst
  the wings of \oiii\ tend to be outflowing.
  The \fevii\ profiles are characterized by blue cores and sometimes
  include even bluer wings.
  The most blueshifted lines
  are found in our highest IP feature, \fexi.
  The widths of the line cores and single-Gaussian line models
  increase monotonically with critical density, with \fexi\ exhibiting
  the broadest lines.
  Thus, we reaffirm earlier studies that found a tendency for the line
  width and blue shifts to increase with increasing IP
  \citep[e.g.,][]{Appenzeller88}.
  Moreover, the FHIL profiles are quantitatively similar to previous
  results.
  According to \citet{Erkens97} the average FWHM ratios
  of the FHILs over \oiii\ are 1.63, 1.58 and 1.86 respectively for
  the \fevii/\oiii, \fex/\oiii\ and \fexi/\oiii\ width ratios; we
  obtain values of 1.60, 1.60 and 2.04 for the width ratios of
  \feviig, \fex\ and \fexi\ over \oi\ (here we use \feviig\ and
  \oi\ instead of \feviic\ and \oiiic\ to be consistent in using
  single-Gaussian models).
  Comparing the different Seyfert types, we find that the NLS1s tend
  to have the highest FHIL outflow velocities, whilst the Sy2s tend to
  have the narrowest FHILs. 
}

\rev{In Table~\ref{tab:profileCorrel} we present the correlation statistics
  between the line profile parameters of various emission features.
  These statistics show that the velocity shift correlations between
  the FHILs are universally stronger than those
  between any other pair of features.
  On the other hand, with the exception of the \fexi--\fex\ pair, the
  strongest line width correlations are found amongst the low-IP
  features.
}

\begin{table*}
\begin{minipage}{175mm}
\caption{\rev{Correlation statistics the line profile parameters of
  selected emission lines.}}
\label{tab:profileCorrel}
{ \centering
\begin{tabular}{lcccccccc}	
\hline
\multicolumn{9}{c}{Line $v_\rmn{sh}$ correlation statistics}	\\
	& \oi		& \han		& \oiiic	& \oiiiw	& \feviig	& \feviic	& \fex		& \fexi 	\\
\hline
\oi	& \nodata  	& \nodata  	& 0.621 (-7.2)	& 0.469 (-4.0)	& 0.390 (-2.8)	& 0.318 (-2.0)	& 0.330 (-2.1)	& 0.508 (-2.8)	\\
\han	& \nodata  	& \nodata  	& 0.321 (-2.0)	& 0.162 (-0.7)	& 0.315 (-1.9)	& 0.344 (-2.2)	& 0.341 (-2.2)	& 0.156 (-0.4)	\\
\oiiic	& 0.621 (-7.2)	& 0.321 (-2.0)	& \nodata  	& 0.237 (-1.2)	& 0.445 (-3.6)	& 0.473 (-4.0)	& 0.302 (-1.8)	& 0.397 (-1.8)	\\
\oiiiw	& 0.469 (-4.0)	& 0.162 (-0.7)	& 0.237 (-1.2)	& \nodata  	& 0.367 (-2.5)	& 0.255 (-1.4)	& 0.308 (-1.9)	& 0.391 (-1.7)	\\
\feviig	& 0.390 (-2.8)	& 0.315 (-1.9)	& 0.445 (-3.6)	& 0.367 (-2.5)	& \nodata  	& \nodata  	& 0.681 (-9.1)	& 0.754 (-7.0)	\\
\feviic	& 0.318 (-2.0)	& 0.344 (-2.2)	& 0.473 (-4.0)	& 0.255 (-1.4)	& \nodata  	& \nodata  	& 0.697 (-9.7)	& 0.702 (-5.7)	\\
\fex	& 0.330 (-2.1)	& 0.341 (-2.2)	& 0.302 (-1.8)	& 0.308 (-1.9)	& 0.681 (-9.1)	& 0.697 (-9.7)	& \nodata  	& 0.761 (-7.1)	\\
\fexi	& 0.508 (-2.8)	& 0.156 (-0.4)	& 0.397 (-1.8)	& 0.391 (-1.7)	& 0.754 (-7.0)	& 0.702 (-5.7)	& 0.761 (-7.1)	& \nodata  	\\
\hline
\multicolumn{9}{c}{Line FWHM correlation statistics} \\
	& \oi		& \han		& \oiiic	& \oiiiw	& \feviig	& \feviic	& \fex		& \fexi 	\\
\hline
\oi	& \nodata	& \nodata  	& 0.810 (-15.1)	& 0.539 (-5.3)	& 0.460 (-3.8)	& 0.356 (-2.4)	& 0.163 (-0.7)	& 0.196 (-0.6)	\\
\han	& \nodata  	& \nodata  	& 0.829 (-16.3)	& 0.567 (-5.9)	& 0.483 (-4.2)	& 0.352 (-2.3)	& 0.259 (-1.4)	& 0.331 (-1.3)	\\
\oiiic	& 0.810 (-15.1)	& 0.829 (-16.3)	& \nodata  	& 0.693 (-9.5)	& 0.422 (-3.2)	& 0.242 (-1.3)	& 0.312 (-1.9)	& 0.332 (-1.3)	\\
\oiiiw	& 0.539 (-5.3)	& 0.567 (-5.9)	& 0.693 (-9.5)	& \nodata  	& 0.331 (-2.1)	& 0.047 (-0.1)	& 0.099 (-0.4)	& 0.132 (-0.4)	\\
\feviig	& 0.460 (-3.8)	& 0.483 (-4.2)	& 0.422 (-3.2)	& 0.331 (-2.1)	& \nodata  	& \nodata  	& 0.411 (-3.1)	& 0.605 (-4.0)	\\
\feviic	& 0.356 (-2.4)	& 0.352 (-2.3)	& 0.242 (-1.3)	& 0.047 (-0.1)	& \nodata  	& \nodata  	& 0.457 (-3.8)	& 0.588 (-3.8)	\\
\fex	& 0.163 (-0.7)	& 0.259 (-1.4)	& 0.312 (-1.9)	& 0.099 (-0.4)	& 0.411 (-3.1)	& 0.457 (-3.8)	& \nodata  	& 0.773 (-7.5)	\\
\fexi	& 0.196 (-0.6)	& 0.331 (-1.3)	& 0.332 (-1.3)	& 0.132 (-0.4)	& 0.605 (-4.0)	& 0.588 (-3.8)	& 0.773 (-7.5)	& \nodata  	\\
\hline
\end{tabular}	
}	

\medskip
\rev{Spearman rank correlation coefficient ($\rho$) and the log of the null
hypothesis probability (in parentheses) for the profile parameters of
selected emission lines.  
Statistics are based upon the full \fex-selected sample with the
exception of \fexi\ correlations, which use only the 36
\fexi\ detections.
\hab\ is omitted from the table because it was not found to correlate
significantly with any of these parameters; its FWHM exhibits no
correlations with confidence above the 90 per cent level, whilst its
velocity showed no correlations with even 50 per cent confidence.
\feviiw\ is omitted because it is only applied to 14 sample members so
the confidence levels of its correlations are low; its only
correlation found to have $> 99$ per cent confidence is between the
FWHM values of \feviiw\ and \oiiiw. 
We do not report correlations between the profiles of \oi\ and
\han\ and between \feviig\ and \feviic\ because these models are often
constrained to have the same profiles; in particular these
\fevii\ models are identical in all but 14 instances. 
As with Table~\ref{tab:fluxCorrel} the values are repeated in the
upper-right and lower-left halves of these tables for convenience.}
\end{minipage}
\end{table*}

\begin{figure}
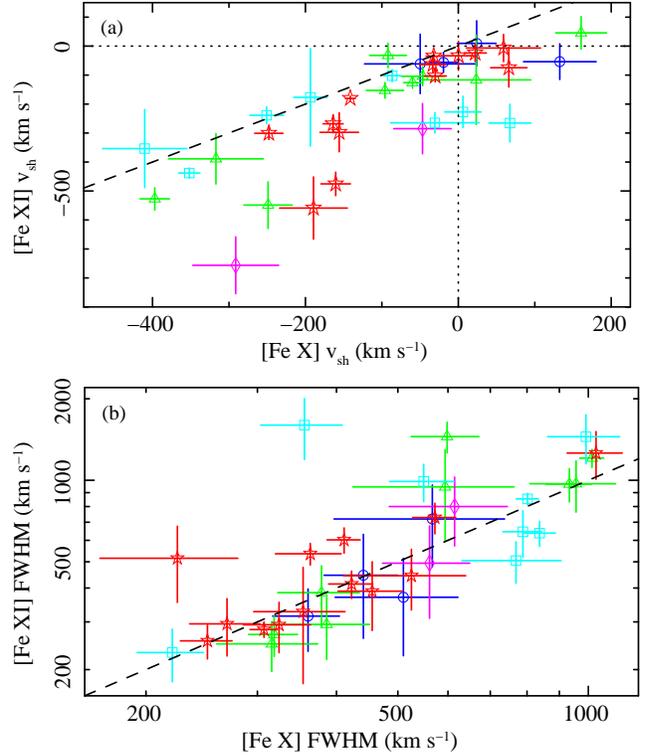

  \centering
  \rotatebox{270}{\resizebox{!}{8.4cm}{\includegraphics{fig9a.eps}}}
  \rotatebox{270}{\resizebox{!}{8.4cm}{\includegraphics{fig9b.eps}}}
  \caption{Velocity shifts (a) and FWHM (b) of \fexi\ and \fex.
    \rev{Point shapes and colours are the same as
      Fig.~\ref{fig:lineDiag}.}
    Dashed $x = y$ lines are included to guide the eye.
    Most \fex\ and \fexi\ lines appear to be outflowing
    ($v_\rmn{sh} < 0 \  \kms$).
    In a little more than half the sample the
    $v_\rmn{sh}$ values are consistent;
    in the remaining cases
    the outflow velocities of \fexi\ are faster
    (by up to 300~\kms).
    The FWHM values of \fexi\ tend to be either as
    broad or broader than those of \fex,
    but the line width uncertainties are large.
 }
  \label{fig:FeXI_Vel-FeX_Vel}
\end{figure}

The profiles of \fex\ and \fexi\ are mutually consistent for much of
the sample, but a subset of the \fexi\ lines appear to be both
broader and bluer (Fig.~\ref{fig:FeXI_Vel-FeX_Vel}a--b).
The velocity shifts 
are almost always
negative, indicating that the emitting clouds are outflowing.
The $v_\rmn{sh}$ values are consistent for roughly half of the sample,
whereas the \fexi\ outflow velocities are significantly faster in
roughly one third of the sources (\fexi\ is faster in 11 out of 36
sources with at least 3-$\sigma$ confidence).
In no instance is \fexi\ significantly slower.
Overall, the average velocity shift difference is 
$v_\rmn{sh} \, (\fexi) - v_\rmn{sh} \, (\fex) = -109 \  \kms$ with
an RMS scatter of 126\ \kms.
The widths of the \fexi\ lines tend to be either consistent with or
broader than \fex\ (Fig.~\ref{fig:FeXI_Vel-FeX_Vel}b), although the
FWHM values are not as well constrained as $v_\rmn{sh}$.
For both ionic species, the fastest 
\rev{outflow velocities are found amongst the broadest lines 
  (although the broadest lines do not always have fast $v_\rmn{sh}$
  values).
}

Unlike \fexi, there is no tendency for the \feviig\ models to have
different widths or velocity shifts from those of \fex.  The average
velocity shift difference is effectively zero with somewhat less
scatter ($v_\rmn{sh} \, (\fex) - v_\rmn{sh} \, (\feviig) = 5 \  \kms$
with $\rmn{RMS} = 108 \  \kms$).
Likewise, there is no significant difference between the widths of
\feviig\ and \fex.  However, in contrast to $v_\rmn{sh}$, the
correlation of the line widths is weaker and lower confidence.

\begin{figure}
  \centering
  \rotatebox{270}{\resizebox{!}{8.4cm}{\includegraphics{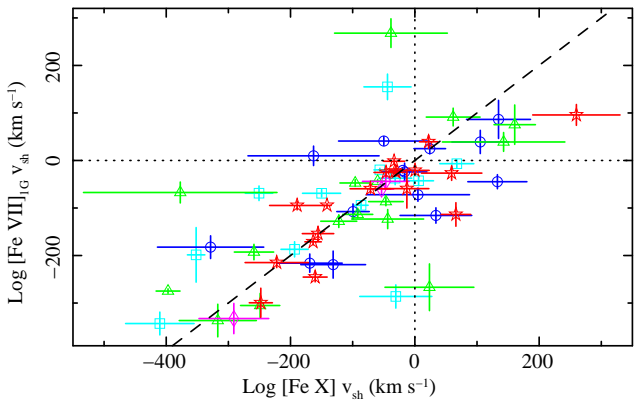}}}
  \caption{\rev{Velocity shifts of \feviig\ and \fex.
    A dashed $x = y$ line is sketched to guide the eye.
    Point shapes and colours are the same as
      Fig.~\ref{fig:lineDiag}.
    Unlike the $v_\rmn{sh}$ values of
    \fexi\ (Fig.~\ref{fig:FeXI_Vel-FeX_Vel}a), there is no
    tendency for the \feviig\ lines to be more or less shifted than
    \fex.  The scatter in this plot is dominated by a few data points
    with large \fex\ $v_\rmn{sh}$ uncertainties.}
  }
  \label{fig:FeVII-1G_Vel-FeX_Vel}
\end{figure}

\rev{Amongst the lower-IP features, the only lines or line components
  that tend to be in outflow are the wings of \oiii.  At best,
  the velocity shifts of \oiiiw\ and the FHILs are weakly correlated
  (only \feviig\ appears to correlate with greater than 99 per cent
  confidence).  However, as noted in \S\ref{sect:OiiiInterp}, there
  is a degeneracy in the process of separating the core and wing
  component models that introduces additional uncertainties in these
  component parameters, possibly causing
  systematic effects and certainly increasing the noise.
  We therefore interpret the \oiiiw\ parameters with caution. 
  It is true that the measured \oiiiw\ outflow velocities tend to
  be higher than those of \feviic, \feviig\ and \fex, but this may be
  an artefact of the deblending process.}
What is clear is that
\oiii\ has (at least) two components, the broader of which is
outflowing whilst the narrower one is not shifted relative to the
lower-ionisation species such as \sii\ and \oi.
\rev{
  The fact that the measured \oiiiw\ velocity shifts are not much
  larger than those of the FHILs leaves open the possibility that this
  component may be kinematically related to the FHIL-emitting region.
}

\subsection{X-ray properties}
\label{sect:XrayProp}

\begin{figure}
  \centering
  \rotatebox{270}{\resizebox{!}{8.4cm}{\includegraphics{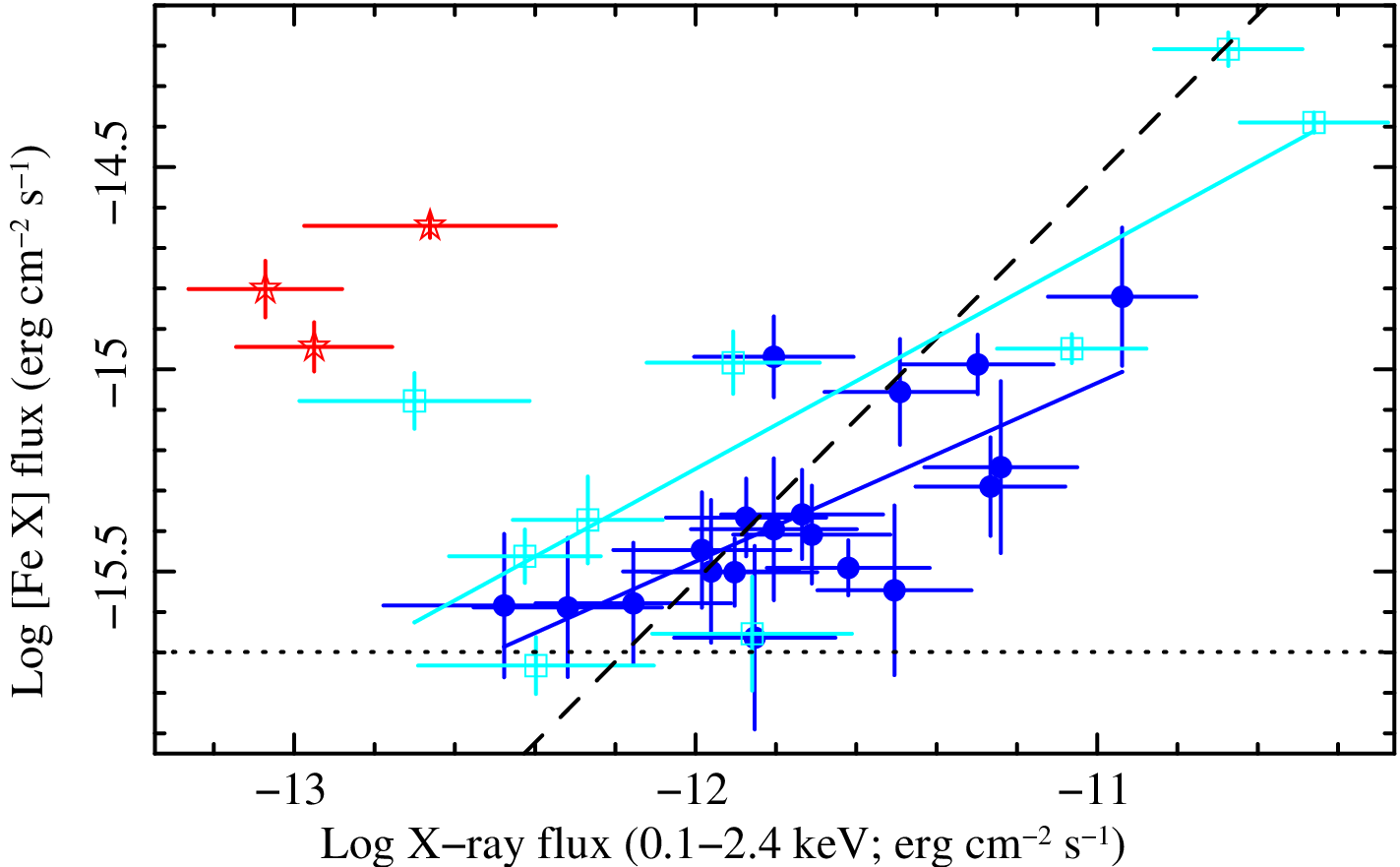}}}
  \caption{\fex\ vs.\ X-ray flux in the full
    \textit{Rosat} band (0.1--2.4~keV).
    Point styles are the same as in Fig.~\ref{fig:FeX_vs_FeVII_flux}.
    The solid diagonal lines are linear
    correlations fitted to the NLS1 (higher, cyan line) and BLS1
    (lower, blue line) populations.
    The horizontal dotted line at
    $F_{\fex} = \scinot{2}{-16} \  \cgsflux$ is the
    approximate flux limit of \fex\ in our sample.
    The black dashed line indicates a constant ratio of $\fex / \mbox{X-ray}
    = \scinot{3}{-4}$, which is representative of the flux ratios
    shown by \citet{Porquet99}.
    Apart from the Seyfert 2s, which are likely affected by X-ray
    absorption,
    the highest \fex/X-ray ratios are generally NLS1s.
    }
  \label{fig:FeX_vs_FxRosat}
\end{figure}

\begin{figure}
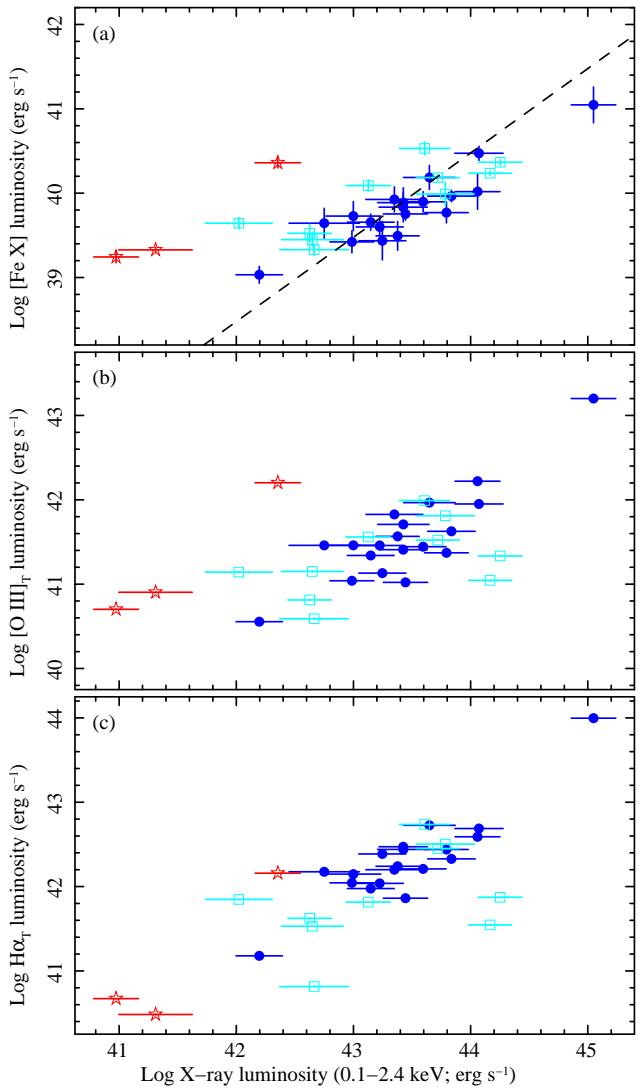

  \centering
  \rotatebox{270}{\resizebox{!}{8.4cm}{\includegraphics{fig12a.eps}}}
  \rotatebox{270}{\resizebox{!}{8.4cm}{\includegraphics{fig12b.eps}}}
  \rotatebox{270}{\resizebox{!}{8.4cm}{\includegraphics{fig12c.eps}}}
  \caption{Luminosities of \fex, \oiiit, and \hatt\ (panels a--c, respectively)
    vs.\ $L_x$ in the observed 0.1--2.4~keV band.
    Point styles and the dashed line are as in Fig.~\ref{fig:FeX_vs_FxRosat}
    (the error bars for \oiiit\ and \hatt\ 
    are smaller than the points).
    Panels a and b demonstrate that the line and
    X-ray powers of type~1 Seyferts are correlated,
    but the Sy2s are offset by more than an order of
    magnitude in $L_x$; in panel c the Sy2s, whose \hatt\ lines
    lack a broad component, lie closer to
    the type 1 objects.
    An inspection of the type~1 Seyferts shows that
    the relationships of \oiiit\ and \hatt\ with $L_x$ exhibit more
    scatter than \fex, although
    in the case of \hatt\ the additional scatter is primarily due to a
    few outliers (see \S\ref{sect:outliers}).
    We note that whilst these plots are affected by
    Malmquist bias, they still offer a fair representation of the
    relative amounts of scatter amongst the different correlations.
    }
  \label{fig:Line_vs_LxRosat}
\end{figure}

In Fig.~\ref{fig:FeX_vs_FxRosat}, we show the correlation between
\fex\ and 0.1--2.4~keV fluxes
amongst the 32 sample members with
\textit{Rosat} detections.
The \fex\ intensities of the type~1 Seyferts (1.0, 1.5 \& NLS1) scale
with power in the X-ray band, consistent with the trend shown by
\citet[fig.~2; represented here in Fig.~\ref{fig:FeX_vs_FxRosat}
  by the diagonal dashed line]{Porquet99}.
The only three \textit{Rosat}-detected Seyfert 2s have by far the
highest \fex /X-ray ratios, by roughly a factor of 30 over the type~1
Seyferts.
Moreover,
as demonstrated by the plot of \fex\ and X-ray luminosities
(Fig.~\ref{fig:Line_vs_LxRosat}, upper panel)
they also have some of the lowest observed X-ray powers.
This is readily interpreted within the Seyfert unification framework
\citep[e.g.,][]{Anton93} as obscuration along the line of sight to the
X-ray emitting regions of the type~2 systems.
If such obscuration also affects our view of the FHIL-emitting clouds, 
then it must cover \rev{only a fraction of this region as these sources are
detected by their \fex\ lines}.

The lower IP lines do not correlate as tightly with
X-ray power.  
This is illustrated by Fig.~\ref{fig:Line_vs_LxRosat}, which plots
the luminosities of \fex, \oiiit, and \hab\ against $L_x$.
Amongst the type~1 
Seyferts the correlations are
all positive, but with differing amounts of scatter: \fex\ has the
least 
\rev{(the RMS scatter about the mean \fex/$F_x$ ratio is 0.38 dex),
  followed by \oiiit\ (0.45 dex) and \hatt\ (0.48 dex; see the last
  column of Table~\ref{tab:fluxRatios} for the RMS scatter of other
  lines with $F_x$; see also Fig.~\ref{fig:many-Rosat_Lum})}. 
It is worth noting that
\rev{of all the measured lines, \feviit\ exhibits the weakest
  correlation with $F_x$ with a larger RMS than any other line.}
We note further that the scatter in the \hatt\ vs.\ X-ray plot (bottom
panel of Fig.~\ref{fig:Line_vs_LxRosat}) is dominated by a few
outliers; if we disregard the three NLS1s with the lowest \hatt/X-ray
ratios the correlation becomes 
\rev{marginally stronger and tighter than
  that of \fex\ ($\rho = 0.667$ with an RMS scatter of 0.34 dex).}

\section{Discussion}
\label{sect:discussion}

\subsection{Line profiles, IPs and FHIL kinematics}
\label{sect:FHILprofiles}

The \fex\ and \fexi\ lines are expected to be well correlated because
they have similar \rev{IPs} (234 and 262~eV,
respectively) and \critdens\ ($10^{9.7}$ and $10^{10.4} \ 
\rmn{cm}^{-3}$).
Indeed, we find that their fluxes are essentially indistinguishable
(\S\ref{sect:lineStrengths}), and they have mutually-consistent
profiles in more than half of the sample members with \fexi\ 
detections.
However, in the remaining sources the \fexi\ lines tend to be both
broader and bluer than \fex.
This suggests that the \rev{distribution of} \fexi-emitting clouds
sometimes extends \rev{closer to the BLR},
probing a region where the outflow velocities are higher.
Similar evidence has been shown for selected objects in smaller
samples \citep[e.g.,][]{Erkens97,Mullaney08}.

The velocity shifts of \feviig\ are similar to, and well correlated
with, those of \fex, suggesting that the clouds responsible for the
core of the \fevii\ profile are part of the same flow as the
\fex-emitting clouds.
However, there is only a weak correlation between the widths of these
components.  It is therefore unlikely that the \feviig\ and \fex\
lines are dominated by the same clouds.
An alternative hypothesis is that they are from spatially distinct
regions of an outflow that is ``coasting'', neither accelerated nor
decelerated significantly between the \fex- and \fevii-emitting
portions of the flow.  In such a scenario the \fevii\ clouds are
expected to be the downstream component due to the lower ionisation
parameter required and may be more readily resolved spatially.
\rev{This is consistent with our recent photoionisation models of FHIL
  emission in the NLS1 Ark 564, from which we infer that clouds in a
  radiatively-driven outflow are accelerated before the FHIL becomes
  strong.  The terminal velocity is approached and Fe is released
  within the clouds as dust grains become sublimated \citep{Mullaney09}.}
We note further that the wings of \oiii\ are outflowing and may be
related to the \fex- and/or \feviig-emitting portion of the outflow,
as the \rev{average \oiiiw\ velocity shifts amongst the NLS1s, Sy1.0s
  and Sy1.5s are consistent with those of these FHILs
  (cf.\ \S\ref{sect:lineProfiles})}.
On the other hand, the velocity shifts of \oiiic\ and the
lower-ionisation narrow lines are consistent with that of \sii\ and
are not part of this flow.

\subsection{Line correlations with X-rays}
\label{sect:FHILxrays}

\begin{figure}
  \centering
  \rotatebox{270}{
    \resizebox{!}{8.4cm}{\includegraphics{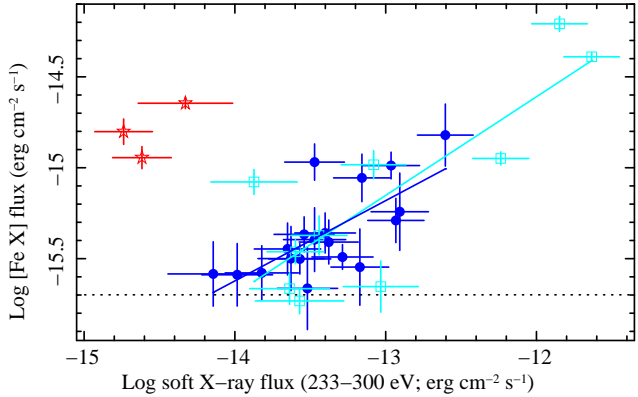}}}
  \caption{\fex\ vs.\ soft X-ray flux.
    The soft band is the portion of the X-ray spectrum
    most directly responsible for photoionising \fex.
    Here we estimate the 233--300~keV flux with the further
    assumption that NLS1s have a softer X-ray spectrum ($\Gamma = 3.0$). 
    As a result the difference between the \fex /X-ray ratios of NLS1s
    and BLS1s essentially disappears
    (compare to Fig.~\ref{fig:FeX_vs_FxRosat}). 
    This is demonstrated by the linear correlations fitted to these two
    populations (cyan and blue solid lines, respectively).
    The horizontal dashed line is once again the
    approximate \fex\ flux limit and the
    point styles are the same as in Fig.~\ref{fig:FeX_vs_FxRosat}.
  }
  \label{fig:FeX_vs_FxSoft}
\end{figure}

In \S\ref{sect:XrayProp} we show that
the \fex\ lines of type~1 Seyferts 
scale rather well with X-ray flux, in agreement with previous studies
\citep{Porquet99}. 
However, we note a systematic difference between the NLS1 and
BLS1s.
This is made clear by the offset between the linear correlations
fitted to these two populations in
Fig.~\ref{fig:FeX_vs_FxRosat}.
One contributor to this is likely to be
our simplistic assumption of a uniform
spectral model to convert \textit{Rosat} count rates to fluxes.
NLS1s have long been known to have, on average, stronger soft excess
and steeper soft X-ray spectra \citep[e.g.,][]{Boller00}.
If we adopt a steeper spectral index for these objects ($\Gamma = 3.0$
instead of 1.5)
and plot \fex\ flux against just the estimated 233--300~eV flux (which
dominates the photoionisation of \fex\ and \fexi), the offset between
the NLS1 and BLS1 distributions disappears
(Fig.~\ref{fig:FeX_vs_FxSoft}).
Thus, it appears that 
a single correlation between the \fex\ and soft X-ray fluxes applies
to both NLS1s and BLS1s.
It follows that observations of coronal lines may
be used to constrain the SED at soft X-ray energies when observations
in this band are either not available or of low sensitivity.
\rev{The average ratio we obtain for $\log (\fex / F_\rmn{233-300 eV})$
    is $-2.03 \pm 0.07$ with an RMS of 0.36.}
Additionally, we note that the NLS1 and BLS1 distributions are not
as clearly separated in the \oiiit-$L_x$ or \hatt-$L_x$ planes
(Fig.~\ref{fig:Line_vs_LxRosat}b--c).
Inaccuracies in the assumed spectral indices of NLS1s should have a
smaller effect upon these correlations because, unlike \fex, the
\oiiit\ and \hatt\ line emission is not
particularly responsive to any specific portion of the X-ray band.
Consequently, they
should scale more directly with the bolometric power, hence the
overall \textit{Rosat} luminosity, the estimate of which is less
sensitive to the model used to interpret PSPC count rates.

The amount of scatter in the emission line vs.\ X-ray correlations
allows us to put constraints on the size of the line emitting regions
and the variability history of the AGN.
In principle the lines should scale with the bolometric
power of the central engine, hence they should correlate with the
X-ray flux.
\rev{However, any variability of the photoionising continuum strength
  will weaken the observed correlation because the}
\textit{Rosat} data provide a snapshot of the AGN power at a moment in
time, whereas the line emitting regions are extended and therefore the
observed lines are a response to the AGN power averaged over the
light-crossing time of these regions.  
\rev{
  The more variable the bolometric power is, on time-scales much
  shorter than either the light-crossing time of the line-emitting
  regions or the time lag between the X-ray and optical spectroscopic
  observations, 
  the more likely it becomes that the instantaneous X-ray power
  observed would not be representative of the time-averaged power
  affecting the measured lines.
  The relatively modest scatter found amongst the type 1 Seyferts in
  Fig.~\ref{fig:FeX_vs_FxSoft} (0.38 dex) indicates that 
  the instantaneous X-ray power measured by \textit{Rosat}
  was within a factor of a few of the time-averaged power to
  which the FHILs were responding when observed by SDSS 10--15 years
  later.  Thus, none of our sources appear to have been caught in
  a short time-scale X-ray fluctuation at the time of the
  \textit{Rosat} observation, or a several-year drift in power since
  that time, by more than a factor of a few.}
A contribution to the scatter in the emission line vs.\ X-ray plots
is the made by the model dependence inherent in converting
\textit{Rosat} count rates to fluxes, which in the most extreme cases
could introduce $F_x$ errors up to a factor of two. 
\rev{We note that the scatter exhibited by the NLS1 is much larger
  than that of the BLS1 distribution in Fig.~\ref{fig:FeX_vs_FxSoft}
  (0.48 vs.\ 0.27 dex).  This could be due to either variability or
  greater inaccuracies in the models used to interpret the count
  rates, as NLS1s are known both to be more variable and to}
exhibit a wider range of spectral indices in the soft X-ray
band.

The fact that the correlation between \feviit\ and X-rays is weaker
\rev{with much more scatter}
than that between \fex\ and X-rays suggests that the \fevii-emitting
clouds are much more extended than those of \fex.
\rev{The longer light-crossing time of a more extended region gives
  the central engine more time to drift away from the time-averaged
  power to which the \fevii\ line is responding.
  Other sources of noise in this correlation are less likely because
  (1) there is not likely to be much photoionised \fevii\ emission
  that is unrelated to the AGN because the stellar continuum power
  above the IP energy of 99 eV is negligible, and (2) the correlation
  between the velocities of \feviig\ and \fex\ 
  (Fig.~\ref{fig:FeVII-1G_Vel-FeX_Vel})
  connect the
  \fevii-emitting clouds to the \fex-bearing outflow, demonstrating
  both that the \fevii-emitting medium likely originates in the AGN
  and showing no sign of the deceleration that would take place if
  this flow were shocked.}
Furthermore, the similarity of the \fevii--X-ray correlation to those
between various narrow lines and X-rays, \rev{and the strong
  correlations between \fevii\ and the NLR lines (notably \oi\ and
  \oiii)} 
suggests that the \fevii-emitting region merges with that of the
traditional NLR.
This is corroborated by previous studies which have shown that unlike
coronal lines with higher IPs and \critdens, the measured profiles of
\fevii\ are often consistent with those of the lower-ionisation
forbidden lines \citep{Veilleux91b}.

\subsection{Structure of the FHIL-emitting region}
\label{sect:FHILstructure}

What is the geometry of the FHIL-emitting region of Seyfert galaxies?
MT98 found that amongst a sample of 35 Seyferts mostly
collected from the literature, type~2 Seyferts have significantly
lower \fevii /\oiii\ flux ratios than type~1s,  suggesting that
at least part of the \fevii-emitting region is obstructed along our
line of sight in Seyfert 2 galaxies.  From this they infer that some
of the \fevii\ emission arises from regions hidden from our view by
the same structure that obscures the BLR in Sy2s,
the circumnuclear torus. 
NTM00 reaffirm this result using an expanded sample 
(Fig.~\ref{fig:FeVII-OIII_rat})
and report a similar,
albeit weaker, disparity in the distributions of $\fex / \oiii$
ratios, therefore 
concluding that the \fex-emitting clouds are less obscured than those
responsible for the \fevii\ emission.
These authors therefore argue that a significant fraction of the
\fevii\ flux arises from material at the inner surface of the torus,
whereas a larger fraction of the \fex\ emission arises
from a larger-scale region.
On the other hand, several other authors have interpreted the
correlation between critical densities, ionisation potentials, and
the width of the line profiles as evidence for a
stratified line-emitting region in which the clouds responsible for
the FHILs extend from the outer reaches of the BLR through the NLR,
with the highest-ionisation lines arising closest to the BLR
\citep{DeRobertis84b,DeRobertis86,Appenzeller88,Rodriguez02}.

\begin{figure}
  \centering
    \resizebox{8.4cm}{!}{\includegraphics{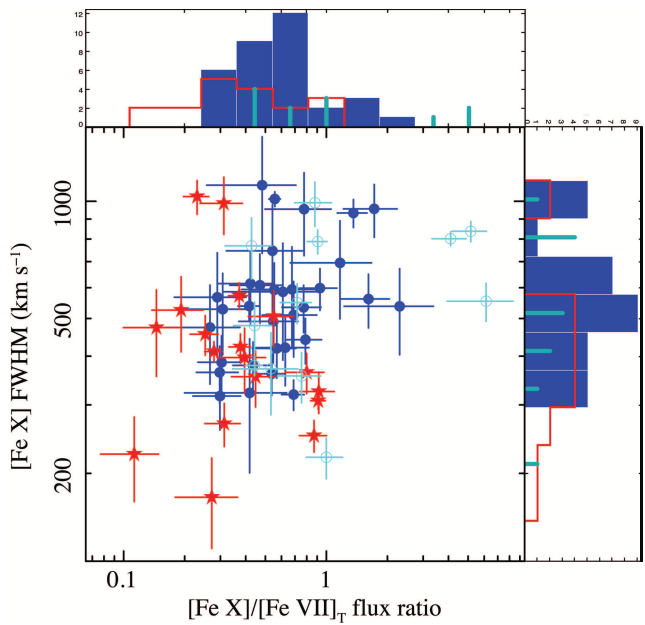}}
  \caption{\fex\ line width vs.\ \fex/\feviit\ flux ratio.  
    Point styles as in
    Fig.~\ref{fig:FeX_vs_FeVII_flux};
    histogram styles as in 
    Fig.~\ref{fig:z} (top).
    The Sy2s on average have less \fex\ power per unit of \fevii\ power
    than either the BLS1s or the NLS1s,
    although the histogram at the top demonstrates that the
    distributions are broad with substantial overlap.
    The histogram at right shows that the \fex\
    lines tend to be narrower in Seyfert 2s.  Taken together, these
    suggest that a part of the \fex-emitting region, the part that
    contributes the broadest line flux, is often hidden from us in Sy2s.
  }
  \label{fig:FeX_FWHM_Lum}
\end{figure}

As described in \S\ref{sect:lineStrengths}, we do not find any
significant difference between the \fevii/\oiii\ ratios of type 1 and
type 2 Seyferts, in contradiction with MT98 and NTM00.
\rev{We do find that the \fex/\feviit}\footnote{Here we
  discuss \fex/\fevii\ ratios instead of Fe-line/\oiii\ to mitigate
  two factors: 
  (1) the effect of reddening, for which we have not corrected, is
  less than 10 per cent for this ratio (\S\ref{sect:reddening}), and
  (2) any bias of the present \fex-selected sample in favor of high
  FHIL/non-FHIL ratios (\S\ref{sect:lineStrengths}).  However, the
  \fex/\oiii\ ratios show similar tendencies as \fex/\fevii.} 
\rev{ratios differ between the
  spectral types, but these differences are the opposite of what
  would be expected from the NTM00 results.}
The mean averages of the log ratios we obtain are 
$0.03 \pm 0.13$, $-0.24 \pm 0.04$, and $-0.44 \pm 0.06$
for NLS1s, BLS1s and Sy2s, respectively.
There is considerable overlap in the distributions (see
the histogram at the top of Fig.~\ref{fig:FeX_FWHM_Lum}), but the
probability 
that the NLS1s and Sy2s are drawn from populations with the same ratio
distribution 
is less than 0.001 per cent
according to a KS test.
The interpretation of these differences could be either an excess of
\fevii\ or a deficiency of \fex\ amongst Sy2s and the opposite
situation amongst NLS1, as will be discussed below.
Note, however, that the average ratio of the NLS1s is
increased significantly by three outliers with 
$\log(\fex/\fevii) > 0.6$.
If these extreme sources are
omitted, the mean ratio amongst the remaining 9 NLS1s ($-0.19 \pm 0.05$)
is consistent with that of the BLS1s.
Thus, the NLS1s may represent a heterogeneous population that includes
some objects with rather extreme properties and some that more closely
resemble broad-lined Sy1s.

Differences between the best-fitting \fex\ model parameters of type~1
and type~2 Seyferts provide evidence that the low \fex/\fevii\ ratios
amongst Sy2s is due to partially-obscured \fex\ emission.
In Seyfert 2s these lines tend to be narrower and have
lower fluxes relative to \fevii\
(Fig.~\ref{fig:FeX_FWHM_Lum}).
This
is consistent with a scenario in which the part of the \fex-emitting
region that contributes the broadest line flux generally cannot be
observed in Seyfert 2 galaxies, and in which the \fevii-emitting
region is less affected by obscuration.

\subsection{NLS1s with extreme 
  [Fe$\,${\sevensize\bf X}]/[O$\,${\sevensize\bf III}] ratios}
\label{sect:outliers}

In the preceding sections it was noted that a few NLS1s have extreme
line ratios that cause them to stand out from both the other NLS1s and the
other sample members in general (see
Figs.~\rev{\ref{fig:FeX_vs_FeVII_flux} and} \ref{fig:Line_vs_LxRosat}
and the discussion of \fex/\fevii\ flux ratios in \S\ref{sect:FHILstructure}).
Two NLS1s in particular, sources 25 and 50, are consistently the most
extreme outliers whenever there are outliers to a correlation.
For instance, these are the only two
sample members with \rev{$\log (\fex/\oiiit) > -1.0$}, whereas the mean for the
rest of the sample is \rev{$-1.66 \pm 0.04$ with an $\rmn{RMS} = 0.28$}.
In addition to their extreme line ratios,
these sources have two of the narrowest \hab\ lines in our sample
(FWHM values of $745 \pm 2$ and $780 \pm 2 \  \kms$ respectively).
What is it about these NLS1s that sets them apart from the rest of the sample?

Not every pair of correlated parameters has significant outliers.
For instance, there are no outliers in the \fex--\fexi\ plane
\rev{(Fig.~\ref{fig:FeXI-FeX_Flux})}, nor if
we plot \oi\ vs.\ \oiiit\
\rev{(Fig.~\ref{fig:many-OIA_Lum}a)}.
As discussed earlier, \fex\ and \fexi\ emission lines both respond to the
continuum in the soft X-ray band.  We therefore expect these three
fluxes to be closely correlated.
This is found to be the case for the present sample, without
exception.
Likewise, we do not find any obvious outliers when we compare fluxes
of pairs of NLR features (such as \oiii, \oi, \nii, or \han).
It is only when we compare fluxes between these two groups that the
outliers tend to stand out, with high ratios of \fex\ (or \fexi\ or
X-rays) over NLR fluxes.
\rev{These groups may be thought of as compact and extended emission
  regions, or as high- and low-energy processes.}

It appears that for the two consistent outliers the relationship
between the smaller-scale/higher-energy processes and the
larger-scale/lower-energy features differs from that of most other
Seyferts.  
Two possible causes for this disconnect are variability and
differences in the broad-band spectral shapes.
If the average power of the AGN has changed relatively recently
(10--100 years or so), then the compact FHIL-emitting region will have
had time to respond but the more extended NLR will not.
Alternatively, these outliers may be affected by a broadband spectral
feature that is not present (or as prominent) in other sample
members.  
\citet{Erkens97} have suggested that AGNs with strong excesses in the
soft X-ray band tend to produce strong FHILs, based upon the
correlation they demonstrated between the \fex\ equivalent widths and
the photon index of \textit{Rosat} data.
An extreme soft excess would provide a disproportionate supply of
$\sim 250 \  \rmn{eV}$ photons, thereby producing extra FHIL emission
through photoionisation; lines that respond to other parts of the
continuum will not be affected so their relative strength will be
lower.  It is worth noting that one of these outliers is KUG~1031+398,
which has one of the strongest soft excesses known\footnote{In
  addition to its extreme soft excess, KUG~1031+398 is of particular
  interest because the strongest evidence for a QPO in an AGN has
  recently been discovered in this object \citep{Gierlinski08}.}
\rev{(\citealp{Puchnarewicz01}; \citealp*{Middleton07})}.
We note further
that the two outliers have the lowest HR1 values of all the
RASS-detected members of the sample (HR1 = hardness ratio for the
low energy end of the \textit{Rosat} band), suggesting that source 50
may also have a strong soft excess.

\rev{The observation that both of our consistent outliers may have
  strong excesses in the soft X-ray band is suggestive that the
  mechanism that causes them to stand out relates to the SED shape,
  but with just two objects this does not rule out the possibility
  that these sources stand out due to variability.  To break the
  degeneracy between an origin relating to the scale size and one
  relating to the SED, we consider \hab.  The BLR is more compact than
  the FHIL-emitting region but \ha\ emission is not specifically
  sensitive to the soft X-ray continuum.  In
  Fig.~\ref{fig:Line_vs_LxRosat}c, the two data points farthest from
  the \hatt--$L_x$ correlation are sources 25 and 50, whereas
  Fig.~\ref{fig:many-HaT_Lum}b shows that there are no egregious
  outliers in the \oiiit--\hatt\ plane\footnote{Note that
    whilst this figure uses \hatt, \hatt\ is a reasonable proxy for
    \hab\ in type 1 Seyferts because the majority of the \ha\ flux is
    contributed by the broad component.}.
  Thus, the \hab\ emission in these objects tracks with the
  larger-scale/lower-energy features and not the
  smaller-scale/higher-energy
  features, indicating that sources 25 and 50 behave differently from
  most of the sample due to a difference in their SEDs, not
  variability.
}

\rev{If sources 25 and 50 do have strong soft excesses, and if this
  feature of their SED affects the emission lines, then we can use the
  emission lines to constrain portions of the soft excess that are
  difficult to observe directly.  Specifically, we can use the
  strength of \fevii\ relative to other features to constrain the
  photoionising continuum around 100~eV.
  We find that these sources are outliers in the \fevii--\fex\ flux
  plane (Fig.~\ref{fig:FeX_vs_FeVII_flux}) but not in the
  \fevii--\oiii\ plane (Fig.~\ref{fig:many-OIIIt_Lum}c).
  Thus it appears that the \fevii\ lines do not respond strongly to
  the soft excess: either this spectral component does not extend down
  to 100~eV or the bulk of the \fevii-emitting region is shielded from
  this photoionising continuum by intrinsic absorption.
}

\section{Summary and conclusions}
\label{sect:conclusions}

We have used \fex\ line emission to select a sample of galaxies from
the SDSS with strong forbidden high-ionisation line (FHIL) emission.
The resulting 63 Seyfert galaxies comprise the first ever FHIL-selected
sample of AGN.  Moreover, it is the most homogeneous and one of the
largest samples of FHIL-emitting galaxies to date. 
In each of these spectra we have measured the strengths and profiles
of several emission lines (including \fex, \fexi, \fevii, \oiii\ and
\ha).  We have also included soft X-ray data for all objects with
available \textit{Rosat} measurements.
From our analysis of these data we conclude:

\begin{itemize}
  \item That the \fex\ and \fexi\ lines are well correlated with both
    each other and the soft X-ray continuum 
    \rev{flux, with a mean \fex/$F_\rmn{233-300 eV}$ ratio of 
      $9.4 \times 10^{-3}$.}
    Their correlation with
    the X-ray strength supports a photoionisation origin of these line
    species, rather than collisional.  The correlation with X-rays
    applies to all type~1 Seyferts and may be used to estimate soft
    X-ray fluxes based upon observed FHIL strengths.
  \item There are significant differences in the \fex/\fevii\ ratios
    between narrow-lined Seyfert 1s (NLS1s), broad-lined Seyfert 1s
    (BLS1s, including both 1.0 and intermediate type 1.5 Seyferts),
    and Seyfert 2s (Sy2s), either because of selective
    absorption of the \fex\ or because of a different ionising continuum.
    In the Sy2s the ratios of \fex/\fevii\ tend to be lower and the
    FWHM of the \fex\ line profiles are narrower. 
    These facts suggest that some fraction of the \fex\ flux (i.e.,
    the broad component) may be hidden in Sy2s, possibly because
    it is emitted in a region of size comparable to the dusty
    molecular torus.
    A subset of NLS1s have high \fex/\fevii\ and \fex/\oiii\ ratios,
    which is consistent 
    with an over production of \fex\ in response to an excess of
    ionising photons in the soft X-ray band.
  \item Using the low ionisation lines of \sii\ as a proxy for the
    systemic velocity of the host galaxy, we find that \fexi, \fex\
    and \fevii\ are
    generally blueshifted with velocities of order 100~\kms.
    The most blueshifted lines tend to have the broadest profiles.
    In particular, about 1/3 of \fexi\ lines are both broader and bluer
    than the other FHILs.
    The \oiii\ lines are found to have blue wings with widths and
    velocities comparable to those of the FHILs and narrow cores at
    the systemic velocity.
  \item In our correlations there are a few sources that
    consistently are far from the general trends.  
    These outliers are type~1 Seyferts that have
    the narrowest \hab\ lines amongst our NLS1s
    and the softest X-ray spectra amongst our entire sample.
    Their deviations from the overall trends may be manifestations of
    important intrinsic differences between examples of very narrow
    line BLRs and the majority of Seyfert 1s.
  \item The good correlation between $F_x$ and \fex\ emission observed
    10--15 years later means that
    the measured X-ray flux (an instantaneous snapshot) is relatively
    representative of average X-ray power to which the FHIL-emitting
    region responds.  It follows that variability over
    \rev{time-scales ranging from that of the \textit{Rosat} exposure
      times to the lag time of the SDSS observations (several minutes
      to several years)}
    is usually no more than a factor of a few.

\end{itemize}

These findings are consistent with a stratified wind model, in which
outflowing photoionised clouds produce \fexi, \fex, \fevii\ and
possibly the wings of \oiii.  The \fexi- and \fex-emitting clouds lie
closest to the BLR at a scale comparable to the obscuring torus, which
\rev{sometimes cover} a significant fraction of these clouds in Seyfert 2s.  
\rev{\fevii, in comparison to \fex\ and \fexi, is found to correlate less
well with $F_x$ and better with \oiii; it is also apparently not
affected when the \fex\ emitting region is partially covered.
These facts suggest} that the \fevii-emitting region is more
extended and may approach the scale of the NLR.
However, the correlation between \fevii\ and \fex\ fluxes and 
\rev{in particular}
the similarity of their velocity shifts implies that the clouds emitting
these features are related to each other and may be different parts of
the same outflow.
The narrow cores of \oiii\ and the other low ionisation species such
as \oi\ do not have a velocity shift relative to \sii;
the clouds emitting these lines are either a decelerated portion of
the outflow or are unrelated to it.

Based upon the strong correlation between \fex\ and soft X-ray flux
(\S\ref{sect:FHILxrays})
and our discussion of the sources that appear to be exceptions to the
overall correlations found in the sample (\S\ref{sect:outliers}), 
we \rev{infer that strong \fex\ emission (relative to other optical
  lines such as \oiii\ and \fevii)} 
is an optical-wavelength indicator of strong soft
excesses in the X-ray band.  Furthermore, the fact that these same
objects do not have high \fevii/\oiii\ ratios suggests that these soft
excesses \rev{either} do not continue through $\sim 100$~eV into the
difficult--to--observe EUV band \rev{or the \fevii-emitting clouds are
  shielded from the soft excess by intrinsic absorption}.
\rev{The connection between FHIL emission and the intrinsic continuum
  in the few hundred eV range needs to be tested with better X-ray
  data.  A more detailed analysis of the available X-ray data would
  be useful, but the number of \fex-selected sources with available
  X-ray spectra is severely limited.  New observations would be
  necessary to better calibrate the FHIL--X-ray correlations.}

Finally, we note that our \fex\ selection
yielded only one object that is not obviously an AGN 
(source 45 here; Ward et~al.\ in preparation).
This confirms that it is exceptionally rare for non-active galaxies to
produce strong FHIL emission.  However, we cannot quantify just how
rare this is (or what fraction of AGN are strong FHIL emitters, or
whether one Seyfert class is more likely than another to have strong
FHILs) because the present study does not include a control group.  
\rev{To remedy this, we are conducting a broader search for FHIL
  emission amongst a much larger sample of SDSS emission line
  galaxies.  This next step will reduce selection effects in the
  present sample and will better enable us to put \fex-selected
  sources into context.}

\section*{Acknowledgements}
\rev{The authors wish to thank Matthew Boulton for his contributions
  in an early stage of this programme, and the anonymous referee whose
  suggestions helped to make this a stronger paper.}

\rev{This research has made use of data obtained from the High Energy
  Astrophysics Science Archive Research Center (HEASARC), provided by
  NASA's Goddard Space Flight Center; NASA's Astrophysics Data System
  (ADS); and the NASA/IPAC Extragalactic Database (NED), which is
  operated by the Jet Propulsion Laboratory, California Institute of
  Technology, under contract with the National Aeronautics and Space
  Administration.}

\rev{Funding for the SDSS and SDSS-II has been provided by the Alfred
  P. Sloan Foundation, the Participating Institutions, the National
  Science Foundation, the U.S. Department of Energy, the National
  Aeronautics and Space Administration, the Japanese Monbukagakusho,
  the Max Planck Society, and the Higher Education Funding Council for
  England. The SDSS Web Site is {http://www.sdss.org/}.}
\rev{The SDSS is managed by the Astrophysical Research Consortium for
  the Participating Institutions. The Participating Institutions are
  the American Museum of Natural History, Astrophysical Institute
  Potsdam, University of Basel, University of Cambridge, Case Western
  Reserve University, University of Chicago, Drexel University,
  Fermilab, the Institute for Advanced Study, the Japan Participation
  Group, Johns Hopkins University, the Joint Institute for Nuclear
  Astrophysics, the Kavli Institute for Particle Astrophysics and
  Cosmology, the Korean Scientist Group, the Chinese Academy of
  Sciences (LAMOST), Los Alamos National Laboratory, the
  Max-Planck-Institute for Astronomy (MPIA), the Max-Planck-Institute
  for Astrophysics (MPA), New Mexico State University, Ohio State
  University, University of Pittsburgh, University of Portsmouth,
  Princeton University, the United States Naval Observatory, and the
  University of Washington.}

\bibliography{references.bib}
\bibliographystyle{mn2e}

\label{lastpage}

\appendix

\section{Online supplementary material}

\rev{Online we provide
Tables~\ref{tab:FexData}--\ref{tab:HaData} in which we present the
parameters of our best-fitting models to the emission lines in the
SDSS spectra.
In Table~\ref{tab:XrayFluxes} we provide the compiled \textit{Rosat}
count data used in this study: count rates from the \textit{Rosat}
All-Sky Survey (RASS; \citealp{RASS-BSC,RASS-FSC}) and the
\citet*{WGACAT2000} Catalogue of sources
found in pointed \textit{Rosat} observations, RASS hardness ratios,
and flux estimates based upon these count rates.}

\begin{figure}
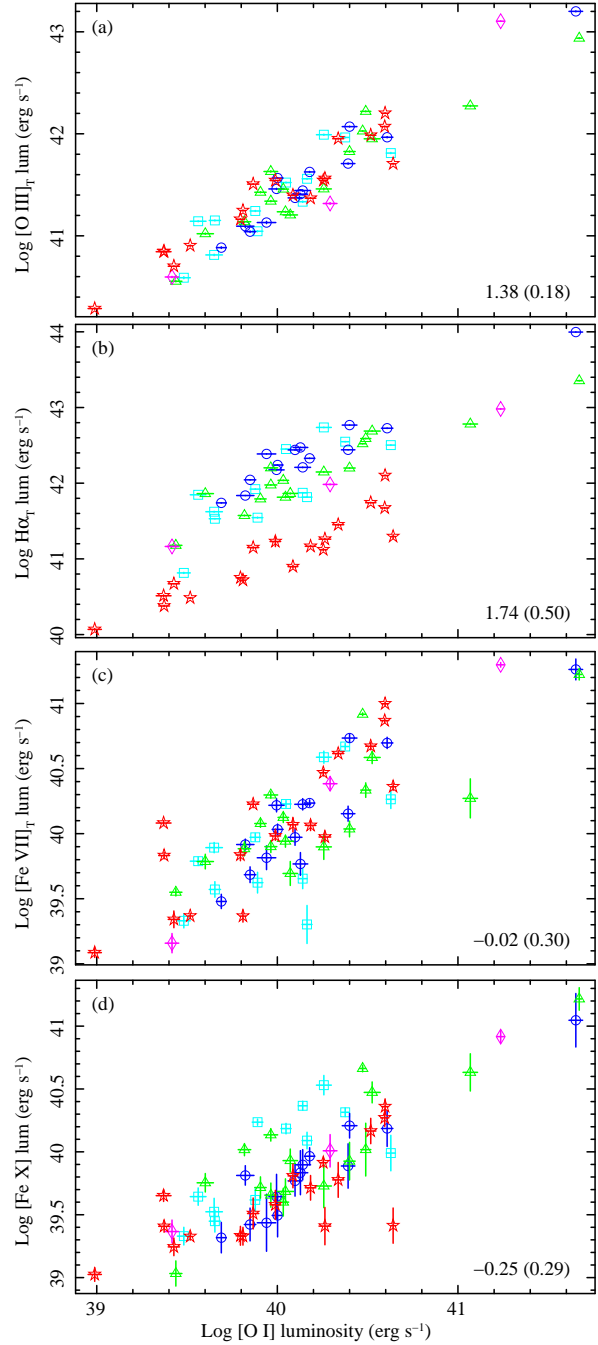

  \centering
  \rotatebox{270}{\resizebox{!}{7.8cm}{\includegraphics{figA1a.eps}}}
  \rotatebox{270}{\resizebox{!}{7.8cm}{\includegraphics{figA1b.eps}}}
  \rotatebox{270}{\resizebox{!}{7.8cm}{\includegraphics{figA1c.eps}}}
  \rotatebox{270}{\resizebox{!}{7.8cm}{\includegraphics{figA1d.eps}}}
  \caption{\rev{Luminosities of (a) \oiiit, (b) \hatt, (c) \feviit,
      and (d) \fex\ vs.\ \oi.
      Spectral classifications are indicated by point colors and
      styles as follows:
      NLS1 = cyan squares, Sy1.0 = blue circles, Sy1.5 = green
      triangles, Sy1.9 = magenta diamonds, and Sy2 = red stars.
      The numbers in the lower-right corners are the mean of the log (
      line / \oi\ ) values, with the RMS about this average ratio
      given in parenthesis.
    }
 }
  \label{fig:many-OIA_Lum}
\end{figure}

\begin{figure}
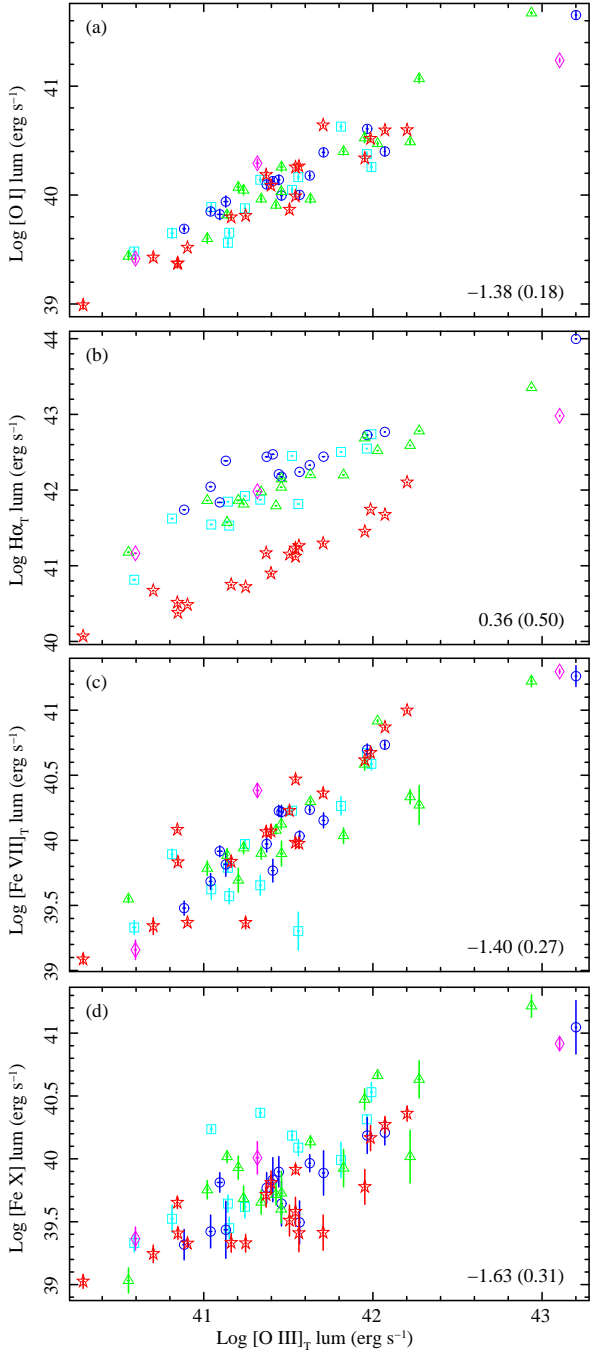

  \centering
  \rotatebox{270}{\resizebox{!}{7.8cm}{\includegraphics{figA2a.eps}}}
  \rotatebox{270}{\resizebox{!}{7.8cm}{\includegraphics{figA2b.eps}}}
  \rotatebox{270}{\resizebox{!}{7.8cm}{\includegraphics{figA2c.eps}}}
  \rotatebox{270}{\resizebox{!}{7.8cm}{\includegraphics{figA2d.eps}}}
  \caption{\rev{Luminosities of (a) \oi, (b) \hatt, (c) \feviit,
      and (d) \fex\ vs.\ \oiiit.
    Spectral classifications are indicated as in
    Fig.\ \ref{fig:many-OIA_Lum} and the numbers are again the mean of
    the logarithmic ratios (this time over \oiiit) and the
    RMS about this average (in parenthesis).}
 }
  \label{fig:many-OIIIt_Lum}
\end{figure}

\begin{figure}
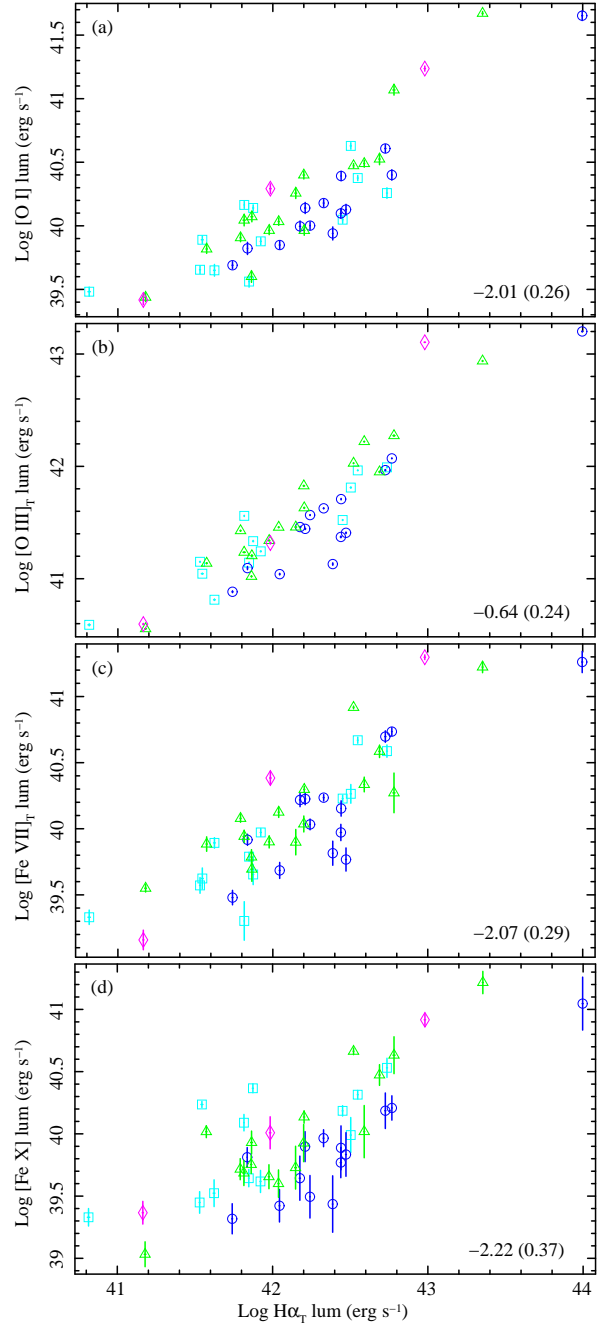

  \centering
  \rotatebox{270}{\resizebox{!}{7.8cm}{\includegraphics{figA3a.eps}}}
  \rotatebox{270}{\resizebox{!}{7.8cm}{\includegraphics{figA3b.eps}}}
  \rotatebox{270}{\resizebox{!}{7.8cm}{\includegraphics{figA3c.eps}}}
  \rotatebox{270}{\resizebox{!}{7.8cm}{\includegraphics{figA3d.eps}}}
  \caption{\rev{Luminosities of (a) \oi, (b) \oiiit, (c) \feviit,
      and (d) \fex\ vs.\ \hatt.
      Sy2 sources are omitted because their \hatt\ luminosities are
      systematically lower because the broad component is not detected
      (see, e.g., Fig.\ \ref{fig:many-OIIIt_Lum}b).
      Apart from the lack of Sy2s, point styles identical to
      Fig.\ \ref{fig:many-OIA_Lum} and the
    numbers are again the mean of the logarithmic ratios (over \hatt)
    and the RMS about this average (in parenthesis).}
 }
  \label{fig:many-HaT_Lum}
\end{figure}

\begin{figure}
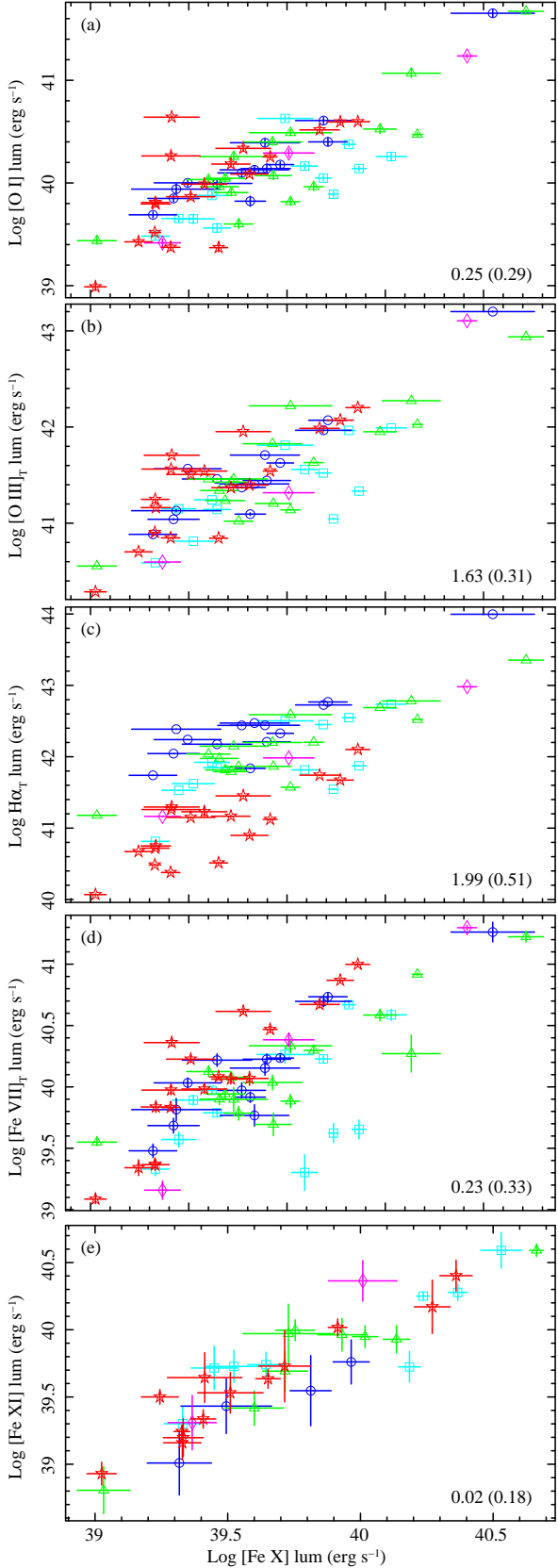

  \centering
  \rotatebox{270}{\resizebox{!}{7.8cm}{\includegraphics{figA4a.eps}}}
  \rotatebox{270}{\resizebox{!}{7.8cm}{\includegraphics{figA4b.eps}}}
  \rotatebox{270}{\resizebox{!}{7.8cm}{\includegraphics{figA4c.eps}}}
  \rotatebox{270}{\resizebox{!}{7.8cm}{\includegraphics{figA4d.eps}}}
  \rotatebox{270}{\resizebox{!}{7.8cm}{\includegraphics{figA4e.eps}}}
  \caption{\rev{Luminosities of (a) \oi, (b) \oiiit, (c) \hatt, (d)
      \feviit, and (e) \fexi\ vs.\ \fex.
    Point styles as in Fig.\ \ref{fig:many-OIA_Lum} and the
    numbers are the mean of the logarithmic ratios (over \fex) and the 
    RMS about this average (in parenthesis).}
 }
  \label{fig:many-FeX_Lum}
\end{figure}

\begin{figure}
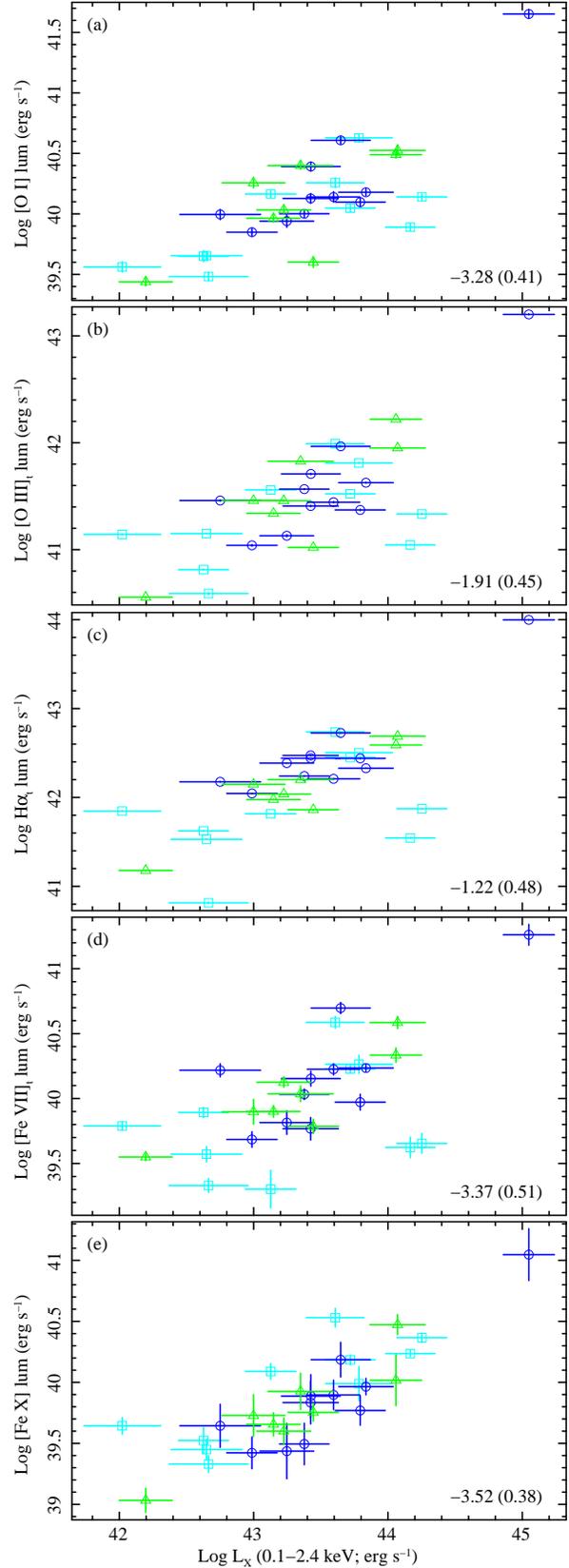

  \centering
  \rotatebox{270}{\resizebox{!}{7.8cm}{\includegraphics{figA5a.eps}}}
  \rotatebox{270}{\resizebox{!}{7.8cm}{\includegraphics{figA5b.eps}}}
  \rotatebox{270}{\resizebox{!}{7.8cm}{\includegraphics{figA5c.eps}}}
  \rotatebox{270}{\resizebox{!}{7.8cm}{\includegraphics{figA5d.eps}}}
  \rotatebox{270}{\resizebox{!}{7.8cm}{\includegraphics{figA5e.eps}}}
  \caption{\rev{Luminosities of (a) \oi, (b) \oiiit, (c) \hatt, (d)
      \feviit, and (e) \fex\ vs.\ $L_X$ in the 0.1--2.4~keV band.
    Point styles as in Fig.\ \ref{fig:many-OIA_Lum} and the
    numbers are again the mean of the logarithmic ratios (over $F_x$)
    and the RMS about this average.  The three Sy2s with
    \textit{Rosat} detections are omitted because they
    are likely to be affected by X-ray absorption.}
 }
  \label{fig:many-Rosat_Lum}
\end{figure}

\rev{Also provided online are a set of figures showing
various line luminosities plotted against the luminosities of
\oi\ (Fig.~\ref{fig:many-OIA_Lum}),
\oiiit\ (\ref{fig:many-OIIIt_Lum}), \hatt\ (\ref{fig:many-HaT_Lum}),
\fex\ (\ref{fig:many-FeX_Lum}), 0.1--2.4~keV X-rays
(\ref{fig:many-Rosat_Lum}).}

\bsp

\newpage   


\newpage   

\newpage   

\newpage   

\newpage   

\clearpage

\begin{table*}
\begin{minipage}{90mm}
\caption{\rev{Best-fitting \fex\ models.}}
\label{tab:FexData}
{ \centering
\begin{tabular}{rr@{$\,\pm\,$}lr@{$\,\pm\,$}lr@{$\,\pm\,$}lr@{$\,\pm\,$}lr}
\hline
ID	& \multicolumn{2}{c}{FWHM}	& \multicolumn{2}{c}{$v_\rmn{sh}$}	& \multicolumn{2}{c}{Flux}	& \multicolumn{2}{c}{EqW}	& $\Delta \chi^2$	\\
	& \multicolumn{2}{c}{(\kms)}	& \multicolumn{2}{c}{(\kms)}		& \multicolumn{2}{c}{($\times 10^{-16}$)}	& \multicolumn{2}{c}{(\AA)}	& 	\\
\hline
1	& 560	& 90	& -50	& 40	& 2.4	& 0.5	& 1.1	& 0.2	& 81.5	\\
2	& 770	& 140	& -30	& 60	& 4.2	& 1.0	& 1.3	& 0.3	& 59.8	\\
3	& 360	& 40	& -248.	& 19.	& 16.	& 3.	& 0.73	& 0.12	& 135.0	\\
4	& 590	& 170	& 20	& 70	& 2.6	& 1.0	& 0.7	& 0.3	& 25.3	\\
5	& 990	& 170	& 260	& 70	& 4.5	& 1.0	& 1.9	& 0.4	& 68.2	\\
6	& 510	& 80	& -10	& 40	& 1.9	& 0.4	& 2.5	& 0.5	& 77.2	\\
7	& 570	& 40	& -160.	& 19.	& 19.	& 2.	& 1.86	& 0.20	& 295.6	\\
8	& 350	& 60	& -160	& 30	& 2.1	& 0.5	& 1.3	& 0.3	& 67.5	\\
9	& 530	& 120	& -130	& 50	& 3.6	& 1.1	& 0.8	& 0.2	& 34.8	\\
10	& 320	& 60	& -100	& 20	& 4.4	& 1.1	& 0.70	& 0.17	& 55.6	\\
11	& 360	& 60	& -100	& 30	& 1.8	& 0.4	& 0.88	& 0.19	& 75.0	\\
12	& 590	& 190	& -330	& 90	& 6.	& 3.	& 0.5	& 0.2	& 15.5	\\
13	& 530	& 110	& 60	& 50	& 4.6	& 1.3	& 0.72	& 0.20	& 40.2	\\
14	& 360	& 40	& -19.	& 19.	& 3.2	& 0.5	& 1.02	& 0.16	& 153.4	\\
15	& 400	& 80	& -70	& 30	& 3.1	& 0.8	& 1.0	& 0.3	& 22.0	\\
16	& 540	& 130	& -40	& 60	& 1.8	& 0.6	& 1.7	& 0.5	& 32.6	\\
17	& 460	& 50	& 0	& 20	& 3.7	& 0.6	& 2.2	& 0.3	& 150.8	\\
18	& 320	& 30	& -60.	& 12.	& 5.0	& 0.6	& 2.6	& 0.3	& 303.8	\\
19	& 380	& 60	& -90	& 20	& 2.8	& 0.5	& 1.7	& 0.3	& 93.6	\\
20	& 790	& 60	& 10	& 20	& 62.	& 6.	& 2.1	& 0.2	& 373.1	\\
21	& 390	& 70	& -50	& 30	& 11.	& 2.	& 0.76	& 0.17	& 68.4	\\
22	& 540	& 40	& -54.	& 19.	& 8.7	& 1.0	& 4.1	& 0.5	& 262.5	\\
23	& 600	& 70	& -250	& 30	& 10.3	& 1.7	& 1.4	& 0.2	& 119.2	\\
24	& 220	& 30	& -87.	& 11.	& 1.9	& 0.3	& 1.5	& 0.2	& 134.7	\\
25	& 800	& 30	& -352.	& 14.	& 41.	& 2.	& 5.5	& 0.3	& 993.4	\\
26	& 1100	& 400	& -380	& 160	& 2.8	& 1.3	& 1.0	& 0.4	& 17.4	\\
27	& 550	& 60	& -30	& 30	& 3.5	& 0.5	& 3.6	& 0.5	& 150.8	\\
28	& 1400	& 300	& -40	& 90	& 8.	& 2.	& 3.3	& 0.9	& 104.3	\\
29	& 530	& 70	& -260	& 30	& 3.2	& 0.6	& 1.9	& 0.4	& 93.9	\\
30	& 320	& 120	& 130	& 50	& 2.2	& 1.0	& 0.41	& 0.20	& 16.3	\\
31	& 420	& 30	& -30.	& 15.	& 10.7	& 1.2	& 2.1	& 0.2	& 290.1	\\
32	& 610	& 130	& -290	& 60	& 3.3	& 0.9	& 1.2	& 0.4	& 39.3	\\
33	& 610	& 120	& -170	& 50	& 3.9	& 1.1	& 1.0	& 0.3	& 46.9	\\
34	& 250	& 20	& -32.	& 10.	& 8.8	& 1.1	& 1.52	& 0.19	& 207.0	\\
35	& 307.	& 15.	& -141.	& 7.	& 22.6	& 1.5	& 2.35	& 0.16	& 715.0	\\
36	& 1000	& 200	& 140	& 100	& 2.6	& 0.9	& 1.3	& 0.4	& 31.0	\\
37	& 320	& 40	& -33.	& 17.	& 7.5	& 1.2	& 1.20	& 0.19	& 133.0	\\
38	& 440	& 60	& 20	& 30	& 3.9	& 0.7	& 1.8	& 0.3	& 112.5	\\
39	& 170	& 50	& -46.	& 19.	& 0.8	& 0.3	& 0.7	& 0.2	& 19.0	\\
40	& 370	& 90	& -40	& 40	& 2.2	& 0.7	& 0.7	& 0.2	& 36.3	\\
41	& 510	& 110	& 130	& 50	& 18.	& 5.	& 0.73	& 0.20	& 53.3	\\
42	& 960	& 150	& -320	& 60	& 5.3	& 1.1	& 2.4	& 0.5	& 86.6	\\
43	& 470	& 120	& -220	& 50	& 2.3	& 0.7	& 1.2	& 0.4	& 40.0	\\
44	& 590	& 100	& 60	& 40	& 4.4	& 1.0	& 1.2	& 0.3	& 72.3	\\
45	& 222.	& 8.	& 24.	& 3.	& 9.0	& 0.4	& 9.0	& 0.4	& 1239.5	\\
46	& 990	& 130	& -410	& 60	& 10.4	& 1.8	& 1.9	& 0.3	& 112.0	\\
47	& 690	& 200	& 0	& 80	& 4.0	& 1.5	& 0.6	& 0.2	& 22.9	\\
48	& 420	& 70	& -120	& 30	& 4.3	& 0.9	& 1.0	& 0.2	& 71.1	\\
49	& 1030	& 100	& -190	& 40	& 11.4	& 1.6	& 2.8	& 0.4	& 171.0	\\
50	& 840	& 50	& -250	& 20	& 11.3	& 0.9	& 4.7	& 0.4	& 472.2	\\
51	& 220	& 60	& 70	& 20	& 2.1	& 0.7	& 0.7	& 0.2	& 41.8	\\
52	& 470	& 130	& 30	& 60	& 2.6	& 1.0	& 0.6	& 0.2	& 23.1	\\
53	& 380	& 20	& -56.	& 9.	& 15.3	& 1.2	& 2.35	& 0.18	& 547.9	\\
54	& 490	& 110	& 110	& 50	& 9.	& 3.	& 0.67	& 0.19	& 31.6	\\
\hline
\end{tabular}
}

\medskip
\rev{Columns are:
(1)	\fex\ sample ID number;
(2--5)	parameters of the single-Gaussian FHIL model: (2) FWHM (\kms),
	(3) velocity shift of the centroid relative to the systemic
	redshift defined by \sii\ 
        (\kms; $v_\rmn{sh} < 0$ 
        indicates that the line is blueshifted relative to
        the host galaxy), (4) model flux ($\times 10^{-16} \ 
        \cgsflux$) and (5) equivalent width (\AA);
(6)	amount by which $\chi^2$ improves when a single Gaussian
        component is included to model the FHIL.}
\end{minipage}
\end{table*}
\begin{table*}
\begin{minipage}{90mm}
\contcaption{}
{ \centering
\begin{tabular}{rr@{$\,\pm\,$}lr@{$\,\pm\,$}lr@{$\,\pm\,$}lr@{$\,\pm\,$}lr}
\hline
ID	& \multicolumn{2}{c}{FWHM}	& \multicolumn{2}{c}{$v_\rmn{sh}$}	& \multicolumn{2}{c}{Flux}	& \multicolumn{2}{c}{EqW}	& $\Delta \chi^2$	\\
	& \multicolumn{2}{c}{(\kms)}	& \multicolumn{2}{c}{(\kms)}		& \multicolumn{2}{c}{($\times 10^{-16}$)}	& \multicolumn{2}{c}{(\AA)}	& 	\\
\hline
55	& 270	& 30	& 22.	& 15.	& 4.4	& 0.7	& 1.10	& 0.19	& 116.4	\\
56	& 570	& 170	& -50	& 70	& 15.	& 6.	& 0.5	& 0.2	& 27.4	\\
57	& 410	& 20	& -164.	& 13.	& 8.5	& 0.7	& 2.7	& 0.2	& 322.7	\\
58	& 420	& 80	& -20	& 40	& 5.1	& 1.4	& 0.8	& 0.2	& 46.9	\\
59	& 550	& 60	& 70	& 30	& 8.4	& 1.3	& 1.22	& 0.19	& 138.1	\\
60	& 930	& 80	& 160	& 30	& 15.2	& 1.7	& 3.5	& 0.4	& 281.4	\\
61	& 360	& 50	& -190	& 20	& 2.2	& 0.4	& 1.6	& 0.3	& 91.4	\\
62	& 480	& 70	& -150	& 30	& 5.0	& 1.0	& 1.2	& 0.2	& 84.5	\\
63	& 700	& 200	& -160	& 110	& 3.2	& 1.2	& 0.8	& 0.3	& 23.9	\\
64	& 1010	& 40	& -397.	& 19.	& 35.	& 2.	& 4.5	& 0.3	& 864.6	\\
\hline
\end{tabular}
}

\end{minipage}
\end{table*}

\begin{table*}
\begin{minipage}{87mm}
\caption{\rev{Best-fitting \fexi\ models.}}
\label{tab:FexiData}
{ \centering
\begin{tabular}{rr@{$\,\pm\,$}lr@{$\,\pm\,$}lr@{$\,\pm\,$}lr@{$\,\pm\,$}lr}
\hline
ID	& \multicolumn{2}{c}{FWHM}	& \multicolumn{2}{c}{$v_\rmn{sh}$}	& \multicolumn{2}{c}{Flux}	& \multicolumn{2}{c}{EqW}	& $\Delta \chi^2$	\\
	& \multicolumn{2}{c}{(\kms)}	& \multicolumn{2}{c}{(\kms)}		& \multicolumn{2}{c}{($\times 10^{-16}$)}	& \multicolumn{2}{c}{(\AA)}	& 	\\
\hline
1	& 490	& 180	& -290	& 90	& 2.1	& 0.9	& 1.1	& 0.5	& 24.9	\\
2	& 510	& 90	& -260	& 40	& 6.8	& 1.8	& 2.1	& 0.6	& 46.4	\\
3	& 540	& 50	& -300	& 20	& 28.	& 4.	& 1.44	& 0.18	& 209.1	\\
4	& 900	& 400	& -120	& 150	& 5.	& 2.	& 1.3	& 0.6	& 17.6	\\
7	& 730	& 90	& -480	& 40	& 18.	& 3.	& 1.8	& 0.3	& 141.7	\\
8	& 330	& 150	& -300	& 70	& 2.2	& 1.2	& 1.5	& 0.8	& 13.8	\\
10	& 250	& 50	& -150	& 30	& 2.9	& 0.8	& 0.51	& 0.14	& 14.4	\\
13	& 440	& 110	& -10	& 50	& 4.8	& 1.6	& 0.8	& 0.3	& 32.4	\\
14	& 320	& 80	& -60	& 30	& 2.0	& 0.7	& 0.7	& 0.3	& 25.8	\\
17	& 390	& 110	& -30	& 50	& 2.9	& 1.3	& 2.0	& 0.9	& 17.3	\\
18	& 270	& 50	& -126.	& 20.	& 3.1	& 0.7	& 2.0	& 0.5	& 65.5	\\
19	& 380	& 100	& -30	& 40	& 2.7	& 0.9	& 1.8	& 0.6	& 31.4	\\
20	& 650	& 130	& -230	& 50	& 21.	& 5.	& 0.9	& 0.2	& 53.9	\\
21	& 290	& 80	& -100	& 30	& 6.	& 2.	& 0.48	& 0.18	& 22.3	\\
23	& 1450	& 190	& -550	& 80	& 18.	& 3.	& 2.9	& 0.5	& 110.1	\\
24	& 230	& 50	& -100	& 20	& 1.7	& 0.5	& 1.6	& 0.5	& 44.0	\\
25	& 850	& 40	& -438.	& 18.	& 42.	& 3.	& 6.6	& 0.4	& 731.6	\\
31	& 410	& 50	& -100	& 20	& 9.1	& 1.4	& 1.9	& 0.3	& 138.0	\\
32	& 800	& 200	& -760	& 100	& 7.	& 2.	& 2.7	& 0.9	& 41.8	\\
34	& 260	& 40	& -35.	& 16.	& 7.1	& 1.4	& 1.3	& 0.3	& 88.5	\\
35	& 281.	& 18.	& -180.	& 8.	& 18.7	& 1.7	& 2.11	& 0.19	& 412.4	\\
37	& 290	& 60	& -60	& 30	& 5.1	& 1.4	& 0.9	& 0.2	& 41.6	\\
38	& 450	& 190	& 10	& 80	& 2.1	& 1.1	& 1.0	& 0.5	& 12.6	\\
41	& 370	& 140	& -50	& 60	& 9.	& 4.	& 0.4	& 0.2	& 14.9	\\
42	& 1000	& 200	& -390	& 90	& 5.7	& 1.6	& 3.4	& 0.9	& 51.9	\\
45	& 220.	& 13.	& 20.	& 6.	& 7.2	& 0.6	& 7.9	& 0.7	& 478.1	\\
46	& 1400	& 300	& -350	& 130	& 12.	& 4.	& 2.6	& 0.8	& 35.3	\\
49	& 1300	& 300	& -560	& 110	& 13.	& 3.	& 3.5	& 0.9	& 55.4	\\
50	& 640	& 70	& -240	& 30	& 9.2	& 1.3	& 4.6	& 0.6	& 198.6	\\
51	& 520	& 160	& -70	& 70	& 3.6	& 1.5	& 1.3	& 0.5	& 24.0	\\
55	& 300	& 70	& -20	& 30	& 3.2	& 1.0	& 0.9	& 0.3	& 36.7	\\
56	& 700	& 200	& -60	& 100	& 13.	& 6.	& 0.6	& 0.3	& 18.5	\\
57	& 600	& 60	& -270	& 30	& 10.8	& 1.6	& 3.7	& 0.5	& 152.8	\\
59	& 990	& 160	& -270	& 70	& 10.	& 2.	& 1.7	& 0.4	& 77.5	\\
60	& 970	& 140	& 50	& 60	& 13.	& 2.	& 3.1	& 0.6	& 101.3	\\
61	& 1600	& 400	& -180	& 170	& 4.0	& 1.4	& 3.6	& 1.3	& 28.0	\\
64	& 1210	& 90	& -530	& 40	& 30.	& 3.	& 4.4	& 0.5	& 299.2	\\
\hline
\end{tabular}
}

\medskip
\rev{The columns are the same as in Table~\ref{tab:FexData}.}

\medskip
\rev{Sample members omitted from this table either had \fexi\ models that
were rejected as not significant ($\Delta \chi^2 < 11.3$; i.e., below
the 99 per cent confidence threshold) or spurious (\S\ref{sect:interpSpurious}),
or did not have data covering $\lambda = 7892$ \AA\ in the emitted frame.}
\end{minipage}
\end{table*}

\begin{table*}
\begin{minipage}{128mm}
\caption{\rev{Best-fitting \fevii\ models.}}
\label{tab:FeviiData}
{ \centering
\begin{tabular}{rccr@{$\,\pm\,$}lr@{$\,\pm\,$}lr@{$\,\pm\,$}lr@{$\,\pm\,$}lr}
\hline
ID	& Model	& Comp.	& \multicolumn{2}{c}{FWHM}	& \multicolumn{2}{c}{$v_\rmn{sh}$}	& \multicolumn{2}{c}{Flux}	& \multicolumn{2}{c}{EqW}	& $\Delta \chi^2$	\\
	& 	& 	& \multicolumn{2}{c}{(\kms)}	& \multicolumn{2}{c}{(\kms)}		& \multicolumn{2}{c}{($\times 10^{-16}$)}	& \multicolumn{2}{c}{(\AA)}	& 	\\
\hline
1	& 1 Gaussian	& total	& 260	& 30	& -44.	& 12.	& 1.5	& 0.3	& 0.70	& 0.12	& 60.9	\\
2	& 1 Gaussian	& total	& 860	& 60	& -290	& 20	& 9.9	& 0.9	& 3.1	& 0.3	& 288.2	\\
3	& 1 Gaussian	& total	& 710	& 80	& -300	& 30	& 20.	& 3.	& 0.88	& 0.13	& 108.9	\\
4	& 1 Gaussian	& total	& 750	& 120	& -270	& 50	& 3.8	& 0.8	& 1.1	& 0.2	& 50.5	\\
5	& 1 Gaussian	& total	& 1030	& 50	& 100	& 20	& 14.3	& 0.9	& 5.9	& 0.4	& 564.7	\\
6	& 1 Gaussian	& total	& 950	& 90	& -60	& 40	& 3.4	& 0.4	& 4.5	& 0.6	& 148.6	\\
7	& 1 Gaussian	& total	& 800	& 20	& -245.	& 10.	& 49.	& 2.	& 4.69	& 0.19	& 1189.3	\\
	& 2 Gaussians	& total	& \multicolumn{2}{c}{\nodata}	& \multicolumn{2}{c}{\nodata}	& 52.	& 2.	& 5.0	& 0.2	& 82.8	\\
	& 		& core	& 318.	& 11.	& -81.	& 3.	& 13.4	& 1.2	& 1.28	& 0.12	& \multicolumn{1}{c}{\nodata}	\\
	& 		& wing	& 860	& 30	& -442	& 20	& 38.3	& 2.0	& 3.67	& 0.19	& \multicolumn{1}{c}{\nodata}	\\
8	& 1 Gaussian	& total	& 570	& 40	& -154.	& 15.	& 4.7	& 0.4	& 3.0	& 0.3	& 266.6	\\
9	& 1 Gaussian	& total	& 850	& 60	& -220	& 30	& 11.6	& 1.2	& 2.6	& 0.3	& 225.7	\\
10	& 1 Gaussian	& total	& 359.	& 18.	& -47.	& 7.	& 12.7	& 0.9	& 2.06	& 0.15	& 354.3	\\
	& 2 Gaussians	& total	& \multicolumn{2}{c}{\nodata}	& \multicolumn{2}{c}{\nodata}	& 14.7	& 1.3	& 2.4	& 0.2	& 18.6	\\
	& 		& core	& 289.	& 11.	& -32.	& 3.	& 8.6	& 0.7	& 1.39	& 0.11	& \multicolumn{1}{c}{\nodata}	\\
	& 		& wing	& 660	& 80	& -260	& 70	& 6.1	& 1.1	& 0.99	& 0.18	& \multicolumn{1}{c}{\nodata}	\\
11	& 1 Gaussian	& total	& 720	& 40	& -108.	& 16.	& 5.9	& 0.4	& 3.3	& 0.2	& 504.7	\\
12	& 1 Gaussian	& total	& 450	& 60	& -180	& 20	& 9.4	& 1.7	& 0.85	& 0.16	& 59.3	\\
13	& 1 Gaussian	& total	& 770	& 40	& -27.	& 15.	& 21.6	& 1.3	& 3.4	& 0.2	& 639.9	\\
	& 2 Gaussians	& total	& \multicolumn{2}{c}{\nodata}	& \multicolumn{2}{c}{\nodata}	& 23.7	& 1.5	& 3.7	& 0.2	& 29.7	\\
	& 		& core	& 340	& 20	& 67.	& 7.	& 4.8	& 0.7	& 0.75	& 0.10	& \multicolumn{1}{c}{\nodata}	\\
	& 		& wing	& 1030	& 50	& -150	& 30	& 18.9	& 1.3	& 3.0	& 0.2	& \multicolumn{1}{c}{\nodata}	\\
14	& 1 Gaussian	& total	& 333.	& 13.	& -21.	& 5.	& 6.0	& 0.3	& 1.92	& 0.11	& 579.9	\\
15	& 1 Gaussian	& total	& 440	& 20	& -60.	& 9.	& 7.8	& 0.6	& 2.46	& 0.19	& 351.8	\\
16	& 1 Gaussian	& total	& 240	& 50	& -120	& 20	& 0.8	& 0.3	& 0.8	& 0.3	& 14.2	\\
17	& 1 Gaussian	& total	& 690	& 20	& -21.	& 9.	& 14.2	& 0.6	& 7.9	& 0.3	& 1432.6	\\
	& 2 Gaussians	& total	& \multicolumn{2}{c}{\nodata}	& \multicolumn{2}{c}{\nodata}	& 14.7	& 0.6	& 8.2	& 0.3	& 114.6	\\
	& 		& core	& 403.	& 8.	& 39.	& 3.	& 9.4	& 0.3	& 5.22	& 0.19	& \multicolumn{1}{c}{\nodata}	\\
	& 		& wing	& 440	& 30	& -447.	& 7.	& 5.3	& 0.4	& 3.0	& 0.2	& \multicolumn{1}{c}{\nodata}	\\
18	& 1 Gaussian	& total	& 262.	& 8.	& -44.	& 3.	& 6.8	& 0.3	& 3.54	& 0.18	& 648.9	\\
	& 2 Gaussians	& total	& \multicolumn{2}{c}{\nodata}	& \multicolumn{2}{c}{\nodata}	& 7.3	& 0.4	& 3.8	& 0.2	& 12.5	\\
	& 		& core	& 243.	& 3.	& -40.	& 2.	& 6.4	& 0.3	& 3.34	& 0.14	& \multicolumn{1}{c}{\nodata}	\\
	& 		& wing	& 260	& 60	& -350	& 20	& 0.9	& 0.3	& 0.47	& 0.16	& \multicolumn{1}{c}{\nodata}	\\
19	& 1 Gaussian	& total	& 401.	& 19.	& -114.	& 7.	& 6.5	& 0.4	& 3.8	& 0.3	& 460.1	\\
20	& 1 Gaussian	& total	& 590	& 30	& -43.	& 10.	& 68.	& 4.	& 2.41	& 0.14	& 626.8	\\
21	& 1 Gaussian	& total	& 470	& 20	& -87.	& 9.	& 35.	& 2.	& 2.49	& 0.16	& 500.7	\\
22	& 1 Gaussian	& total	& 710	& 20	& -60.	& 9.	& 21.0	& 0.9	& 11.2	& 0.5	& 1319.2	\\
23	& 1 Gaussian	& total	& 670	& 60	& -310	& 30	& 11.1	& 1.4	& 1.46	& 0.19	& 131.9	\\
24	& 1 Gaussian	& total	& 260	& 20	& -94.	& 8.	& 1.9	& 0.2	& 1.51	& 0.19	& 106.5	\\
25	& 1 Gaussian	& total	& 940	& 130	& -200	& 60	& 9.9	& 1.8	& 1.4	& 0.2	& 73.1	\\
26	& 1 Gaussian	& total	& 590	& 60	& -70	& 20	& 5.9	& 0.8	& 2.1	& 0.3	& 140.8	\\
27	& 1 Gaussian	& total	& 170	& 30	& -37.	& 14.	& 0.56	& 0.18	& 0.57	& 0.18	& 14.3	\\
28	& 1 Gaussian	& total	& 1000	& 80	& 270	& 30	& 9.6	& 0.9	& 5.0	& 0.5	& 293.9	\\
29	& 1 Gaussian	& total	& 520	& 40	& -193.	& 16.	& 4.1	& 0.4	& 2.6	& 0.3	& 188.8	\\
30	& 1 Gaussian	& total	& 650	& 100	& 90	& 40	& 5.2	& 1.1	& 1.0	& 0.2	& 51.9	\\
31	& 1 Gaussian	& total	& 534.	& 20.	& -25.	& 8.	& 22.7	& 1.2	& 4.5	& 0.2	& 858.1	\\
	& 2 Gaussians	& total	& \multicolumn{2}{c}{\nodata}	& \multicolumn{2}{c}{\nodata}	& 28.4	& 1.7	& 5.6	& 0.3	& 115.7	\\
	& 		& core	& 372.	& 8.	& 32.	& 2.	& 13.5	& 0.7	& 2.65	& 0.14	& \multicolumn{1}{c}{\nodata}	\\
	& 		& wing	& 910	& 70	& -380	& 30	& 14.9	& 1.5	& 2.9	& 0.3	& \multicolumn{1}{c}{\nodata}	\\
\hline
\end{tabular}
}

\medskip
\rev{Columns are:
(1)	\fex\ sample ID number;
(2)	emission line model used (either single- or double-Gaussian);
(3)	line component (``total'' for the overall properties of
	single- and double-Gaussian models, ``core'' and ``wing'' for
        the narrower and broader components within the double-Gaussian
        models);
(4)	Gaussian FWHM (in \kms);
(5)	velocity shift of the Gaussian centroid, measured relative to the
        systemic redshift defined by \sii\ (in \kms; $v_\rmn{sh} < 0$ when the
        emission line is less redshifted than the host galaxy);
(6)	line model flux ($\times 10^{-16} \  \cgsflux$) and 
(7)	equivalent width (\AA); and
(8)	improvement in $\chi^2$ over a model with one less line
        component (i.e., comparing single-Gaussian models to
        spectral models with no \fevii\ line and comparing
        double-Gaussian models to single-Gaussian models).}

\medskip
\rev{Details are provided for double-Gaussian models only when the
additional component provides a significantly improved fit ($\Delta
\chi^2 > 11.35$, hence $>99$ per cent confidence with the addition of three
unconstrained model parameters) and does not produce a spurious result
(\S\ref{sect:interpSpurious}).}
\end{minipage}
\end{table*}
\begin{table*}
\begin{minipage}{128mm}
\contcaption{}
{ \centering
\begin{tabular}{rccr@{$\,\pm\,$}lr@{$\,\pm\,$}lr@{$\,\pm\,$}lr@{$\,\pm\,$}lr}
\hline
ID	& Model	& Comp.	& \multicolumn{2}{c}{FWHM}	& \multicolumn{2}{c}{$v_\rmn{sh}$}	& \multicolumn{2}{c}{Flux}	& \multicolumn{2}{c}{EqW}	& $\Delta \chi^2$	\\
	& 	& 	& \multicolumn{2}{c}{(\kms)}	& \multicolumn{2}{c}{(\kms)}		& \multicolumn{2}{c}{($\times 10^{-16}$)}	& \multicolumn{2}{c}{(\AA)}	& 	\\
\hline
32	& 1 Gaussian	& total	& 890	& 80	& -330	& 30	& 7.8	& 0.9	& 3.0	& 0.4	& 166.6	\\
33	& 1 Gaussian	& total	& 670	& 50	& -216.	& 19.	& 8.3	& 0.8	& 2.2	& 0.2	& 207.2	\\
34	& 1 Gaussian	& total	& 257.	& 14.	& -23.	& 6.	& 10.1	& 0.9	& 1.69	& 0.14	& 225.9	\\
35	& 1 Gaussian	& total	& 331.	& 11.	& -95.	& 4.	& 24.8	& 1.2	& 2.49	& 0.12	& 749.7	\\
36	& 1 Gaussian	& total	& 490	& 50	& 39	& 20	& 3.4	& 0.5	& 1.7	& 0.2	& 108.7	\\
37	& 1 Gaussian	& total	& 197.	& 14.	& -2.	& 6.	& 8.2	& 0.9	& 1.28	& 0.13	& 160.3	\\
38	& 1 Gaussian	& total	& 400	& 30	& 25.	& 11.	& 4.9	& 0.5	& 2.3	& 0.2	& 211.0	\\
39	& 1 Gaussian	& total	& 390	& 20	& -25.	& 10.	& 3.1	& 0.3	& 2.4	& 0.2	& 248.1	\\
40	& 1 Gaussian	& total	& 570	& 70	& 150	& 30	& 4.2	& 0.7	& 1.2	& 0.2	& 84.1	\\
41	& 1 Gaussian	& total	& 430	& 40	& -45.	& 15.	& 26.	& 3.	& 1.05	& 0.13	& 127.5	\\
42	& 1 Gaussian	& total	& 570	& 90	& -340	& 30	& 3.1	& 0.6	& 1.4	& 0.3	& 49.1	\\
43	& 1 Gaussian	& total	& 720	& 30	& -215.	& 11.	& 14.6	& 0.7	& 7.1	& 0.3	& 963.1	\\
	& 2 Gaussians	& total	& \multicolumn{2}{c}{\nodata}	& \multicolumn{2}{c}{\nodata}	& 15.9	& 0.8	& 7.7	& 0.4	& 17.1	\\
	& 		& core	& 238.	& 12.	& -177.	& 3.	& 1.9	& 0.3	& 0.92	& 0.15	& \multicolumn{1}{c}{\nodata}	\\
	& 		& wing	& 940	& 30	& -229.	& 7.	& 14.0	& 0.7	& 6.8	& 0.3	& \multicolumn{1}{c}{\nodata}	\\
44	& 1 Gaussian	& total	& 590	& 50	& 91.	& 19.	& 8.0	& 0.8	& 2.3	& 0.2	& 197.7	\\
45	& 1 Gaussian	& total	& 204.	& 6.	& 0.	& 3.	& 5.2	& 0.3	& 4.9	& 0.3	& 560.3	\\
46	& 1 Gaussian	& total	& 760	& 60	& -340	& 20	& 11.8	& 1.3	& 2.2	& 0.2	& 190.5	\\
47	& 1 Gaussian	& total	& 290	& 40	& -72.	& 14.	& 3.4	& 0.7	& 0.51	& 0.10	& 44.6	\\
48	& 1 Gaussian	& total	& 460	& 40	& -128.	& 13.	& 7.6	& 0.8	& 1.76	& 0.18	& 201.1	\\
49	& 1 Gaussian	& total	& 909.	& 20.	& -95.	& 9.	& 44.0	& 1.3	& 10.9	& 0.3	& 2407.1	\\
	& 2 Gaussians	& total	& \multicolumn{2}{c}{\nodata}	& \multicolumn{2}{c}{\nodata}	& 49.4	& 1.7	& 12.3	& 0.4	& 137.3	\\
	& 		& core	& 290.	& 6.	& -8.	& 2.	& 9.7	& 0.7	& 2.40	& 0.17	& \multicolumn{1}{c}{\nodata}	\\
	& 		& wing	& 1290	& 30	& -251.	& 15.	& 39.7	& 1.5	& 9.9	& 0.4	& \multicolumn{1}{c}{\nodata}	\\
50	& 1 Gaussian	& total	& 310	& 40	& -69.	& 15.	& 2.2	& 0.4	& 0.90	& 0.16	& 62.6	\\
51	& 1 Gaussian	& total	& 1130	& 50	& -110	& 20	& 18.7	& 1.1	& 6.0	& 0.4	& 631.4	\\
	& 2 Gaussians	& total	& \multicolumn{2}{c}{\nodata}	& \multicolumn{2}{c}{\nodata}	& 19.0	& 1.2	& 6.1	& 0.4	& 38.4	\\
	& 		& core	& 273.	& 10.	& 55.	& 3.	& 4.4	& 0.5	& 1.41	& 0.15	& \multicolumn{1}{c}{\nodata}	\\
	& 		& wing	& 1240	& 70	& -430	& 30	& 14.6	& 1.2	& 4.7	& 0.4	& \multicolumn{1}{c}{\nodata}	\\
52	& 1 Gaussian	& total	& 540	& 40	& -116.	& 16.	& 7.9	& 0.8	& 2.0	& 0.2	& 195.8	\\
	& 2 Gaussians	& total	& \multicolumn{2}{c}{\nodata}	& \multicolumn{2}{c}{\nodata}	& 9.8	& 1.1	& 2.5	& 0.3	& 11.7	\\
	& 		& core	& 159.	& 8.	& -104.	& 2.	& 1.5	& 0.4	& 0.37	& 0.10	& \multicolumn{1}{c}{\nodata}	\\
	& 		& wing	& 950	& 70	& -169.	& 16.	& 8.3	& 1.1	& 2.1	& 0.3	& \multicolumn{1}{c}{\nodata}	\\
53	& 1 Gaussian	& total	& 368.	& 8.	& -20.	& 3.	& 30.1	& 0.9	& 4.93	& 0.15	& 1817.7	\\
	& 2 Gaussians	& total	& \multicolumn{2}{c}{\nodata}	& \multicolumn{2}{c}{\nodata}	& 34.7	& 1.9	& 5.7	& 0.3	& 96.2	\\
	& 		& core	& 310.	& 6.	& -3.	& 1.	& 23.5	& 0.9	& 3.85	& 0.14	& \multicolumn{1}{c}{\nodata}	\\
	& 		& wing	& 650	& 80	& -330	& 40	& 11.2	& 1.6	& 1.8	& 0.3	& \multicolumn{1}{c}{\nodata}	\\
54	& 1 Gaussian	& total	& 600	& 60	& 40	& 20	& 16.	& 2.	& 1.24	& 0.17	& 120.3	\\
55	& 1 Gaussian	& total	& 570	& 30	& 39.	& 14.	& 11.1	& 0.9	& 2.7	& 0.2	& 349.3	\\
	& 2 Gaussians	& total	& \multicolumn{2}{c}{\nodata}	& \multicolumn{2}{c}{\nodata}	& 14.0	& 1.4	& 3.4	& 0.3	& 34.1	\\
	& 		& core	& 294.	& 12.	& 51.	& 4.	& 4.3	& 0.6	& 1.06	& 0.14	& \multicolumn{1}{c}{\nodata}	\\
	& 		& wing	& 1130	& 90	& -170	& 50	& 9.6	& 1.3	& 2.3	& 0.3	& \multicolumn{1}{c}{\nodata}	\\
56	& 1 Gaussian	& total	& 440	& 20	& 41.	& 10.	& 49.	& 4.	& 1.79	& 0.14	& 351.6	\\
	& 2 Gaussians	& total	& \multicolumn{2}{c}{\nodata}	& \multicolumn{2}{c}{\nodata}	& 52.	& 5.	& 1.91	& 0.18	& 16.7	\\
	& 		& core	& 307.	& 11.	& 70.	& 7.	& 34.	& 3.	& 1.26	& 0.10	& \multicolumn{1}{c}{\nodata}	\\
	& 		& wing	& 440	& 80	& -260	& 40	& 18.	& 4.	& 0.65	& 0.15	& \multicolumn{1}{c}{\nodata}	\\
57	& 1 Gaussian	& total	& 617.	& 15.	& -171.	& 7.	& 26.0	& 0.9	& 8.2	& 0.3	& 1691.4	\\
	& 2 Gaussians	& total	& \multicolumn{2}{c}{\nodata}	& \multicolumn{2}{c}{\nodata}	& 30.5	& 1.8	& 9.7	& 0.6	& 71.5	\\
	& 		& core	& 390.	& 9.	& -157.	& 2.	& 12.5	& 0.8	& 4.0	& 0.2	& \multicolumn{1}{c}{\nodata}	\\
	& 		& wing	& 1250	& 80	& -280	& 20	& 18.0	& 1.6	& 5.7	& 0.5	& \multicolumn{1}{c}{\nodata}	\\
58	& 1 Gaussian	& total	& 470	& 50	& -24.	& 19.	& 8.2	& 1.2	& 1.28	& 0.18	& 109.9	\\
59	& 1 Gaussian	& total	& 400	& 20	& -7.	& 8.	& 11.7	& 0.9	& 1.72	& 0.13	& 349.2	\\
60	& 1 Gaussian	& total	& 1050	& 100	& 80	& 40	& 11.2	& 1.4	& 2.6	& 0.3	& 148.3	\\
61	& 1 Gaussian	& total	& 440	& 40	& -187.	& 16.	& 2.9	& 0.4	& 2.1	& 0.3	& 105.7	\\
62	& 1 Gaussian	& total	& 420	& 20	& -69.	& 8.	& 11.4	& 0.8	& 2.73	& 0.19	& 426.1	\\
63	& 1 Gaussian	& total	& 520	& 50	& 10	& 20	& 5.8	& 0.8	& 1.6	& 0.2	& 126.7	\\
64	& 1 Gaussian	& total	& 653.	& 11.	& -275.	& 5.	& 63.8	& 1.5	& 8.36	& 0.20	& 3669.3	\\
\hline
\end{tabular}
}

\end{minipage}
\end{table*}

\begin{table*}
\begin{minipage}{175mm}
\caption{\rev{Best-fitting \oiii\ models.}}
\label{tab:OiiiData}
{ \centering
\begin{tabular}{rr@{$\,\pm\,$}lr@{$\,\pm\,$}lr@{$\,\pm\,$}lr@{$\,\pm\,$}lr@{$\,\pm\,$}lr@{$\,\pm\,$}lr@{$\,\pm\,$}lr@{$\,\pm\,$}lr}
\hline
	& \multicolumn{6}{c}{\oiiic\ component parameters}	& \multicolumn{6}{c}{\oiiiw\ component parameters}	& \multicolumn{5}{c}{Combined \oiiit\ model}	\\
ID	& \multicolumn{2}{c}{FWHM}	& \multicolumn{2}{c}{$v_\rmn{sh}$}	& \multicolumn{2}{c}{Flux}			& \multicolumn{2}{c}{FWHM}	& \multicolumn{2}{c}{$v_\rmn{sh}$}	& \multicolumn{2}{c}{Flux}			& \multicolumn{2}{c}{Flux}			& \multicolumn{2}{c}{EqW}	& $\Delta \chi^2$	\\
	& \multicolumn{2}{c}{(\kms)}	& \multicolumn{2}{c}{(\kms)}		& \multicolumn{2}{c}{($\times 10^{-16}$)}	& \multicolumn{2}{c}{(\kms)}	& \multicolumn{2}{c}{(\kms)}		& \multicolumn{2}{c}{($\times 10^{-16}$)}	& \multicolumn{2}{c}{($\times 10^{-16}$)}	& \multicolumn{2}{c}{(\AA)}	& 			\\
\hline
1	& 212.5	& 1.4	& 0.5	& 1.3	& 36.1	& 0.5	& 208.	& 9.	& -186.	& 8.	& 5.2	& 0.4	& 41.3	& 0.6	& 20.2	& 0.3	& 103.4	\\
2	& 304.	& 3.	& 2.4	& 1.0	& 30.8	& 0.8	& 650.	& 3.	& -210.	& 3.	& 51.9	& 0.9	& 82.7	& 1.2	& 26.6	& 0.4	& 315.9	\\
3	& 291.8	& 0.6	& -24.6	& 0.2	& 412.	& 3.	& 629.	& 12.	& -349.	& 10.	& 41.	& 2.	& 453.	& 4.	& 20.14	& 0.18	& 263.7	\\
4	& 433.	& 4.	& -17.8	& 0.9	& 72.	& 2.	& 819.	& 9.	& 45.	& 2.	& 67.	& 2.	& 139.	& 3.	& 42.2	& 0.9	& 134.9	\\
5	& 363.6	& 1.6	& 32.4	& 0.5	& 111.	& 2.	& 1182.	& 3.	& -21.5	& 1.0	& 183.7	& 2.0	& 295.	& 3.	& 105.3	& 1.0	&1248.2	\\
6	& 272.1	& 0.7	& 18.8	& 0.3	& 47.2	& 0.5	& 1032.	& 6.	& -275.	& 3.	& 25.9	& 0.7	& 73.1	& 0.9	& 96.5	& 1.1	& 612.0	\\
7	& 273.0	& 0.5	& -40.5	& 0.2	& 207.	& 2.	& 746.	& 4.	& -413.	& 2.	& 91.9	& 1.7	& 299.	& 3.	& 33.0	& 0.3	&2335.7	\\
8	& 331.2	& 1.1	& -25.1	& 0.4	& 49.8	& 0.8	& 1047.	& 4.	& -339.	& 3.	& 45.7	& 0.9	& 95.5	& 1.1	& 63.9	& 0.8	&1583.4	\\
9	& 365.3	& 1.3	& -9.0	& 0.5	& 146.	& 2.	& 796.	& 4.	& -236.	& 4.	& 69.3	& 1.8	& 216.	& 3.	& 39.1	& 0.5	& 656.5	\\
10	& 286.2	& 0.5	& -4.1	& 0.2	& 284.	& 3.	& 877.	& 12.	& -111.	& 5.	& 31.3	& 2.0	& 316.	& 4.	& 48.6	& 0.6	& 169.0	\\
11	& 216.7	& 0.4	& -1.91	& 0.14	& 70.1	& 0.6	& 681.	& 2.	& -45.5	& 0.6	& 58.3	& 0.7	& 128.4	& 1.0	& 59.4	& 0.5	&2620.5	\\
12	& 416.8	& 0.8	& 2.0	& 0.3	& 602.	& 6.	& 1239.	& 8.	& -23.	& 2.	& 214.	& 5.	& 816.	& 8.	& 51.1	& 0.5	&1055.5	\\
13	& 332.1	& 0.7	& 18.3	& 0.3	& 263.	& 3.	& 929.	& 2.	&-163.3	& 1.1	& 190.	& 2.	& 452.	& 3.	& 73.4	& 0.5	&2943.5	\\
14	& 305.5	& 0.7	& -0.6	& 0.2	& 110.4	& 1.1	& 695.	& 5.	& -84.2	& 1.9	& 37.7	& 1.1	& 148.1	& 1.5	& 45.4	& 0.5	& 498.7	\\
15	& 198.6	& 1.2	& 145.7	& 0.8	& 96.3	& 1.5	& 316.1	& 1.7	& -45.4	& 1.6	& 188.6	& 2.0	& 285.	& 2.	& 96.2	& 0.8	& 660.0	\\
16	& 407.7	& 1.5	& 33.7	& 0.4	& 60.8	& 0.9	& 948.	& 14.	& -9.	& 3.	& 19.7	& 0.9	& 80.5	& 1.2	& 67.2	& 1.0	& 206.3	\\
17	& 389.4	& 1.1	& -3.2	& 0.4	& 128.9	& 1.6	& 931.	& 3.	&-134.4	& 1.2	& 105.5	& 1.5	& 234.	& 2.	& 117.7	& 1.1	&1451.4	\\
18	& 211.4	& 0.8	& -6.3	& 0.3	& 100.0	& 1.7	& 452.	& 2.	& -55.2	& 0.9	& 57.5	& 1.4	& 158.	& 2.	& 69.6	& 1.0	& 464.6	\\
19	& 251.0	& 1.2	& -13.5	& 0.4	& 110.2	& 1.3	& 387.	& 5.	& -166.	& 7.	& 35.4	& 1.0	& 145.6	& 1.7	& 87.6	& 1.0	& 435.0	\\
20	& 337.1	& 1.3	& 3.0	& 0.4	& 518.	& 9.	& 955.	& 3.	& -78.3	& 0.8	& 827.	& 10.	& 1345.	& 13.	& 37.2	& 0.4	&1389.4	\\
21	& 303.4	& 1.9	& -2.8	& 0.7	& 168.	& 3.	& 614.	& 3.	&-115.2	& 1.7	& 189.	& 3.	& 358.	& 4.	& 26.4	& 0.3	& 323.8	\\
22	& 467.5	& 0.5	& -6.03	& 0.17	& 1130.	& 6.	& 1102.	& 4.	& -98.8	& 1.3	& 214.	& 3.	& 1344.	& 6.	& 618.	& 3.	&2301.1	\\
23	& 196.6	& 1.0	& -39.7	& 0.4	& 89.9	& 1.7	& 518.	& 2.	&-133.8	& 1.1	& 100.2	& 1.9	& 190.	& 3.	& 21.5	& 0.3	& 589.5	\\
24	& 208.8	& 0.9	& -18.3	& 0.3	& 26.7	& 0.5	& 548.	& 8.	& -153.	& 6.	& 7.0	& 0.5	& 33.7	& 0.7	& 26.8	& 0.6	& 103.9	\\
25	& 288.9	& 0.9	& -48.9	& 0.4	& 132.	& 2.	& 1026.	& 4.	& -332.	& 2.	& 130.	& 2.	& 262.	& 3.	& 32.7	& 0.4	&1761.7	\\
26	& 282.8	& 0.5	& -9.32	& 0.18	& 345.	& 3.	& 819.	& 4.	& -53.0	& 1.1	& 109.	& 2.	& 453.	& 4.	& 122.8	& 1.0	&1317.0	\\
27	& 173.9	& 0.3	&-10.22	& 0.14	& 89.4	& 1.0	& 529.	& 5.	& -137.	& 3.	& 12.3	& 0.6	& 101.7	& 1.2	& 96.3	& 1.1	& 255.0	\\
28	& 716.	& 2.	& 85.9	& 0.4	& 412.	& 4.	& 1340	& 20	& 84.	& 3.	& 83.	& 3.	& 495.	& 5.	& 177.1	& 1.8	& 233.3	\\
29	& 292.8	& 1.5	& -76.1	& 0.3	& 55.2	& 1.0	& 640.	& 5.	& -84.3	& 1.1	& 39.5	& 1.2	& 94.8	& 1.5	& 50.1	& 0.8	& 267.1	\\
30	& 306.9	& 1.6	& -46.2	& 0.5	& 71.6	& 1.6	& 856.	& 12.	& 80.	& 5.	& 36.	& 2.	& 107.	& 3.	& 17.6	& 0.4	& 163.4	\\
31	& 298.2	& 0.9	& -3.6	& 0.3	& 209.	& 2.	& 628.	& 4.	& -245.	& 4.	& 85.4	& 1.7	& 294.	& 3.	& 62.8	& 0.6	&1480.5	\\
32	& 302.6	& 1.8	& -81.4	& 0.6	& 42.6	& 0.8	& 742.	& 8.	& -230.	& 5.	& 24.3	& 0.9	& 66.9	& 1.2	& 29.1	& 0.5	& 170.6	\\
33	& 267.9	& 1.1	& 11.4	& 0.4	& 83.8	& 1.5	& 610.	& 3.	& -129.	& 2.	& 53.7	& 1.5	& 138.	& 2.	& 34.7	& 0.5	& 523.4	\\
34	& 178.0	& 1.0	& -8.4	& 0.3	& 92.7	& 1.7	& 376.	& 2.	& -65.6	& 1.2	& 68.0	& 1.7	& 161.	& 2.	& 28.4	& 0.4	& 239.8	\\
35	& 210.9	& 0.5	& 1.71	& 0.19	& 705.	& 8.	& 322.	& 2.	& -54.	& 2.	& 148.	& 5.	& 853.	& 9.	& 90.9	& 1.0	& 250.1	\\
36	& 553.5	& 1.6	& 22.1	& 0.5	& 183.	& 2.	& 1400	& 30	& 91.	& 6.	& 27.5	& 1.7	& 210.	& 3.	& 101.6	& 1.3	& 138.7	\\
37	& 212.6	& 0.5	& 5.32	& 0.16	& 506.	& 6.	& 417.	& 2.	& -41.5	& 0.9	& 117.	& 3.	& 623.	& 7.	& 104.1	& 1.1	& 478.9	\\
38	& 304.	& 3.	& -26.6	& 0.7	& 53.0	& 1.6	& 502.	& 15.	& 1.	& 3.	& 21.0	& 1.7	& 74.	& 2.	& 36.3	& 1.2	& 23.8	\\
39	& 360.9	& 1.9	& 69.8	& 0.6	& 51.9	& 0.9	& 692.	& 2.	& -145.	& 2.	& 67.9	& 0.9	& 119.8	& 1.3	& 100.4	& 1.1	& 874.7	\\
40	& 278.6	& 0.8	& 18.8	& 0.3	& 103.4	& 1.4	& 854.	& 6.	& 71.1	& 1.9	& 43.7	& 1.3	& 147.2	& 1.9	& 40.5	& 0.5	& 578.9	\\
41	& 273.2	& 1.5	& -34.4	& 0.4	& 385.	& 8.	& 587.	& 5.	& -26.7	& 1.0	& 276.	& 8.	& 661.	& 12.	& 25.4	& 0.5	& 242.8	\\
42	& 414.	& 3.	&-147.6	& 1.0	& 48.2	& 1.4	& 910.	& 6.	& -323.	& 4.	& 51.9	& 1.6	& 100.	& 2.	& 33.6	& 0.7	& 233.1	\\
43	& 397.2	& 1.1	& 12.7	& 0.4	& 224.	& 3.	& 805.	& 3.	&-145.2	& 1.7	& 121.	& 2.	& 345.	& 3.	& 169.5	& 1.7	&1115.2	\\
44	& 533.7	& 1.0	& -22.3	& 0.6	& 150.2	& 1.4	& 870	& 60	& -860	& 20	& 6.3	& 0.8	& 156.6	& 1.6	& 46.5	& 0.5	& 72.5	\\
45	& 202.1	& 1.6	& 8.5	& 0.5	& 18.5	& 0.6	& 520	& 20	& -73.	& 10.	& 3.5	& 0.7	& 22.0	& 0.9	& 23.4	& 0.9	& 14.3	\\
46	& 303.4	& 0.6	& -14.8	& 0.2	& 240.	& 3.	& 1129.	& 11.	& -52.	& 3.	& 60.	& 2.	& 300.	& 4.	& 46.1	& 0.6	& 528.9	\\
47	& 280.9	& 0.8	& -15.1	& 0.3	& 122.7	& 1.7	& 824.	& 11.	& -149.	& 5.	& 28.2	& 1.7	& 151.	& 2.	& 18.9	& 0.3	& 163.9	\\
48	& 308.4	& 0.8	& 7.7	& 0.7	& 160.	& 2.	& 388.	& 6.	& -168.	& 6.	& 48.1	& 1.6	& 208.	& 3.	& 50.0	& 0.6	& 182.0	\\
49	& 357.8	& 0.7	& 20.1	& 0.3	& 449.	& 5.	& 977.9	& 1.5	&-250.8	& 1.0	& 341.	& 3.	& 790.	& 6.	& 162.6	& 1.2	&4959.5	\\
50	& 231.5	& 0.6	& -24.8	& 0.2	& 71.3	& 1.0	& 874.	& 6.	& -138.	& 2.	& 33.2	& 1.0	& 104.5	& 1.5	& 37.7	& 0.5	& 682.0	\\
51	& 353.	& 2.	& 18.2	& 0.9	& 73.	& 2.	&1139.7	& 1.7	&-232.7	& 0.9	& 347.	& 3.	& 420.	& 4.	& 117.6	& 1.0	& 869.0	\\
52	& 301.1	& 1.3	& -40.4	& 0.4	& 110.0	& 1.5	& 712.	& 6.	& -57.3	& 1.5	& 60.5	& 1.5	& 170.	& 2.	& 41.1	& 0.5	& 286.8	\\
53	& 259.0	& 0.5	& -0.5	& 0.2	& 364.	& 4.	& 738.6	& 1.3	&-143.4	& 0.6	& 319.	& 3.	& 683.	& 5.	& 87.4	& 0.7	&3531.1	\\
54	& 296.6	& 0.9	& -6.5	& 0.3	& 294.	& 4.	& 750.	& 12.	& 40.	& 3.	& 71.	& 4.	& 365.	& 6.	& 25.0	& 0.4	& 142.4	\\
55	& 387.7	& 0.9	& 57.3	& 0.3	& 201.	& 2.	& 1060.	& 4.	& -301.	& 3.	& 94.9	& 1.8	& 296.	& 3.	& 69.8	& 0.7	&1838.2	\\
\hline
\end{tabular}
}

\medskip
\rev{Columns are:
(1)	\fex\ sample ID number;
(2--4)	parameters of the \oiii\ core component (defined as the
	Gaussian whose centroid is closer to the peak of the
        overall line profile): FWHM (\kms),
	velocity shift of the centroid relative to the systemic redshift
        defined by \sii\ 
        (\kms; $v_\rmn{sh} < 0$ indicates blueshifted line
        components), and model flux ($\times 10^{-16} \ 
        \cgsflux$);
(5--7)	Gaussian FWHM, velocity shift and flux of the \oiii\ wing component
        (same units as columns 2--4);
(8)	flux ($\times 10^{-16} \  \cgsflux$) and 
(9)	equivalent width (\AA) of the two components combined; and
(10)	improvement in $\chi^2$ over the single-Gaussian \oiii\ models.}
\end{minipage}
\end{table*}
\begin{table*}
\begin{minipage}{175mm}
\contcaption{}
{ \centering
\begin{tabular}{rr@{$\,\pm\,$}lr@{$\,\pm\,$}lr@{$\,\pm\,$}lr@{$\,\pm\,$}lr@{$\,\pm\,$}lr@{$\,\pm\,$}lr@{$\,\pm\,$}lr@{$\,\pm\,$}lr}
\hline
	& \multicolumn{6}{c}{\oiiic\ component parameters}	& \multicolumn{6}{c}{\oiiiw\ component parameters}	& \multicolumn{5}{c}{Combined \oiiit\ model}	\\
ID	& \multicolumn{2}{c}{FWHM}	& \multicolumn{2}{c}{$v_\rmn{sh}$}	& \multicolumn{2}{c}{Flux}			& \multicolumn{2}{c}{FWHM}	& \multicolumn{2}{c}{$v_\rmn{sh}$}	& \multicolumn{2}{c}{Flux}			& \multicolumn{2}{c}{Flux}			& \multicolumn{2}{c}{EqW}	& $\Delta \chi^2$	\\
	& \multicolumn{2}{c}{(\kms)}	& \multicolumn{2}{c}{(\kms)}		& \multicolumn{2}{c}{($\times 10^{-16}$)}	& \multicolumn{2}{c}{(\kms)}	& \multicolumn{2}{c}{(\kms)}		& \multicolumn{2}{c}{($\times 10^{-16}$)}	& \multicolumn{2}{c}{($\times 10^{-16}$)}	& \multicolumn{2}{c}{(\AA)}	& 			\\
\hline
56	& 344.0	& 1.0	& 26.3	& 0.3	& 1323.	& 13.	& 683.	& 4.	& -61.5	& 1.8	& 461.	& 10.	& 1784.	& 17.	& 49.3	& 0.5	& 439.5	\\
57	& 474.9	& 1.4	& -76.2	& 0.5	& 202.	& 2.	& 1138.	& 4.	& -300.	& 2.	& 160.	& 2.	& 362.	& 3.	& 110.3	& 0.9	&1524.5	\\
58	& 271.8	& 1.0	& 15.4	& 0.3	& 147.	& 2.	& 796.	& 9.	& -24.	& 2.	& 58.	& 2.	& 205.	& 3.	& 25.9	& 0.4	& 268.6	\\
59	& 314.2	& 0.6	& 23.8	& 0.2	& 231.	& 2.	& 910.	& 13.	& -47.	& 4.	& 32.7	& 1.9	& 264.	& 3.	& 37.7	& 0.4	& 185.8	\\
60	& 305.2	& 1.7	& -18.9	& 0.6	& 78.7	& 1.6	& 853.	& 3.	&-134.2	& 1.4	& 122.4	& 1.8	& 201.	& 2.	& 46.7	& 0.6	& 608.3	\\
61	& 273.4	& 0.8	& -32.1	& 0.3	& 74.3	& 0.8	& 838.	& 11.	& -42.	& 9.	& 34.6	& 0.9	& 108.9	& 1.2	& 67.8	& 0.7	& 516.0	\\
62	& 232.8	& 0.9	& 8.2	& 0.3	& 121.3	& 1.8	& 566.	& 2.	& -77.5	& 1.1	& 91.3	& 1.7	& 213.	& 2.	& 45.9	& 0.5	& 651.5	\\
63	& 455.	& 2.	& 16.9	& 0.7	& 140.	& 2.	& 876.	& 6.	& -195.	& 5.	& 69.	& 2.	& 209.	& 3.	& 51.6	& 0.7	& 367.8	\\
64	& 309.1	& 0.7	& -27.8	& 0.3	& 549.	& 6.	& 604.0	& 1.6	&-148.9	& 1.8	& 273.	& 4.	& 822.	& 7.	& 89.3	& 0.7	&1513.2	\\
\hline
\end{tabular}
}

\end{minipage}
\end{table*}

\begin{table*}
\begin{minipage}{172mm}
\caption{\rev{Best-fitting models for forbidden emission lines with
    lower ionisation potentials.}}
\label{tab:NLRData}
{ \centering
\begin{tabular}{rr@{$\,\pm\,$}l@{\ \ \ }r@{$\,\pm\,$}l@{\ \ \ }r@{$\,\pm\,$}l@{\ \ \ }r@{$\,\pm\,$}lr@{$\,\pm\,$}l@{\ \ \ }r@{$\,\pm\,$}l@{\ \ \ }r@{$\,\pm\,$}lr@{$\,\pm\,$}lr@{$\,\pm\,$}l}
\hline
	& \multicolumn{8}{c}{\oi$\lambda$6300 Gaussian properties}	& \multicolumn{4}{c}{\sii\ doublet profile}	& \multicolumn{2}{c}{\sii$\lambda$6716}	& \multicolumn{2}{c}{\sii$\lambda$6731}	& \multicolumn{2}{c}{\sii$_T$}	\\
ID	& \multicolumn{2}{c}{FWHM}	& \multicolumn{2}{c}{$v_\rmn{sh}$}	& \multicolumn{2}{c}{Flux}	& \multicolumn{2}{c}{EqW}	& \multicolumn{2}{c}{FWHM}	& \multicolumn{2}{c}{$v$}	& \multicolumn{2}{c}{Flux}	& \multicolumn{2}{c}{Flux}	& \multicolumn{2}{c}{EqW}	\\
	& \multicolumn{2}{c}{(\kms)}	& \multicolumn{2}{c}{(\kms)}		& \multicolumn{2}{c}{($\times 10^{-16}$)}	& \multicolumn{2}{c}{(\AA)}	& \multicolumn{2}{c}{(\kms)}	& \multicolumn{2}{c}{(\kms)}	& \multicolumn{2}{c}{($\times 10^{-16}$)}	& \multicolumn{2}{c}{($\times 10^{-16}$)}	& \multicolumn{2}{c}{(\AA)}	\\
\hline
1	& 244.	& 13.	& 6.	& 5.	& 2.7	& 0.2	& 1.29	& 0.11	& 227.	& 6.	& 47.	& 3.	& 9.4	& 0.4	& 7.8	& 0.4	& 8.0	& 0.2	\\
2	& 420	& 30	& 12.	& 13.	& 5.7	& 0.6	& 1.78	& 0.19	& 288.	& 13.	& 66.	& 6.	& 7.1	& 0.4	& 5.8	& 0.4	& 3.89	& 0.18	\\
3	& 380	& 20	& -18.	& 8.	& 24.0	& 1.7	& 1.13	& 0.08	& 323.	& 5.	& -18.	& 3.	& 79.9	& 1.8	& 79.4	& 1.8	& 7.42	& 0.12	\\
4	& 570	& 40	& 19.	& 16.	& 8.7	& 0.8	& 2.5	& 0.2	& 513.	& 15.	& 5.	& 8.	& 19.9	& 0.8	& 18.8	& 0.8	& 10.6	& 0.3	\\
5	& 500	& 20	& 23.	& 9.	& 10.0	& 0.6	& 4.3	& 0.3	& 356.	& 6.	& -36.	& 3.	& 21.2	& 0.5	& 17.4	& 0.5	& 16.3	& 0.3	\\
6	& 273.	& 11.	&0.$^a$	& 4.	& 3.6	& 0.2	& 4.8	& 0.3	& \multicolumn{2}{c}{\nodata}	& \multicolumn{2}{c}{\nodata}	& \multicolumn{2}{c}{\nodata}	& \multicolumn{2}{c}{\nodata}	& \multicolumn{2}{c}{\nodata}	\\
7	& 257.	& 15.	& -24.	& 5.	& 10.0	& 0.9	& 1.01	& 0.09	& 275.	& 4.	& 21.	& 2.	& 20.2	& 0.4	& 21.6	& 0.4	& 4.07	& 0.05	\\
8	& 440	& 20	& -26.	& 8.	& 6.2	& 0.4	& 4.1	& 0.3	& 390.	& 8.	& 96.	& 4.	& 13.4	& 0.4	& 12.1	& 0.4	& 16.5	& 0.4	\\
9	& 366.	& 18.	& -5.	& 7.	& 9.4	& 0.7	& 2.15	& 0.16	& 403.	& 10.	& 95.	& 5.	& 21.9	& 0.8	& 20.8	& 1.0	& 9.5	& 0.3	\\
10	& 289.	& 13.	& -2.	& 5.	& 11.9	& 0.8	& 1.96	& 0.13	& 301.	& 10.	& 39.	& 5.	& 23.3	& 1.1	& 21.5	& 1.0	& 7.2	& 0.2	\\
11	& 330	& 20	& -7.	& 8.	& 2.7	& 0.3	& 1.55	& 0.15	& 147.	& 9.	& 49.	& 5.	& 2.39	& 0.19	& 2.7	& 0.2	& 2.48	& 0.13	\\
12	& 520	& 30	& 28.	& 13.	& 23.	& 2.	& 2.05	& 0.18	& 410	& 20	& -15.	& 12.	& 23.5	& 1.7	& 23.0	& 1.6	& 3.81	& 0.19	\\
13	& 308.	& 15.	& 2.	& 5.	& 10.3	& 0.7	& 1.65	& 0.12	& 333.	& 9.	& -2.	& 4.	& 31.1	& 1.2	& 25.7	& 1.1	& 9.1	& 0.3	\\
14	& 313.	& 14.	& 13.	& 5.	& 5.3	& 0.4	& 1.75	& 0.12	& 298.	& 15.	& 22.	& 7.	& 8.1	& 0.6	& 7.4	& 0.5	& 4.8	& 0.2	\\
15	& 388.	& 18.	& 25.	& 7.	& 8.0	& 0.5	& 2.61	& 0.17	& 345.	& 6.	& 32.	& 3.	& 25.1	& 0.6	& 20.9	& 0.6	& 14.7	& 0.3	\\
16	& 410	& 30	& 32.	& 10.	& 5.0	& 0.5	& 4.9	& 0.5	& 409.5	& 1.3	& -45.4	& 0.7	& 4.123	& 0.017	& 3.617	& 0.018	& 7.54	& 0.02	\\
17	& 452.	& 17.	& 28.	& 6.	& 7.8	& 0.4	& 4.6	& 0.2	& 414.	& 10.	& 15.	& 5.	& 13.6	& 0.4	& 12.7	& 0.4	& 15.4	& 0.3	\\
18	& 233.	& 12.	& -17.	& 5.	& 3.4	& 0.3	& 1.81	& 0.16	& 201.	& 19.	& 38.	& 10.	& 2.0	& 0.3	& 2.0	& 0.3	& 2.04	& 0.19	\\
19	& 269.	& 12.	& -7.	& 5.	& 4.4	& 0.3	& 2.66	& 0.19	& 223.3	& 0.3	& 43.60	& 0.15	& 8.201	& 0.014	& 7.077	& 0.015	& 9.140	& 0.012	\\
20	& 334.	& 15.	& 21.	& 5.	& 45.	& 3.	& 1.59	& 0.10	& 417.	& 12.	& -51.	& 6.	& 82.	& 3.	& 84.	& 3.	& 5.52	& 0.14	\\
21	& 450	& 20	& 18.	& 10.	& 27.	& 2.	& 1.98	& 0.15	& 316.	& 13.	& 50.	& 7.	& 38.	& 2.	& 40.	& 2.	& 5.5	& 0.2	\\
22	& 528.	& 18.	& 3.	& 7.	& 18.2	& 0.8	& 9.4	& 0.4	& 467.	& 13.	& -16.	& 6.	& 24.9	& 0.9	& 25.9	& 0.9	& 19.8	& 0.5	\\
23	& 226.	& 14.	& -4.	& 6.	& 7.2	& 0.8	& 0.97	& 0.10	& 230.	& 9.	& 66.	& 5.	& 17.5	& 1.0	& 15.8	& 1.0	& 4.58	& 0.19	\\
24	& 244.	& 12.	& -4.	& 5.	& 2.6	& 0.2	& 2.23	& 0.18	& 199.	& 4.	& 84.4	& 1.9	& 3.12	& 0.08	& 2.95	& 0.08	& 5.15	& 0.09	\\
25	& 412.	& 19.	& -19.	& 8.	& 18.4	& 1.2	& 2.52	& 0.17	& 283.	& 7.	& 107.	& 4.	& 25.8	& 0.9	& 24.2	& 0.9	& 7.00	& 0.17	\\
26	& 295.	& 12.	& 9.	& 5.	& 8.4	& 0.5	& 3.1	& 0.2	& 292.	& 8.	& 27.	& 4.	& 25.8	& 1.1	& 21.7	& 1.0	& 15.6	& 0.5	\\
27	& 215.	& 8.	& -16.	& 3.	& 4.1	& 0.3	& 4.3	& 0.3	& 211.	& 5.	& 19.	& 3.	& 10.9	& 0.4	& 8.2	& 0.4	& 19.5	& 0.6	\\
28	& 820	& 20	& 108.	& 11.	& 25.5	& 1.0	& 12.9	& 0.5	& 350	& 20	& -102.	& 13.	& 7.2	& 0.8	& 5.5	& 0.6	& 4.9	& 0.4	\\
29	& 330	& 20	& 0.	& 9.	& 3.6	& 0.4	& 2.2	& 0.2	& 311.	& 16.	& 45.	& 7.	& 5.8	& 0.4	& 4.9	& 0.4	& 6.6	& 0.3	\\
30	& 390	& 30	& -44.	& 12.	& 6.9	& 0.8	& 1.39	& 0.16	& 430	& 20	& -3.	& 12.	& 10.7	& 0.8	& 8.7	& 0.8	& 3.6	& 0.2	\\
31	& 291.	& 13.	& 5.	& 5.	& 9.9	& 0.7	& 2.02	& 0.14	& 240.	& 5.	& 77.	& 3.	& 21.0	& 0.6	& 19.5	& 0.6	& 8.20	& 0.18	\\
32	& 317.	& 19.	& -67.	& 7.	& 6.3	& 0.6	& 2.4	& 0.2	& 335.	& 12.	& 78.	& 5.	& 18.1	& 0.9	& 17.8	& 0.9	& 13.0	& 0.5	\\
\hline
\end{tabular}
}

~$^a$ \sii\ data were not available for source 6 so \oi 6300 is used to
define the redshift of this one object.

\medskip
\rev{Columns are:
(1)	\fex\ sample ID number;
(2--5)	parameters of the best-fitting Gaussian for \oi 6300:
	(2) FWHM (\kms),
	(3) velocity shift of the centroid relative to the systemic redshift
        defined by \sii\ 
        (\kms; $v_\rmn{sh} < 0$ when the emission line is less redshifted than
        the host galaxy),
	(4) model flux ($\times 10^{-16} \  \cgsflux$), and
	(5) equivalent width (\AA);
(6--7)	parameters of the best-fitting \sii\ line profile:
	(6) FWHM (\kms) and
	(7) velocity of the centroid relative to the redshift
        reported by SDSS;
	(8) \sii 6716 and
        (9) \sii 6731 Gaussian model fluxes (both $\times 10^{-16} \ 
        \cgsflux$); and
(10) combined equivalent width of the \sii\ doublet.}

\medskip
\rev{The \sii\ doublet line models were forced to have the same profile
(FWHM and $v$) with only the normalizations left independent.
\oi 6364 is not listed because its model had no free parameters (it
was constrained to have the same profile and 1/3 as much flux as
\oi 6300).
The \nii\ doublet parameters are likewise not reported because the
profiles were also tied to \oi 6300; the normalizations were allowed
to vary but are often dominated by the residuals of \ha.}
\end{minipage}
\end{table*}
\begin{table*}
\begin{minipage}{175mm}
\contcaption{}
{ \centering
\begin{tabular}{rr@{$\,\pm\,$}lr@{$\,\pm\,$}lr@{$\,\pm\,$}lr@{$\,\pm\,$}lr@{$\,\pm\,$}lr@{$\,\pm\,$}lr@{$\,\pm\,$}lr@{$\,\pm\,$}lr@{$\,\pm\,$}l}
\hline
	& \multicolumn{8}{c}{\oi$\lambda$6300 Gaussian properties}	& \multicolumn{4}{c}{\sii\ doublet profile}	& \multicolumn{2}{c}{\sii$\lambda$6716}	& \multicolumn{2}{c}{\sii$\lambda$6731}	& \multicolumn{2}{c}{\sii$_T$}	\\
ID	& \multicolumn{2}{c}{FWHM}	& \multicolumn{2}{c}{$v_\rmn{sh}$}	& \multicolumn{2}{c}{Flux}	& \multicolumn{2}{c}{EqW}	& \multicolumn{2}{c}{FWHM}	& \multicolumn{2}{c}{$v$}	& \multicolumn{2}{c}{Flux}	& \multicolumn{2}{c}{Flux}	& \multicolumn{2}{c}{EqW}	\\
	& \multicolumn{2}{c}{(\kms)}	& \multicolumn{2}{c}{(\kms)}		& \multicolumn{2}{c}{($\times 10^{-16}$)}	& \multicolumn{2}{c}{(\AA)}	& \multicolumn{2}{c}{(\kms)}	& \multicolumn{2}{c}{(\kms)}	& \multicolumn{2}{c}{($\times 10^{-16}$)}	& \multicolumn{2}{c}{($\times 10^{-16}$)}	& \multicolumn{2}{c}{(\AA)}	\\
\hline
33	& 360	& 30	& -5.	& 9.	& 6.8	& 0.7	& 1.83	& 0.18	& 337.	& 16.	& -2.	& 7.	& 15.7	& 1.0	& 14.3	& 1.0	& 7.7	& 0.4	\\
34	& 203.	& 10.	& -6.	& 4.	& 8.1	& 0.7	& 1.42	& 0.12	& 229.	& 6.	& 11.	& 3.	& 19.2	& 0.8	& 18.7	& 0.7	& 6.72	& 0.19	\\
35	& 223.	& 4.	& -6.1	& 1.5	& 35.1	& 1.0	& 3.70	& 0.11	& 210.	& 4.	& 48.4	& 1.7	& 65.5	& 1.8	& 56.1	& 1.7	& 12.7	& 0.3	\\
36	& 520	& 30	& 13.	& 11.	& 7.9	& 0.6	& 4.1	& 0.3	& 498.1	& 0.6	& -1.5	& 0.3	& 14.22	& 0.02	& 14.86	& 0.02	& 14.375	& 0.016	\\
37	& 248.	& 6.	& 6.	& 2.	& 22.7	& 0.9	& 3.72	& 0.14	& 259.	& 5.	& 39.	& 2.	& 53.0	& 1.6	& 46.8	& 1.5	& 16.1	& 0.4	\\
38	& 320	& 20	& 7.	& 9.	& 4.0	& 0.4	& 1.9	& 0.2	& 240	& 20	& -2.	& 11.	& 5.0	& 0.8	& 4.7	& 0.6	& 4.4	& 0.4	\\
39	& 690	& 40	& 31.	& 16.	& 6.0	& 0.4	& 4.9	& 0.4	& 459.	& 18.	& 29.	& 8.	& 11.8	& 0.6	& 11.8	& 0.6	& 19.2	& 0.7	\\
40	& 357.	& 15.	& 52.	& 6.	& 9.7	& 0.6	& 2.90	& 0.17	& 365.	& 13.	& -97.	& 7.	& 16.0	& 0.8	& 14.0	& 1.0	& 8.6	& 0.4	\\
41	& 290.	& 12.	& 6.	& 5.	& 42.	& 3.	& 1.76	& 0.11	& 258.	& 7.	& 43.	& 4.	& 66.	& 2.	& 65.	& 2.	& 5.19	& 0.14	\\
42	& 580	& 40	& -92.	& 16.	& 7.4	& 0.7	& 3.3	& 0.3	& 190	& 40	& 170	& 20	& 1.0	& 0.3	& 1.4	& 0.3	& 1.1	& 0.2	\\
43	& 450	& 20	& 40.	& 9.	& 8.4	& 0.6	& 4.3	& 0.3	& 383.	& 7.	& 39.	& 4.	& 19.2	& 0.5	& 17.1	& 0.5	& 18.2	& 0.4	\\
44	& 960	& 80	& -100	& 30	& 10.1	& 1.0	& 2.9	& 0.3	& 458.	& 6.	& 39.	& 3.	& 13.3	& 0.3	& 12.5	& 0.2	& 7.15	& 0.10	\\
45	& 200.	& 10.	& 10.	& 4.	& 3.1	& 0.3	& 3.1	& 0.3	& 160	& 50	& -20	& 20	& 0.7	& 0.3	& 1.0	& 0.3	& 1.7	& 0.5	\\
46	& 241.	& 17.	& -7.	& 7.	& 5.5	& 0.5	& 1.02	& 0.10	& 296.	& 1.	& 61.5	& 0.7	& 16.07	& 0.10	& 14.04	& 0.09	& 5.67	& 0.02	\\
47	& 289.	& 18.	& 10.	& 7.	& 7.9	& 0.7	& 1.18	& 0.11	& 294.	& 12.	& 21.	& 6.	& 12.6	& 0.7	& 12.2	& 0.7	& 3.68	& 0.15	\\
48	& 358.	& 19.	& -3.	& 7.	& 8.7	& 0.7	& 2.08	& 0.16	& 311.	& 14.	& 66.	& 6.	& 16.6	& 1.0	& 15.5	& 1.0	& 7.0	& 0.3	\\
49	& 390.	& 12.	& -5.	& 5.	& 19.6	& 0.9	& 5.1	& 0.2	& 356.	& 7.	& 55.	& 3.	& 50.4	& 1.4	& 43.5	& 1.3	& 24.0	& 0.5	\\
50	& 238.	& 8.	& -6.	& 3.	& 6.7	& 0.4	& 2.83	& 0.16	& 184.5	& 0.7	& 69.4	& 0.4	& 9.56	& 0.05	& 8.54	& 0.04	& 7.88	& 0.03	\\
51	& 852.	& 18.	& -36.	& 8.	& 36.1	& 1.0	& 11.8	& 0.3	& 570.	& 3.	& 36.6	& 1.6	& 44.3	& 0.3	& 42.7	& 0.3	& 27.65	& 0.15	\\
52	& 266.	& 16.	& -19.	& 6.	& 5.9	& 0.6	& 1.48	& 0.14	& 280	& 30	& 44.	& 12.	& 8.5	& 1.1	& 7.6	& 0.9	& 3.7	& 0.3	\\
53	& 325.	& 10.	& 25.	& 4.	& 17.6	& 0.8	& 2.78	& 0.13	& 349.	& 12.	& 42.	& 6.	& 15.8	& 0.7	& 19.2	& 0.9	& 5.46	& 0.19	\\
54	& 350.	& 18.	& -1.	& 6.	& 23.5	& 1.7	& 1.86	& 0.14	& 280.	& 17.	& 8.	& 7.	& 26.	& 2.	& 27.	& 2.	& 4.1	& 0.2	\\
55	& 366.	& 14.	& 8.	& 5.	& 12.7	& 0.7	& 3.22	& 0.18	& 358.	& 3.	& 50.6	& 1.5	& 24.4	& 0.3	& 22.1	& 0.3	& 11.77	& 0.09	\\
56	& 353.	& 17.	& 26.	& 6.	& 49.	& 3.	& 1.80	& 0.12	& 264.	& 16.	& -15.	& 7.	& 49.	& 4.	& 47.	& 4.	& 3.5	& 0.2	\\
57	& 570.	& 18.	& -2.	& 8.	& 18.6	& 0.9	& 6.1	& 0.3	& 386.	& 16.	& 91.	& 7.	& 15.2	& 0.8	& 15.3	& 0.9	& 10.0	& 0.4	\\
58	& 340	& 20	& 19.	& 7.	& 10.9	& 1.0	& 1.72	& 0.15	& 276.2	& 1.4	& -11.5	& 0.8	& 15.68	& 0.11	& 15.61	& 0.11	& 4.48	& 0.02	\\
59	& 278.	& 18.	& 36.	& 7.	& 6.9	& 0.7	& 1.04	& 0.10	& 241.	& 18.	& -8.	& 8.	& 9.4	& 1.0	& 10.2	& 1.0	& 2.9	& 0.2	\\
60	& 339.	& 19.	& 15.	& 7.	& 9.6	& 0.8	& 2.25	& 0.18	& 340.	& 12.	& 2.	& 7.	& 18.3	& 0.9	& 15.6	& 0.9	& 7.8	& 0.3	\\
61	& 249.	& 11.	& -20.	& 4.	& 3.5	& 0.2	& 2.56	& 0.18	& 234.	& 8.	& 58.	& 3.	& 8.5	& 0.4	& 8.3	& 0.4	& 12.6	& 0.4	\\
62	& 276.	& 12.	& 9.	& 4.	& 9.2	& 0.6	& 2.21	& 0.15	& 338.	& 8.	& 6.	& 4.	& 16.9	& 0.6	& 17.3	& 0.6	& 8.3	& 0.2	\\
63	& 470	& 30	& -45.	& 12.	& 10.1	& 0.9	& 2.7	& 0.2	& 475.	& 6.	& 30.	& 3.	& 20.1	& 0.3	& 21.9	& 0.3	& 10.27	& 0.11	\\
64	& 336.	& 9.	& -9.	& 4.	& 22.9	& 0.9	& 2.98	& 0.12	& 334.	& 7.	& 77.	& 3.	& 57.4	& 1.7	& 56.9	& 1.7	& 14.3	& 0.3	\\
\hline
\end{tabular}
}

\end{minipage}
\end{table*}

\begin{table*}
\begin{minipage}{175mm}
\caption{\rev{The parameters of the narrow and broad components and
    the overall \ha\ models.}}
\label{tab:HaData}
{ \centering
\begin{tabular}{r@{\ \ \ }cc@{\ \ }r@{$\,\pm\,$}l@{\ \ \ }r@{$\,\pm\,$}l@{\ \ \ }r@{$\,\pm\,$}lc@{\ \ }r@{$\,\pm\,$}l@{\ \ \ }r@{$\,\pm\,$}l@{\ \ \ }r@{$\,\pm\,$}lr@{$\,\pm\,$}l@{\ \ \ }r@{$\,\pm\,$}l}
\hline
	& 	& \multicolumn{7}{c}{\han\ parameters}	& \multicolumn{7}{c}{\hab\ parameters}	& \multicolumn{4}{c}{\hatt\ parameters}	\\
ID	& Sy	& $N_N$	& \multicolumn{2}{c}{FWHM}	& \multicolumn{2}{c}{$v_\rmn{sh}$}	& \multicolumn{2}{c}{Flux}			& $N_B$	& \multicolumn{2}{c}{FWHM}	& \multicolumn{2}{c}{$v_\rmn{sh}$}	& \multicolumn{2}{c}{Flux}			& \multicolumn{2}{c}{Flux}			& \multicolumn{2}{c}{EqW}	\\
	& type	& (ct)	& \multicolumn{2}{c}{(\kms)}	& \multicolumn{2}{c}{(\kms)}		& \multicolumn{2}{c}{($\times 10^{-16}$)}	& (ct)	& \multicolumn{2}{c}{(\kms)}	& \multicolumn{2}{c}{(\kms)}		& \multicolumn{2}{c}{($\times 10^{-16}$)}	& \multicolumn{2}{c}{($\times 10^{-16}$)}	& \multicolumn{2}{c}{(\AA)}	\\
\hline
1	& Sy1.9	& 2	& 211.2	& 0.5	& 22.8	& 0.3	& 48.4	& 0.9	& 1	& 2424.	& 4.	& 49.1	& 1.9	& 104.2	& 1.0	& 152.6	& 1.3	& 72.5	& 0.6	\\
2	& NLS1	& 1	& 493.	& 12.	& -62.	& 4.	& 12.	& 3.	& 2	& 1388.	& 7.	& 6.	& 3.	& 522.	& 6.	& 534.	& 7.	& 169.	& 2.	\\
3	& Sy2	& 1	& 332.7	& 0.6	& -24.5	& 0.3	& 421.8	& 1.2	& 0	& \multicolumn{2}{c}{\nodata}	& \multicolumn{2}{c}{\nodata}	& \multicolumn{2}{c}{\nodata}	& 421.8	& 1.2	& 19.9	& 0.1	\\
4	& Sy1.5	& 1	& 624.	& 17.	& -41.	& 6.	& 106.	& 4.	& 2	& 2388.	& 11.	& 61.	& 3.	& 571.	& 8.	& 677.	& 9.	& 199.	& 3.	\\
5	& Sy2	& 2	& 432.0	& 0.3	& 13.0	& 0.2	& 168.0	& 1.6	& 0	& \multicolumn{2}{c}{\nodata}	& \multicolumn{2}{c}{\nodata}	& \multicolumn{2}{c}{\nodata}	& 168.0	& 1.6	& 72.0	& 0.7	\\
6	& Sy2	& 1	& 236.	& 5.	& 0.	& 4.	& 23.1	& 1.0	& 0	& \multicolumn{2}{c}{\nodata}	& \multicolumn{2}{c}{\nodata}	& \multicolumn{2}{c}{\nodata}	& 23.1	& 1.0	& 28.8	& 1.2	\\
7	& Sy2	& 1	& 311.0	& 0.7	& -36.6	& 0.4	& 96.2	& 1.6	& 1$^a$	& 2385.	& 13.	& -504.	& 6.	& 104.	& 3.	& 139.	& 4.	& 13.9	& 0.4	\\
8	& Sy2	& 1	& 458.	& 8.	& -105.	& 3.	& 45.7	& 1.0	& 2$^a$	& 1197.	& 13.	& 489.	& 8.	& 26.8	& 1.0	& 59.9	& 1.4	& 39.9	& 1.0	\\
9	& Sy1.0	& 1	& 405.	& 7.	& -74.	& 3.	& 62.	& 2.	& 2	& 4075.	& 4.	& -135.	& 2.	& 1178.	& 9.	& 1240.	& 9.	& 296.	& 2.	\\
\hline
\end{tabular}
}

~$^a$ The best-fitting models for some Sy2 spectra include broad
\ha\ components.
Each of these has strong wings on
its \oiii\ lines.
The \han\ and \nii\ lines are likely to have similar wings in their profiles,
so the \hab\ models are assumed to represent a blend
of these wings and not a genuine broad \ha\ component.
In these cases, the tabulated values of \hatt\ flux and EqW are
the sum of the \han\ and \textit{rescaled} \hab\ values, where the
\han/(\han+\nii) flux ratio is used to estimate the \ha\ contribution
to the \hab\ model.

\medskip
\rev{Columns are:
(1)	\fex\ sample ID number;
(2)	spectroscopic classification (as defined in \S\ref{sect:specTypes});
(3--6)	parameters relating to the narrow \ha\ model (\han):
	(3) the number of Gaussian components included in \han\ ($N_N$),
	(4) FWHM (\kms) and
	(5) velocity shift (\kms\ relative to \sii) of the
	\han\ component with the most flux, and
	(6) the combined flux ($\times 10^{-16} \  \cgsflux$) of the
        \han\ component(s);
(7--10)	parameters relating to the broad \ha\ model (\hab):
	(7) the number of Gaussian components included in \hab\ ($N_B$),
	(8) FWHM (\kms) and
	(9) velocity shift (\kms\ relative to \sii) of the
	\hab\ component with the most flux, and
	(10) the combined flux ($\times 10^{-16} \  \cgsflux$) of the
        \hab\ component(s);
(11) the flux ($\times 10^{-16} \  \cgsflux$) and
(12) equivalent width (\AA) of the overall \ha\ model (including both
        \han\ and \hab).}
\end{minipage}
\end{table*}
\begin{table*}
\begin{minipage}{175mm}
\contcaption{}
{ \centering
  \begin{tabular}{r@{\ \ }c@{\ \ }c@{\ }r@{$\,\pm\,$}l@{\ \ }r@{$\,\pm\,$}l@{\ \ }r@{$\,\pm\,$}lc@{\ }r@{$\,\pm\,$}l@{\ \ }r@{$\,\pm\,$}l@{\ \ }r@{$\,\pm\,$}lr@{$\,\pm\,$}l@{\ \ \ }r@{$\,\pm\,$}l}
\hline
	& 	& \multicolumn{7}{c}{\han\ parameters}	& \multicolumn{7}{c}{\hab\ parameters}	& \multicolumn{4}{c}{\hatt\ parameters}	\\
ID	& Sy	& $N_N$	& \multicolumn{2}{c}{FWHM}	& \multicolumn{2}{c}{$v_\rmn{sh}$}	& \multicolumn{2}{c}{Flux}			& $N_B$	& \multicolumn{2}{c}{FWHM}	& \multicolumn{2}{c}{$v_\rmn{sh}$}	& \multicolumn{2}{c}{Flux}			& \multicolumn{2}{c}{Flux}			& \multicolumn{2}{c}{EqW}	\\
	& type	& (ct)	& \multicolumn{2}{c}{(\kms)}	& \multicolumn{2}{c}{(\kms)}		& \multicolumn{2}{c}{($\times 10^{-16}$)}	& (ct)	& \multicolumn{2}{c}{(\kms)}	& \multicolumn{2}{c}{(\kms)}		& \multicolumn{2}{c}{($\times 10^{-16}$)}	& \multicolumn{2}{c}{($\times 10^{-16}$)}	& \multicolumn{2}{c}{(\AA)}	\\
\hline
10	& Sy1.5	& 1	& 322.	& 4.	& -72.4	& 1.7	& 106.	& 4.	& 2	& 4426.	& 15.	& 141.	& 3.	& 1093.	& 9.	& 1199.	& 10.	& 203.1	& 1.6	\\
11	& Sy1.0	& 1	& 325.	& 7.	& -90.	& 3.	& 29.3	& 0.7	& 1	& 8460	& 20	& -325.	& 12.	& 610.	& 2.	& 640.	& 2.	& 374.3	& 1.4	\\
12	& Sy1.0	& 1	& 553.	& 13.	& -42.	& 4.	& 392.	& 13.	& 2	& 3072.	& 11.	& 18.	& 2.	& 4720	& 30	& 5110	& 40	& 478.	& 3.	\\
13	& Sy2	& 2	& 332.	& 2.	& 45.4	& 1.7	& 143.	& 3.	& 1$^a$	& 2072.	& 13.	& 23.	& 6.	& 86.	& 3.	& 198.	& 4.	& 32.1	& 0.7	\\
14	& Sy1.0	& 1	& 349.	& 5.	& -60.7	& 1.8	& 49.2	& 1.2	& 2	& 6959.	& 13.	& 535.	& 5.	& 696.	& 4.	& 745.	& 4.	& 251.2	& 1.4	\\
15	& Sy2	& 3	& 364.7	& 0.4	& 4.6	& 0.2	& 140	& 20	& 0	& \multicolumn{2}{c}{\nodata}	& \multicolumn{2}{c}{\nodata}	& \multicolumn{2}{c}{\nodata}	& 140	& 20	& 45.	& 7.	\\
16	& Sy1.5	& 1	& 456.	& 10.	& -31.	& 3.	& 31.5	& 1.0	& 1	& 7150	& 80	& -60	& 30	& 228.	& 4.	& 260.	& 4.	& 276.	& 4.	\\
17	& Sy2	& 1	& 434.	& 6.	& -53.	& 2.	& 73.0	& 1.3	& 2$^a$	& 890.	& 10.	& 359.	& 6.	& 44.	& 1.	& 93.7	& 1.6	& 56.0	& 1.0	\\
18	& Sy1.5	& 1	& 264.	& 4.	& -94.3	& 1.8	& 52.3	& 1.5	& 2	& 4890	& 40	& -44.	& 6.	& 535.	& 5.	& 588.	& 5.	& 326.	& 3.	\\
19	& Sy1.5	& 2	& 331.1	& 0.4	& -44.7	& 0.2	& 68.	& 2.	& 1	& 3233.	& 2.	& 91.1	& 1.2	& 269.0	& 1.5	& 337.	& 3.	& 207.0	& 1.6	\\
20	& NLS1	& 1	& 401.	& 5.	& -52.4	& 1.9	& 560	& 20	& 2	& 1656.	& 3.	& 102.0	& 1.1	& 10850	& 60	& 11410	& 60	& 419.	& 2.	\\
21	& Sy1.5	& 1	& 454.	& 8.	& -65.	& 3.	& 256.	& 7.	& 1	& 3740	& 30	& -49.	& 11.	& 1248.	& 18.	& 1504.	& 19.	& 109.6	& 1.4	\\
22	& Sy1.9	& 1	& 546.	& 7.	& -66.	& 2.	& 302.	& 6.	& 2	& 5950	& 30	& -98.	& 7.	& 710.	& 9.	& 1012.	& 11.	& 502.	& 5.	\\
23	& Sy1.5	& 1	& 267.	& 5.	& -83.	& 2.	& 115.	& 5.	& 2	& 3242.	& 10.	& 68.	& 3.	& 1205.	& 10.	& 1319.	& 11.	& 186.7	& 1.6	\\
24	& NLS1	& 2	& 185.3	& 1.9	& -2.8	& 1.1	& 12.8	& 0.6	& 1	& 1933.	& 7.	& -34.	& 3.	& 43.8	& 0.8	& 56.5	& 1.0	& 48.5	& 0.9	\\
25	& NLS1	& 1	& 473.	& 7.	& -263.	& 3.	& 124.	& 7.	& 2	& 745.	& 2.	& 17.8	& 1.6	& 705.	& 9.	& 829.	& 12.	& 120.2	& 1.7	\\
26	& Sy1.5	& 1	& 317.	& 4.	& -66.4	& 1.7	& 117.	& 3.	& 1	& 4426.	& 17.	& -14.	& 7.	& 945.	& 7.	& 1062.	& 7.	& 392.	& 3.	\\
27	& NLS1	& 2	& 252.99	& 0.11	& -6.01	& 0.07	& 82.6	& 0.4	& 1	& 1682.3	& 0.9	& -3.9	& 0.5	& 100.9	& 0.3	& 183.5	& 0.5	& 189.2	& 0.5	\\
28	& Sy1.5	& 1	& 833.	& 14.	& 8.	& 4.	& 272.	& 5.	& 1	& 7760	& 30	& -659.	& 14.	& 1020.	& 8.	& 1292.	& 9.	& 648.	& 5.	\\
29	& Sy1.5	& 1	& 342.	& 8.	& -81.	& 3.	& 53.9	& 1.7	& 2	& 4800	& 30	& 214.	& 8.	& 464.	& 5.	& 518.	& 5.	& 341.	& 3.	\\
30	& Sy1.0	& 2	& 599.	& 5.	& 154.	& 3.	& 127.	& 5.	& 1	& 5905.	& 7.	& -115.	& 2.	& 1811.	& 7.	& 1938.	& 8.	& 399.6	& 1.7	\\
31	& Sy2	& 1	& 386.	& 5.	& 7.0	& 0.4	& 99.9	& 1.9	& 0	& \multicolumn{2}{c}{\nodata}	& \multicolumn{2}{c}{\nodata}	& \multicolumn{2}{c}{\nodata}	& 99.9	& 1.9	& 20.6	& 0.4	\\
32	& Sy1.9	& 1	& 412.7	& 1.0	& 10.6	& 0.3	& 158.1	& 1.9	& 1	& 2415.	& 10.	& 127.	& 4.	& 153.	& 3.	& 311.	& 4.	& 117.4	& 1.3	\\
33	& Sy1.0	& 1	& 387.	& 8.	& -75.	& 3.	& 57.	& 2.	& 2	& 2327.	& 2.	& 158.3	& 1.1	& 744.	& 6.	& 800.	& 7.	& 218.8	& 1.8	\\
34	& Sy2	& 3	& 218.	& 4.	& -81.8	& 1.5	& 97.	& 2.	& 0	& \multicolumn{2}{c}{\nodata}	& \multicolumn{2}{c}{\nodata}	& \multicolumn{2}{c}{\nodata}	& 97.	& 2.	& 17.3	& 0.4	\\
35	& Sy2	& 2	& 196.6	& 1.6	& 15.3	& 1.1	& 325.	& 4.	& 0	& \multicolumn{2}{c}{\nodata}	& \multicolumn{2}{c}{\nodata}	& \multicolumn{2}{c}{\nodata}	& 325.	& 4.	& 34.5	& 0.4	\\
36	& Sy1.5	& 2	& 602.	& 3.	& 49.	& 7.	& 88.	& 5.	& 1	& 3664.	& 8.	& 344.	& 2.	& 408.	& 3.	& 497.	& 6.	& 264.	& 3.	\\
37	& Sy2	& 1	& 254.0	& 0.4	& -12.34	& 0.19	& 185.	& 2.	& 0	& \multicolumn{2}{c}{\nodata}	& \multicolumn{2}{c}{\nodata}	& \multicolumn{2}{c}{\nodata}	& 185.	& 2.	& 30.5	& 0.3	\\
38	& Sy1.0	& 1	& 356.	& 8.	& -65.	& 3.	& 23.2	& 0.6	& 1	& 2917.	& 3.	& 44.4	& 1.9	& 384.9	& 0.7	& 408.1	& 0.9	& 193.5	& 0.4	\\
39	& Sy2	& 1	& 636.	& 14.	& -46.	& 5.	& 59.7	& 1.5	& 0	& \multicolumn{2}{c}{\nodata}	& \multicolumn{2}{c}{\nodata}	& \multicolumn{2}{c}{\nodata}	& 59.7	& 1.5	& 49.2	& 1.2	\\
40	& NLS1	& 1	& 391.	& 5.	& -21.	& 2.	& 61.	& 2.	& 2	& 1788.	& 5.	& 92.5	& 1.9	& 663.	& 6.	& 724.	& 6.	& 223.	& 2.	\\
41	& Sy1.0	& 1	& 312.	& 4.	& -69.0	& 1.7	& 242.	& 10.	& 2	& 6082.	& 14.	& 125.	& 4.	& 4500	& 30	& 4740	& 30	& 201.4	& 1.2	\\
42	& Sy1.5	& 2	& 686.	& 2.	& 43.	& 3.	& 295.	& 6.	& 1	& 3309.	& 15.	& -85.	& 4.	& 162.	& 3.	& 457.	& 6.	& 217.	& 3.	\\
43	& Sy2	& 1	& 479.	& 9.	& -33.	& 3.	& 109.	& 2.	& 0	& \multicolumn{2}{c}{\nodata}	& \multicolumn{2}{c}{\nodata}	& \multicolumn{2}{c}{\nodata}	& 109.	& 2.	& 56.1	& 1.2	\\
44	& Sy1.5	& 1	& 960	& 50	& -133.	& 14.	& 88.	& 5.	& 1	& 4750	& 40	& 92.	& 13.	& 507.	& 9.	& 595.	& 10.	& 173.	& 3.	\\
45	& gal	& 2	& 195.7	& 1.3	& 8.4	& 0.3	& 32.0	& 1.0	& 0	& \multicolumn{2}{c}{\nodata}	& \multicolumn{2}{c}{\nodata}	& \multicolumn{2}{c}{\nodata}	& 32.0	& 1.0	& 32.4	& 1.0	\\
46	& NLS1	& 1	& 279.	& 6.	& -84.	& 2.	& 88.	& 4.	& 2	& 1571.	& 4.	& 60.0	& 1.2	& 1582.	& 11.	& 1670.	& 11.	& 327.	& 2.	\\
47	& Sy1.0	& 1	& 336.	& 6.	& -61.	& 2.	& 40.	& 2.	& 2	& 4772.	& 11.	& 440.	& 6.	& 1709.	& 11.	& 1749.	& 11.	& 272.0	& 1.7	\\
48	& Sy1.5	& 1	& 374.	& 6.	& -79.	& 2.	& 116.	& 3.	& 2	& 5330	& 30	& -90.	& 6.	& 785.	& 8.	& 901.	& 9.	& 213.	& 2.	\\
49	& Sy2	& 2	& 540.3	& 1.3	& 88.	& 2.	& 470.	& 6.	& 1$^a$	& 3475.	& 13.	& -294.	& 5.	& 244.	& 3.	& 627.	& 7.	& 166.3	& 1.9	\\
50	& NLS1	& 1	& 256.	& 3.	& -86.3	& 1.2	& 50.6	& 1.5	& 2	& 780.	& 2.	& -34.6	& 0.6	& 311.	& 2.	& 361.	& 3.	& 161.6	& 1.3	\\
51	& Sy2	& 1	& 715.	& 8.	& -85.	& 3.	& 86.	& 4.	& 1$^a$	& 1970	& 30	& -7.	& 17.	& 209.	& 9.	& 163.	& 10.	& 53.	& 3.	\\
52	& Sy1.0	& 1	& 346.	& 7.	& -96.	& 3.	& 42.1	& 1.9	& 2	& 6410	& 40	& -302.	& 7.	& 845.	& 7.	& 887.	& 8.	& 229.	& 2.	\\
53	& NLS1	& 1	& 373.	& 4.	& -55.1	& 1.4	& 264.	& 6.	& 2	& 2106.	& 4.	& 65.7	& 1.1	& 2357.	& 14.	& 2621.	& 15.	& 441.	& 2.	\\
54	& Sy1.0	& 1	& 371.	& 6.	& 329.	& 3.	& 114.	& 4.	& 2	& 5460.	& 14.	& -178.	& 4.	& 3569	& 20	& 3680	& 20	& 301.1	& 1.6	\\
55	& Sy2	& 1	& 432.	& 5.	& 20.4	& 0.4	& 114.	& 2.	& 0	& \multicolumn{2}{c}{\nodata}	& \multicolumn{2}{c}{\nodata}	& \multicolumn{2}{c}{\nodata}	& 114.	& 2.	& 29.5	& 0.5	\\
56	& Sy1.0	& 1	& 393.	& 6.	& -47.	& 2.	& 403.	& 7.	& 2	& 3175.	& 7.	& -41.	& 2.	& 8020	& 30	& 8420	& 30	& 329.6	& 1.0	\\
57	& Sy2	& 1	& 549.	& 7.	& -86.	& 3.	& 103.	& 2.	& 1$^a$	& 1800	& 20	& -242.	& 7.	& 70.	& 4.	& 138.	& 4.	& 45.4	& 1.4	\\
58	& Sy1.0	& 1	& 397.	& 7.	& -63.	& 3.	& 58.	& 3.	& 2	& 4123	& 20	& -191.	& 5.	& 2340	& 20	& 2400	& 20	& 386.	& 3.	\\
59	& NLS1	& 1	& 343.	& 6.	& -38.	& 2.	& 68.	& 2.	& 2	& 2008.	& 5.	& 215.1	& 1.6	& 1265.	& 7.	& 1333.	& 7.	& 205.7	& 1.1	\\
60	& Sy1.5	& 1	& 768.	& 2.	& 263.5	& 0.8	& 227.	& 3.	& 1	& 3175.	& 11.	& 120.	& 3.	& 320.	& 4.	& 548.	& 5.	& 130.1	& 1.2	\\
61	& NLS1	& 1	& 257.	& 4.	& -95.2	& 1.5	& 39.3	& 0.9	& 2	& 730.8	& 1.5	& 39.3	& 0.5	& 221.5	& 1.5	& 260.9	& 1.7	& 204.3	& 1.4	\\
62	& NLS1	& 1	& 310.	& 4.	& -68.1	& 1.5	& 78.	& 3.	& 2	& 1791.	& 5.	& 104.2	& 1.4	& 937.	& 8.	& 1014.	& 8.	& 259.	& 2.	\\
63	& Sy1.0	& 1	& 541.	& 11.	& -116.	& 4.	& 35.8	& 0.8	& 2	&2319.8	& 0.2	& -115.62	& 0.11	&1095.8	& 0.2	&1131.6	& 0.8	& 308.1	& 0.2	\\
64	& Sy1.5	& 1	& 374.	& 3.	& -86.8	& 1.3	& 295.	& 5.	& 2	& 5591.	& 7.	& 255.	& 3.	& 2264.	& 12.	& 2559.	& 13.	& 347.4	& 1.8	\\
\hline
\end{tabular}
}

\end{minipage}
\end{table*}

\begin{table*}
\begin{minipage}{133mm}
\caption{\textit{Rosat} X-ray data.}
\label{tab:XrayFluxes}
{
\begin{tabular}{rccccccc}
\hline
\multicolumn{1}{c}{ID}
	 & PSPC			 & \multicolumn{2}{c}{RASS}			 & \multicolumn{4}{c}{Estimated fluxes ($\times 10^{-14} \  \cgsflux$)} \\
\multicolumn{1}{c}{num}
	 & count rate		 & HR1			 & HR2			 &\multicolumn{2}{c}{0.1--2.4~keV}&\multicolumn{2}{c}{233--300~eV} \\
	 & (cps)		 & 			 & 			 & $\Gamma=1.5$	 & $\Gamma=3.0$	 & $\Gamma=1.5$	 & $\Gamma=3.0$	\\
\hline
2	 & 0.0413 $\pm$ 0.0027	 & 	\nodata		 & 	\nodata		 &   57.8	 &   45.7	 &  1.25	 &   1.67	\\
3	 & 0.0065 $\pm$ 0.0007	 & 	\nodata		 & 	\nodata		 &    9.1	 &    7.2	 &  0.20	 &   0.72	\\
4	 & 0.0368 $\pm$ 0.0107	 &\ 1.00 $\pm$ 0.27	 & -0.09 $\pm$ 	0.30	 &   48.0	 &   29.9	 &  1.04	 &   3.22	\\
9	 & 0.0795 $\pm$ 0.0193	 & -0.43 $\pm$ 0.20	 &\ 0.50 $\pm$ 0.38	 &  103.6	 &   64.5	 &  2.23	 &   6.96	\\
10	 & 0.1412 $\pm$ 0.0236	 &\ 0.12 $\pm$ 0.16	 &\ 0.21 $\pm$ 0.20	 &  184.1	 &  114.7	 &  3.97	 &  12.36	\\
12	 & 0.4414 $\pm$ 0.0441	 & -0.04 $\pm$ 0.09	 & -0.12 $\pm$ 0.13	 &  575.6	 &  358.5	 & 12.41	 &  38.64	\\
14	 & 0.1840 $\pm$ 0.0319	 &\ 0.03 $\pm$ 0.16	 &\ 0.13 $\pm$ 0.22	 &  239.9	 &  149.5	 &  5.17	 &  16.11	\\
20	 & 1.6260 $\pm$ 0.0643	 &\ 0.38 $\pm$ 0.03	 &\ 0.14 $\pm$ 0.04	 & 2120.3	 & 1320.7	 & 45.70	 & 142.35	\\
21	 & 0.1201 $\pm$ 0.0181	 &\ 0.17 $\pm$ 0.15	 &\ 0.36 $\pm$ 0.18	 &  156.6	 &   97.6	 &  3.38	 &  10.51	\\
23	 & 0.3864 $\pm$ 0.0344	 & -0.24 $\pm$ 0.08	 & -0.03 $\pm$ 0.14	 &  503.9	 &  313.9	 & 10.86	 &  33.83	\\
24	 & 0.0307 $\pm$ 0.0133	 &\ 0.79 $\pm$ 0.56	 & -0.84 $\pm$ 1.01	 &   40.0	 &   24.9	 &  0.86	 &   2.69	\\
25	 & 2.6580 $\pm$ 0.0879	 & -0.74 $\pm$ 0.02	 & -0.37 $\pm$ 0.08	 & 3466.0	 & 2159.0	 & 74.71	 & 232.70	\\
26	 & 0.2398 $\pm$ 0.0266	 & -0.30 $\pm$ 0.10	 & -0.15 $\pm$ 0.18	 &  312.7	 &  194.8	 &  6.74	 &  20.99	\\
27	 & 0.0288 $\pm$ 0.0025	 & 	\nodata		 & 	\nodata		 &   40.3	 &   31.9	 &  0.87	 &   3.43	\\
29	 & 0.0959 $\pm$ 0.0179	 & -0.39 $\pm$ 0.16	 & -0.25 $\pm$ 0.31	 &  125.0	 &   77.9	 &  2.69	 &   8.39	\\
30	 & 0.1076 $\pm$ 0.0175	 & -0.09 $\pm$ 0.16	 &\ 0.33 $\pm$ 0.22	 &  140.3	 &   87.4	 &  3.02	 &   9.42	\\
33	 & 0.1492 $\pm$ 0.0203	 &\ 0.10 $\pm$ 0.13	 &\ 0.35 $\pm$ 0.16	 &  194.6	 &  121.2	 &  4.19	 &  13.06	\\
35	 & 0.0167 $\pm$ 0.0079	 &\ 0.35 $\pm$ 0.47	 &\ 0.40 $\pm$ 0.48	 &   21.8	 &   13.6	 &  0.47	 &   1.46	\\
36	 & 0.0537 $\pm$ 0.0169	 & -0.15 $\pm$ 0.27	 & -0.01 $\pm$ 0.46	 &   70.0	 &   43.6	 &  1.51	 &   4.70	\\
40	 & 0.1060 $\pm$ 0.0348	 &\ 0.39 $\pm$ 0.32	 &\ 0.05 $\pm$ 0.78	 &  138.2	 &   86.1	 &  2.98	 &   9.28	\\
46	 & 0.0951 $\pm$ 0.0214	 & -0.46 $\pm$ 0.18	 & -0.05 $\pm$ 0.39	 &  124.0	 &   77.2	 &  2.67	 &   8.33	\\
47	 & 0.1201 $\pm$ 0.0227	 &\ 0.04 $\pm$ 0.19	 & -0.09 $\pm$ 0.26	 &  156.6	 &   97.6	 &  3.38	 &  10.51	\\
48	 & 0.1024 $\pm$ 0.0161	 &\ 0.03 $\pm$ 0.15	 &\ 0.17 $\pm$ 0.18	 &  133.5	 &   83.2	 &  2.88	 &   8.96	\\
49	 & 0.0086 $\pm$ 0.0011	 & 	\nodata		 & 	\nodata		 &   12.0	 &    9.5	 &  0.26	 &   1.03	\\
50	 & 0.6629 $\pm$ 0.0386	 & -0.52 $\pm$ 0.05	 & -0.05 $\pm$ 0.11	 &  864.4	 &  538.5	 & 18.63	 &   5.80	\\
52	 & 0.0256 $\pm$ 0.0114	 &\ 0.28 $\pm$ 0.37	 & -0.26 $\pm$ 0.45	 &   33.3	 &   20.8	 &  0.72	 &   2.24	\\
54	 & 0.2476 $\pm$ 0.0197	 &\ 0.33 $\pm$ 0.07	 &\ 0.00 $\pm$ 0.09	 &  322.9	 &  201.1	 &  6.96	 &  21.68	\\
56	 & 0.8849 $\pm$ 0.0306	 &\ 0.03 $\pm$ 0.03	 &\ 0.16 $\pm$ 0.04	 & 1153.9	 &  718.8	 & 24.87	 &  77.47	\\
58	 & 0.4159 $\pm$ 0.0260	 & -0.31 $\pm$ 0.06	 &\ 0.02 $\pm$ 0.10	 &  542.3	 &  337.8	 & 11.69	 &  36.41	\\
59	 & 0.0153 $\pm$ 0.0064	 &\ 0.17 $\pm$ 0.51	 & -1.00 $\pm$ 0.37	 &   19.9	 &   12.4	 &  0.43	 &   1.34	\\
61	 & 0.0263 $\pm$ 0.0097	 &\ 0.04 $\pm$ 0.35	 & -1.00 $\pm$ 1.27	 &   34.3	 &   21.4	 &  0.74	 &   2.30	\\
63	 & 0.0838 $\pm$ 0.0201	 &\ 0.68 $\pm$ 0.20	 &\ 0.53 $\pm$ 0.22	 &  109.3	 &   68.1	 &  2.36	 &   7.34	\\
\hline
\end{tabular}
}

\medskip
Columns are:
(1)	\fex\ sample ID number;
(2)	\textit{Rosat} PSPC count rates in the 0.1--2.4~keV band
	reported by RASS (or 0.24--2.0~keV count rates from the WGA
	catalogue when undetected by RASS);
(3--4)	RASS hardness ratios (no value entered for the WGA-only
	objects);
(5--8)	\textit{Rosat} fluxes ($\times 10^{-14} \  \cgsflux$)
        estimated from PSPC count rates, 
	assuming power law spectra absorbed by $N_H = 3 \times 10^{20}
	\, \rmn{cm}^{-2}$ and $z = 0$:
(5)	0.1--2.4~keV flux, assuming $\Gamma = 1.5$;
(6)	0.1--2.4~keV flux, assuming $\Gamma = 3.0$;
(7)	233--300~eV flux, assuming $\Gamma = 1.5$; and
(8)	233--300~eV flux, assuming $\Gamma = 3.0$.
The values in columns (7--8) are based upon the same models as used in
columns (5--6), the only difference being that the fluxes are
integrated from 233 to 300~eV.  
Note that the conversion from \textit{Rosat} count rates to
233--300~eV fluxes is more strongly model dependent than the
0.1--2.4~eV flux estimates.

\end{minipage}
\end{table*}


\end{document}